\documentclass[12pt]{article}
\usepackage{graphics}
\usepackage{amssymb}
\usepackage{epsfig,graphicx}
\usepackage{color}
\usepackage{amsmath}
\usepackage{amsthm}
\usepackage{amsopn}
\usepackage{subfigure}
\usepackage{xcolor}
\usepackage{hyperref}
\hypersetup{
    colorlinks=false,
    pdfborder={0 0 1},
    linkbordercolor=red,
    urlbordercolor=cyan
}
\usepackage{caption}
\usepackage{afterpage}
\usepackage{float}
\usepackage{booktabs}
\usepackage{url}
\def\uno{\mbox{1 \kern-.59em {\rm l}}}

\def\be{\begin{equation}}
	\def\ee{\end{equation}}
\def\bea{\begin{eqnarray}}
	\def\eea{\end{eqnarray}}

\usepackage[numbers, sort&compress]{natbib}
\begin{document}
\begin{center}
{\bf{\large  Quantum Correlations of Neutrinos in the Kerr-Newman Space-time
}}
\vskip 1em
\begingroup
\renewcommand{\thefootnote}{\fnsymbol{footnote}}
\setcounter{footnote}{0}
Ze-Wen Li\textsuperscript{1,\,}\footnote{\href{mailto:1020241985@glut.edu.cn}{1020241985@glut.edu.cn}} and
Shu-Jun Rong\textsuperscript{1,\,}\footnote{\href{mailto:rongshj@glut.edu.cn}{rongshj@glut.edu.cn}}\\
\endgroup
\vskip 1em
 $^{1}$\textit{College of Physics and Electronic Information Engineering,}\\
\textit{Guilin University of Technology, Guilin, Guangxi 541004, China}
 \end{center}
\vspace*{0.5cm}

\begin{abstract}

 Quantum phases establish a connection between gravitation and quantum information, offering a novel avenue for exploring the properties of space-time.
In this paper, we investigate the quantum correlations (QCs) of neutrinos in the Kerr--Newman space-time for both zero- and nonzero-angular-momentum propagation. The results show that, for zero-angular-momentum propagation, the oscillation periods of the survival probability and QCs progressively decrease with propagation distance in the inward direction. In the outward direction, increasing \(M\) lengthens the oscillation periods of \(P_{\nu_e\rightarrow\nu_e}\), entanglement, and the monogamy of nonlocality, whereas increasing the angular momentum \(a\) or charge \(Q\) shortens them. For nonzero-angular-momentum propagation, the metric parameters also generate local profile modulations through additional two-path interference terms, rather than merely rescaling the oscillation period. Furthermore, we find that, despite differences in their ranges of variation, entanglement and coherence exhibit highly consistent oscillatory behavior in both propagation cases. These findings provide a comprehensive understanding of neutrino-based relativistic quantum information.
\end{abstract}

\newpage
\newpage
\section{Introduction}

Owing to their extremely weak interactions with matter, neutrinos are expected to be less susceptible to environmental decoherence than electrons or photons, making them promising candidates for quantum information processing. In this context, quantum correlations (QCs) in neutrinos, such as entanglement, coherence, and steering \cite{Blasone:2007vw,Blasone:2007wp,Bittencourt:2023asd,Blasone:2014jea,Blasone:2015lya,Alok:2014gya,Banerjee:2015mha,Formaggio:2016cuh,Ettefaghi:2020otb,Banerjee:2025gau,Dixit:2019swl,Siwach:2022xhx,Dixit:2023fke,Konwar:2024pkh,Alok:2024xeg,SinghKoranga:2024rum,Wang:2024tfh,
Nambiar:2026zhs,Burrage:2025xac} have become active topics in recent years. An important feature of these QCs is that they can often be expressed in terms of flavor oscillation probabilities, thereby connecting quantum-resource measures with neutrino oscillation observables~\cite{Yadav:2026lsx,Chundawat:2026jjd,Huang:2026bws,Chundawat:2026lcm}.

Among these QC quantities, entanglement and coherence provide two fundamental perspectives for characterizing three-flavor neutrino oscillations \cite{Kurashvili:2020nwb,Li:2021fft,Blasone:2007vw,Blasone:2014jea,Banerjee:2015mha,Formaggio:2016cuh,Ettefaghi:2020otb,Banerjee:2025gau,Blasone:2015lya,Dixit:2019swl}. Entanglement captures nonclassical correlations among different flavor modes. Bipartite entanglement can be quantified by concurrence \cite{Hill:1997pfa}, negativity \cite{Vidal:2002zz}, and entanglement of formation \cite{Wootters:1997id}. Coherence, on the other hand, describes the superposition structure of flavor states and can be quantified by the \(l_1\)-norm \cite{Song:2018bma,Ettefaghi:2022nsq} or examined through the Leggett-Garg inequality as a qualitative criterion \cite{Formaggio:2016cuh}. For multipartite systems such as three-flavor neutrinos, it is also important to understand how QCs are distributed among different subsystems \cite{Gour:2018uts,Terhal:2003pzu,Seevinck:2010hzk,Koashi:2003pgf,Guo:2019yyl}. This distribution is described by monogamy relations, which express the restricted shareability of quantum correlations. Conventional CKW-type monogamy relations are mainly formulated in terms of bipartite entanglement measures and therefore provide only a partial description of multipartite correlations \cite{Coffman:1999jd}. To overcome this limitation, Guo and Zhang established a stricter framework for defining multipartite entanglement measures by introducing the unification and hierarchy conditions, and they further proposed complete and tightly complete monogamy relations \cite{Guo:2019mue}. These developments provide a more comprehensive basis for characterizing the distribution of multipartite entanglement. Consequently, monogamy properties have been investigated in neutrino oscillations as a useful way to analyze the multipartite structure of flavor correlations \cite{Wang:2023rbf,Wang:2024tfh}.

Quantum phases may provide a direct route for connecting gravitational effects with QCs, offering a novel avenue for probing the properties of space-time. Colella et al.\ observed a gravitationally induced quantum phase shift in neutron interferometry \cite{Colella:1975dq}. Subsequently, Stodolsky developed a semiclassical description of the phase accumulated by a particle propagating in an external gravitational field and applied it to matter- and light-wave interferometry \cite{Stodolsky:1978ks}. These phase-based studies indicate that quantum systems governed by interference can be sensitive to gravitational backgrounds. Neutrino flavor oscillation offers a natural extension of this idea, since a flavor eigenstate is a coherent superposition of different mass eigenstates and the oscillation is governed by their relative propagation phases \cite{Bilenky:2018hbz,Forero:2014bxa,Esteban:2018azc,Esteban:2020cvm,SNO:2001kpb,Gonzalez-Garcia:2002bkq,KamLAND:2004mhv,Bilenky:2005mx,Super-Kamiokande:1998kpq}. In curved space-time, gravitational effects on neutrino oscillations are therefore encoded in the modification of the phases of different mass eigenstates \cite{Fornengo:1996ef,Pereira:2000kq,Crocker:2003cw,Ahluwalia:1996ev,Godunov:2009ce,Visinelli:2014xsa,Turimov:2023itj}. A covariant expression for the neutrino phase in a general curved space-time was obtained by Fornengo et al. \cite{Fornengo:1996ef}, and the WKB approximation for radial propagation was derived by Visinelli \cite{Visinelli:2014xsa}. For non-radial propagation, gravitational lensing further enriches the phase structure by introducing path-dependent contributions \cite{Chakrabarty:2021bpr,Shi:2025xkd,Chakrabarty:2023kld,Alexandre:2025qip,Alloqulov:2024sns,Shi:2024flw,Shi:2025ywa,Shi:2025plr}.

Based on gravitational effects on the neutrino-oscillation phase, Ettefaghi et al. studied quantum coherence for two-flavor neutrinos propagating in the Schwarzschild metric by using both the Leggett-Garg inequality (LGI) and the \(l_1\)-norm \cite{Ettefaghi:2022nsq}. They found that gravity can damp the maximum value of the Leggett-Garg parameter \(K_3\) and even remove LGI violation in certain energy intervals, whereas the maximum value of the \(l_1\)-norm coherence remains unchanged. Wang et al. extended this issue to three-flavor neutrinos and investigated entanglement, nonlocality, and their monogamy properties in the Schwarzschild background \cite{Wang:2024tfh}. Their results show that, under specific gravitational parameters, entanglement can be either enhanced or suppressed, and the effective bound of the maximum bipartite nonlocality of the neutrino system is modified. These works indicate that classical gravitational fields can induce nontrivial modifications in the QCs of neutrinos.

However, these studies of gravitationally affected neutrino QCs have mainly focused on radial propagation in Schwarzschild space-time, corresponding to a nonrotating source and zero-angular-momentum neutrinos. The effects of source rotation and gravitational lensing involving nonzero-angular-momentum propagation have therefore remained largely unexplored. To go beyond this limitation, we consider the Kerr--Newman space-time, which provides a more general background characterized by the mass, angular momentum, and charge of the gravitational source. We investigate three-flavor neutrino oscillations and analyze QCs, including tripartite entanglement, quantum coherence, and the monogamy property of nonlocality, in the zero-angular-momentum case. We also consider nonzero-angular-momentum propagation to illustrate gravitational lensing effects. We examine the influence of space-time parameters on the oscillation probabilities and demonstrate the properties of QCs beyond those in Schwarzschild space-time.

The paper is organized as follows: In Sec.\ref{2}, we derive the neutrino phases for zero- and nonzero-angular-momentum propagation in Kerr-Newman space-time. In Sec.\ref{3}, we calculate and plot the variations in neutrino oscillation probabilities and QCs for zero-angular-momentum propagation. In Sec.~\ref{4}, we present the results for nonzero-angular-momentum propagation, including gravitational lensing effects. Finally, the conclusions are given in Sec.\ref{5}. Throughout the paper, we use units with $G = \hbar = c = 1$.
	
\section{The oscillation probability of neutrinos in the Kerr-Newman metric }\label{2}

In this section, we derive the oscillation probabilities of neutrinos in the Kerr-Newman metric within the three-flavor scenario. We first introduce the oscillation phases and probabilities in flat space-time and then extend them to the Kerr-Newman metric.\\

\subsection*{A. Oscillation probability in flat space-time}

 A neutrino flavor state $\vert\nu_{\alpha}\rangle$ ($\alpha = e, \mu, \tau$) is a coherent superposition of the mass eigenstates $\nu_k$ ($k = 1, 2, 3$), i.e.,
\begin{equation}\label{eq:formula1}
\vert\nu_{\alpha}\rangle=\sum_{k}U_{\alpha k}^{*}\vert\nu_{k}\rangle,
\end{equation}
where $U$ is the Pontecorvo--Maki--Nakagawa--Sakata (PMNS) matrix \cite{Maki:1962mu,Pontecorvo:1957qd}. The matrix is parameterized by three mixing angles, $\theta_{ij}$, and one CP-violating phase, namely
\begin{equation}
	U =
	\begin{pmatrix}
	c_{12}c_{13} & s_{12}c_{13} & s_{13}e^{-i\delta_{cp}} \\
	-s_{12}c_{23} - c_{12}s_{13}s_{23}e^{-i\delta_{cp}} & c_{12}c_{23} - s_{12}s_{13}s_{23}e^{i\delta_{cp}} & c_{13}s_{23} \\
	s_{12}s_{23} - c_{12}s_{13}c_{23}e^{i\delta_{cp}} & -c_{12}c_{23} - s_{12}s_{13}c_{23}e^{i\delta_{cp}} & c_{13}c_{23}
	\end{pmatrix},
	\label{eq:formula2}
	\end{equation}
	where \(   c_{ij} = \cos\theta_{ij}   \) and \(   s_{ij} = \sin\theta_{ij}   \) (\(   i,j = 1,2,3   \)).
The mass eigenstate is expressed as follows:
\begin{equation}\label{eq:formula3}
	\vert\nu_{k}(t, \vec{x})\rangle = e^{-i\varPhi_{k}}\vert\nu_{k}\rangle.
	\end{equation}
For a neutrino produced at point \( A \) (\( t_A, \vec{x}_A \)) and detected at point \( B \) (\( t_B, \vec{x}_B \)), the phase in the plane-wave ansatz is
\begin{equation}\label{eq:formula4}
\varPhi_{k} = E_{k}(t_{B} - t_{A}) - \vec{p}_{k} \cdot (\vec{x}_{B} - \vec{x}_{A}).
\end{equation}
Accordingly, the flavor oscillation probability can be written as \cite{Bilenky:2018hbz}:
\begin{equation}
\begin{split}
P_{\alpha\rightarrow\beta}
&=\left|\left\langle\nu_{\beta}\mid\nu_{\alpha}(t_B,\vec{x}_B)\right\rangle\right|^2 \\
&=\delta_{\alpha\beta}
-4\sum_{k>j}\mathrm{Re}\!\left(U_{\alpha k}^{*}U_{\beta k}U_{\alpha j}U_{\beta j}^{*}\right)
\sin^2\!\left(\frac{\varPhi_{kj}}{2}\right) \\
&\quad+2\sum_{k>j}\mathrm{Im}\!\left(U_{\alpha k}^{*}U_{\beta k}U_{\alpha j}U_{\beta j}^{*}\right)
\sin\!\left(\varPhi_{kj}\right),
\label{eq:formula5}
\end{split}
\end{equation}
where $\varPhi_{kj} = \varPhi_{k} - \varPhi_{j}$. For relativistic neutrinos, one obtains
\begin{equation}
\Phi_{kj} \simeq \dfrac{\Delta m_{kj}^2}{2E_0} \vert \vec{x}_B - \vec{x}_A \vert,
\label{eq:formula8}
\end{equation}
with $\Delta m_{kj}^2 = m_k^2 - m_j^2$.\\

\subsection*{B. Oscillation probability for zero-angular-momentum propagation}

The neutrino oscillation phase in curved spacetime can be expressed in the covariant form \cite{Stodolsky:1978ks}:
\begin{equation}\label{eq:formula9}
\Phi_k = \int_{\mathcal{A}}^{\mathcal{B}} \left[ E_k dt - p_k(r) dr - J_k d\phi \right],
\end{equation}
where \( E_k \equiv p_t^{(k)} \), $p_k(r) \equiv -p_r^{(k)}$, and \( J_k \equiv -p_\phi^{(k)} \), with the momentum
\begin{equation}\label{eq:formula10}
p_\mu^{(k)} = m_k g_{\mu\nu} \frac{dx^\nu}{ds}.
\end{equation}
Here, $g_{\mu\nu}$ denotes the metric, $ds$ the line element, $x^\mu$ the coordinate, and $m_k$ satisfies the mass-shell relation:
\begin{equation}\label{eq:formula11}
m_k^2 = g^{\mu\nu} p_\mu^{(k)} p_\nu^{(k)}.
\end{equation}
The line element of the Kerr--Newman metric is given by
\begin{equation}\label{eq:formula12}
\begin{split}
ds^{2} &= \left(1 - \frac{\Lambda}{\rho^{2}}\right)dt^{2} + \frac{2\Lambda a \sin^{2}\theta}{\rho^{2}}dt\,d\phi - \frac{\rho^{2}}{\Delta}dr^{2} - \rho^{2}d\theta^{2} \\
&\quad - \left[a^{2} + r^{2} + \frac{a^{2}\Lambda \sin^{2}\theta}{\rho^{2}}\right]\sin^{2}\theta d\phi^{2},
\end{split}
\end{equation}
where \(M\), \(a\), and \(Q\) denote the mass, specific angular momentum, and electric charge of the central object, respectively. The auxiliary metric functions are defined as \(\Lambda=r_s r-r_Q^2\), \(\Delta=r^2+a^2-\Lambda\), and \(\rho^2=r^2+a^2\cos^2\theta\), with \(r_s=2M\) and \(r_Q=Q\).

In this paper, we consider neutrino propagation in the equatorial plane, where $\theta=\pi/2$. Owing to the stationarity and axisymmetry of the Kerr--Newman spacetime, the covariant momentum components $p_t^{(k)}$ and $p_\phi^{(k)}$ are conserved along the trajectory. Consequently, for zero angular momentum, the phase is given by
\begin{equation}\label{eq:formula13}
\Phi_k = \int_{r_A}^{r_B} \left[ E_k \left( \frac{dt}{dr} \right)_0 - p_k(r) \right] dr,
\end{equation}
where the subscript ``0'' denotes the corresponding quantity for a massless particle \cite{Fornengo:1996ef}.

From Eqs.~(\ref{eq:formula10}) and~(\ref{eq:formula12}), we obtain
\begin{align}
p_t^{(k)}
&=m_k\left[
\left(1-\frac{r_s}{r}+\frac{Q^2}{r^2}\right)
\frac{dt}{ds}
+a\left(\frac{r_s}{r}-\frac{Q^2}{r^2}\right)
\frac{d\phi}{ds}
\right]
\label{eq:formula14},\\
p_r^{(k)}
&=-m_k\frac{r^2}{r^2-r_s r+a^2+Q^2}
\frac{dr}{ds}
\label{eq:formula15},\\
p_\phi^{(k)}
&=m_k\left[
a\left(\frac{r_s}{r}-\frac{Q^2}{r^2}\right)
\frac{dt}{ds}
-\left(
r^2+a^2+\frac{a^2r_s}{r}
-\frac{a^2Q^2}{r^2}
\right)
\frac{d\phi}{ds}
\right]
=0
\label{eq:formula16}.
\end{align}
Since $E_k \equiv p_t^{(k)}$, the mass-shell relation in Eq.~(\ref{eq:formula11}) yields the following expression:
\begin{equation}
p_k(r)=\sqrt{\frac{E_k^2(r^4+a^2r^2+a^2r_s r-a^2Q^2)-m_k^2(r^2+a^2-r_s r+Q^2)r^2}{(r^2+a^2-r_s r+Q^2)^2}}.
\end{equation}
For relativistic neutrinos, we  have
\begin{equation}\label{eq:formula17}
p_k(r)=\frac{E_k\sqrt{r^4+a^2r^2+a^2r_s r-a^2Q^2}}{r^2+a^2-r_s r+Q^2}\left(1-\frac{1}{2}\frac{m_k^2(r^2+a^2-r_s r+Q^2)r^2}{(r^4+a^2r^2+a^2r_s r-a^2Q^2)E_k^2}\right).
\end{equation}
From Eqs.~(\ref{eq:formula14})--(\ref{eq:formula16}) together with Eq.~(\ref{eq:formula17}), we obtain:
\begin{equation}\label{eq:formula18}
\left(\frac{dt}{dr}\right)_0
=\frac{\sqrt{r^4+a^2r^2+a^2r_s r-a^2Q^2}}
{r^2+a^2-r_s r+Q^2}.
\end{equation}
Using the approximation \cite{Bilenky:2018hbz}
\begin{equation}\label{eq:formula19}
E_k\simeq E_0+\frac{m_k^2}{2E_0},
\end{equation} we obtain the phase difference as follows \cite{Ren:2010yf}
\begin{equation}\label{eq:formula20}
\Phi_{kj}\simeq
\frac{\Delta m_{kj}^2}{2E_0}
\int_{r_A}^{r_B}
\frac{r^2\,dr}
{\sqrt{r^4+a^2r^2+a^2r_s r-a^2Q^2}}.
\end{equation}
Here, $E_0$ denotes the energy of a neutrino at infinity.
Equation~(\ref{eq:formula20}) is evaluated numerically, and the resulting phase difference is used to calculate the radial oscillation probability from Eq.~(\ref{eq:formula5}) in the Kerr--Newman metric.

\subsection*{C. Oscillation probability for nonzero-angular-momentum propagation}

For non-radial propagation, we must account for angular momentum. The phase in Eq.~(\ref{eq:formula9}) is written as follows:
\begin{equation}\label{eq:formula21}
\Phi_k= \int_{r_A}^{r_B} \left[ E_k \left( \frac{dt}{dr} \right)_0 - p_k(r) - J_k \left( \frac{d\phi}{dr} \right)_0 \right] dr.
\end{equation}
The angular momentum can be written as
\begin{equation}\label{eq:formula22}
J_k=E_k b v_k^{(\infty)},
\end{equation}
 where \(v_k^{(\infty)}\) is the velocity at infinity and $b$ is the impact parameter \cite{Weinberg:1972kfs}.
For relativistic neutrinos, we obtain
\begin{equation}
v_k^{(\infty)} = \frac{\sqrt{E_k^2 - m_k^2}}{E_k} \approx 1 - \frac{m_k^2}{2E_k^2}.
\label{eq:formula23}
\end{equation}

For a given $J_k$, $p_k$ obtained from the mass-shell relation in Eq.~(\ref{eq:formula11}) is expressed as
\begin{equation}\label{eq:formula24}
p_k(r)
=
\frac{E_k\sqrt{R(r,b)}}{\Delta}
\left[
1+\frac{m_k^2C(r,b)}
{2E_k^2R(r,b)}
\right],
\end{equation}
where the auxiliary functions $R(r,b)$ and $C(r,b)$ are defined as
\begin{align*}
R(r,b)
&=
\left(r^2+a^2-ab\right)^2
-\Delta\left(a-b\right)^2,\\
C(r,b)
&=
b^2\left(\Delta-a^2\right)
+ab\left(r_s r-Q^2\right)
-r^2\Delta.
\end{align*}
Substituting $p_\phi^{(k)}=-J_k$ into Eq.~(\ref{eq:formula16}) yields the following coordinate derivatives along the massless trajectory:
\begin{align}
\left(\frac{dt}{dr}\right)_0
&=
\frac{
r^2\left(r^2+a^2\right)
-a(a-b)\left(Q^2-r_s r\right)
}{
\Delta\sqrt{R(r,b)}
}
\label{eq:formula27},\\
\left(\frac{d\phi}{dr}\right)_0
&=
\frac{
-a\left(Q^2-r_s r\right)
+b\left(r^2+Q^2-r_s r\right)
}{
\Delta\sqrt{R(r,b)}
}
\label{eq:formula28}.
\end{align}
Using the results above, the neutrino phase for propagation with nonzero angular momentum in the Kerr--Newman spacetime can be written as
\begin{equation}\label{eq:formula30}
\Phi_k
=
\int_{r_A}^{r_B}
\frac{m_k^2r^2}
{2E_0\sqrt{R(r,b)}}\,dr.
\end{equation}
In the following analysis, the phase is evaluated directly using numerical integration.\\

\section{Quantum correlations for zero-angular-momentum propagation}\label{3}
As discussed above, QCs can be expressed in terms of oscillation probabilities. We first present the oscillation probabilities in the Kerr--Newman spacetime and then investigate the corresponding QCs.

\subsection{Numerical results on oscillation probabilities}

Using Eqs.~(\ref{eq:formula5}) and~(\ref{eq:formula20}), we numerically evaluate the oscillation probabilities for zero-angular-momentum neutrinos. We first consider outward propagation. Owing to frame dragging, the condition $p_\phi^{(k)}=0$ does not imply
$d\phi=0$ in Boyer--Lindquist coordinates; rather, it corresponds to locally
radial propagation as measured by a zero-angular-momentum observer
(ZAMO). Accordingly, we express the results in terms of the local
energy and proper distance measured in the ZAMO frame.
A ZAMO has the four-velocity $u_{\mathrm{Z}}^\mu=\alpha^{-1}(1,0,0,\omega)$, where $\omega=-g_{t\phi}/g_{\phi\phi}$ and $\alpha=1/\sqrt{g^{tt}}$ \cite{Cheng:2025qlc}. Thus, the local energy at $r_B$ is
\begin{equation}\label{eq:formula39}
E^{\mathrm{loc}}(r_B)=p_\mu u_{\mathrm{Z}}^\mu=\frac{E_0}{\alpha(r_B)}=\frac{E_0\sqrt{(r_B^2+a^2)^2-a^2\Delta(r_B)}}{r_B\sqrt{\Delta(r_B)}}.
\end{equation}

Because the propagation is locally radial in the ZAMO frame, the corresponding proper distance for outward propagation is
\begin{equation}\label{eq:formula37}
L_p=\int_{r_A}^{r_B}\sqrt{-g_{rr}}\,dr=\int_{r_A}^{r_B}\frac{r}{\sqrt{\Delta(r)}}\,dr=\left[\sqrt{\Delta(r)}+\frac{r_s}{2}\ln\left(2r-r_s+2\sqrt{\Delta(r)}\right)\right]_{r_A}^{r_B}.
\end{equation}
For neutrinos propagating inward toward the gravitational source, we interchange $r_A$ and $r_B$ to obtain the proper distance. The same modification is applied to the phase in Eq.~(\ref{eq:formula20}) for inward propagation.

To present the numerical results for the oscillation probability, we set $E^{\mathrm{loc}}(r_B)=2\times10^3~\mathrm{GeV}$. For outward propagation, we take $r_A=1\times10^8~\mathrm{km}$ and $0\leq L_p\leq3\times10^8~\mathrm{km}$. For inward propagation, we take $r_A=4\times10^8~\mathrm{km}$ and $0\leq L_p\leq3.5\times10^8~\mathrm{km}$.
The leptonic mixing parameters are taken as follows \cite{Esteban:2024eli}:
\begin{align*}
\Delta m_{21}^2 &= 7.49 \times 10^{-5} \, \text{eV}^2, & \Delta m_{31}^2 &= 2.513 \times 10^{-3} \text{eV}^2, & \Delta m_{32}^2 &= 2.438 \times 10^{-3} \, \text{eV}^2, \\
\theta_{12} &= 33.68^\circ ,& \theta_{23} &= 43.3^\circ ,& \theta_{13} &= 8.56^\circ,\\
\delta_{\text{CP}} &= 212^\circ.
\end{align*}

In the numerical calculations, we impose the nonextremal Kerr--Newman horizon condition.
\[
M^{2}-a^{2}-Q^{2}>0.
\]
When \(a\) is varied, \(M\) is held fixed and \(Q=0.20M\), yielding
\[
a_{\max}=\sqrt{M^{2}-Q^{2}}.
\]
Similarly, when \(Q\) is varied, \(M\) is held fixed and \(a=0.20M\), yielding
\[
Q_{\max}=\sqrt{M^{2}-a^{2}}.
\]
All other parameter choices also satisfy the horizon condition. The corresponding parameter values are summarized in Table~\ref{tab:zero_angular_parameters}. For a Kerr--Newman black hole, the outer event-horizon radius is
\begin{equation}
r_{+}=M+\sqrt{M^{2}-a^{2}-Q^{2}}.
\end{equation}
Physical consistency further requires that the entire propagation path satisfy \(r>r_{+}\). For the zero-angular-momentum radial case, this exterior-path condition is verified in Appendix~\ref{app:exterior-horizon-zero-angular}, and the minimum path radii and corresponding outer-horizon radii are summarized in Table~\ref{tab:radial-horizon-check}.

\begin{table}[H]
    \centering
    \caption{Metric parameters used for propagation with zero angular momentum.
    All values are given in units of \(10^{7}\,\mathrm{km}\).}
    \label{tab:zero_angular_parameters}
    \renewcommand{\arraystretch}{1.15}
    \begin{tabular}{lccc}
        \hline
        Case & \(M\) & \(a\) & \(Q\) \\
        \hline
        Flat spacetime                  & 0               & 0                     & 0                     \\
        Schwarzschild spacetime         & 4.8             & 0                     & 0                     \\
        Kerr--Newman spacetime           & 4.8             & 3.12                  & 3.12                  \\
        Variation of \(M\)              & 2.0,\;3.3,\;4.6 & 1.0                   & 1.0                   \\
        Variation of \(a\)              & 4.8             & 0,\;2.82,\;4.47     & 0.96                  \\
        Variation of \(Q\)              & 4.8             & 0.96                  & 0,\;2.82,\;4.47     \\
        \hline
    \end{tabular}
\end{table}

\begin{figure}[H]
  \centering
  \begin{minipage}{0.48\textwidth}
    \centering
    \includegraphics[width=\linewidth]{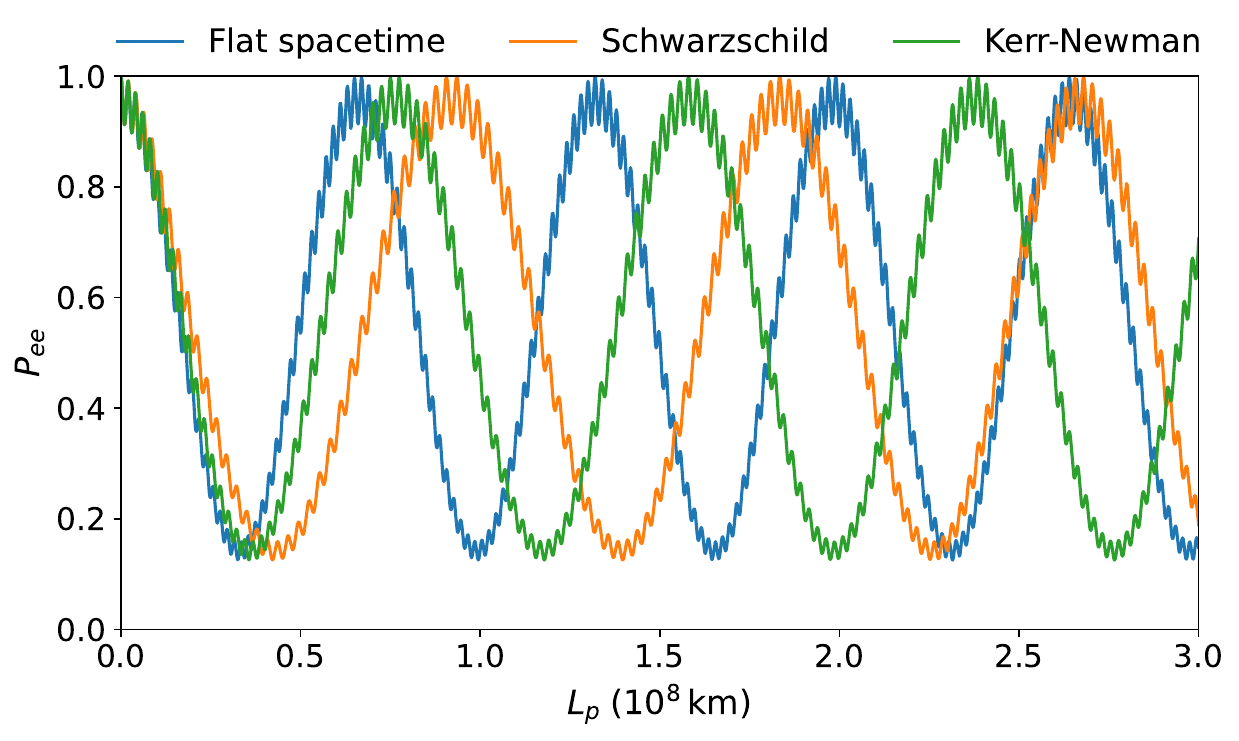}
  \end{minipage}
  \hfill
  \begin{minipage}{0.48\textwidth}
    \centering
    \includegraphics[width=\linewidth]{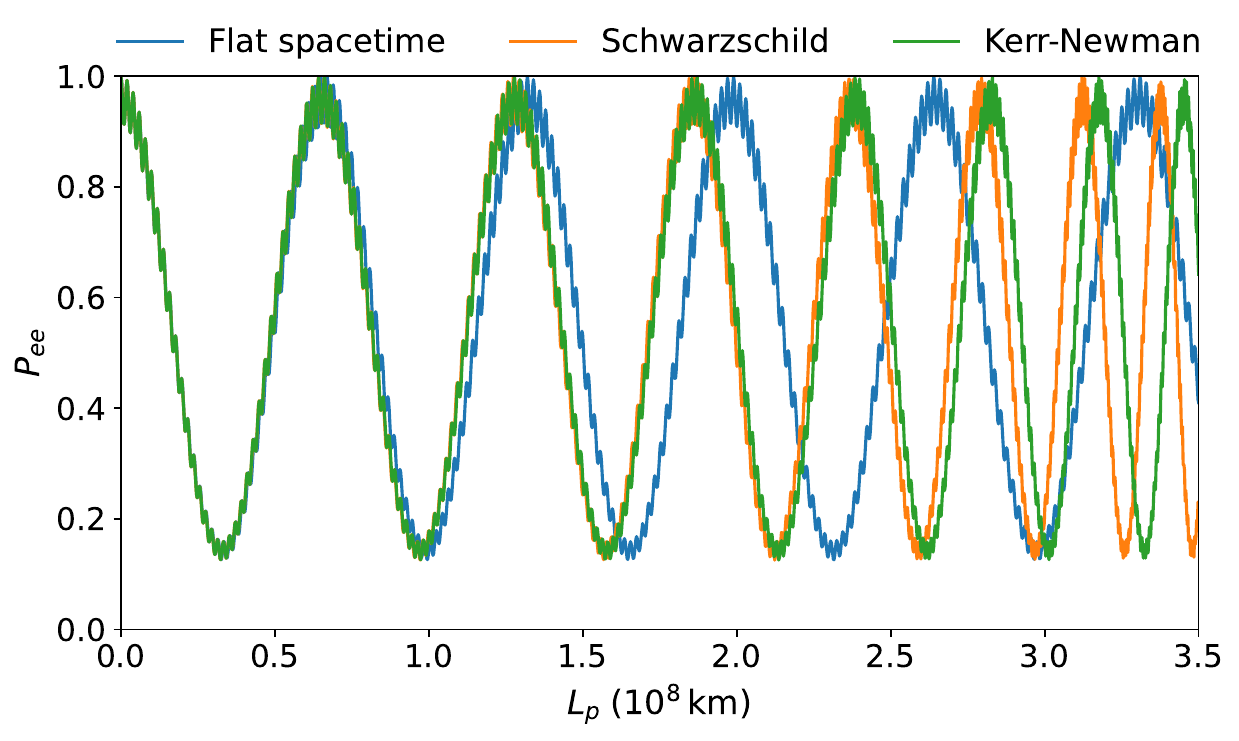}
  \end{minipage}
  \caption{Neutrino oscillation probability \(P_{\nu_e \to \nu_e}\) for different Kerr-Newman metric parameters. Left panel: outward propagation; right panel: inward propagation.}
  \label{fig:survival_prob_1}
  \end{figure}

\begin{figure}[H]
    \centering
    \begin{minipage}{0.32\textwidth}
        \centering
        \includegraphics[width=\linewidth]{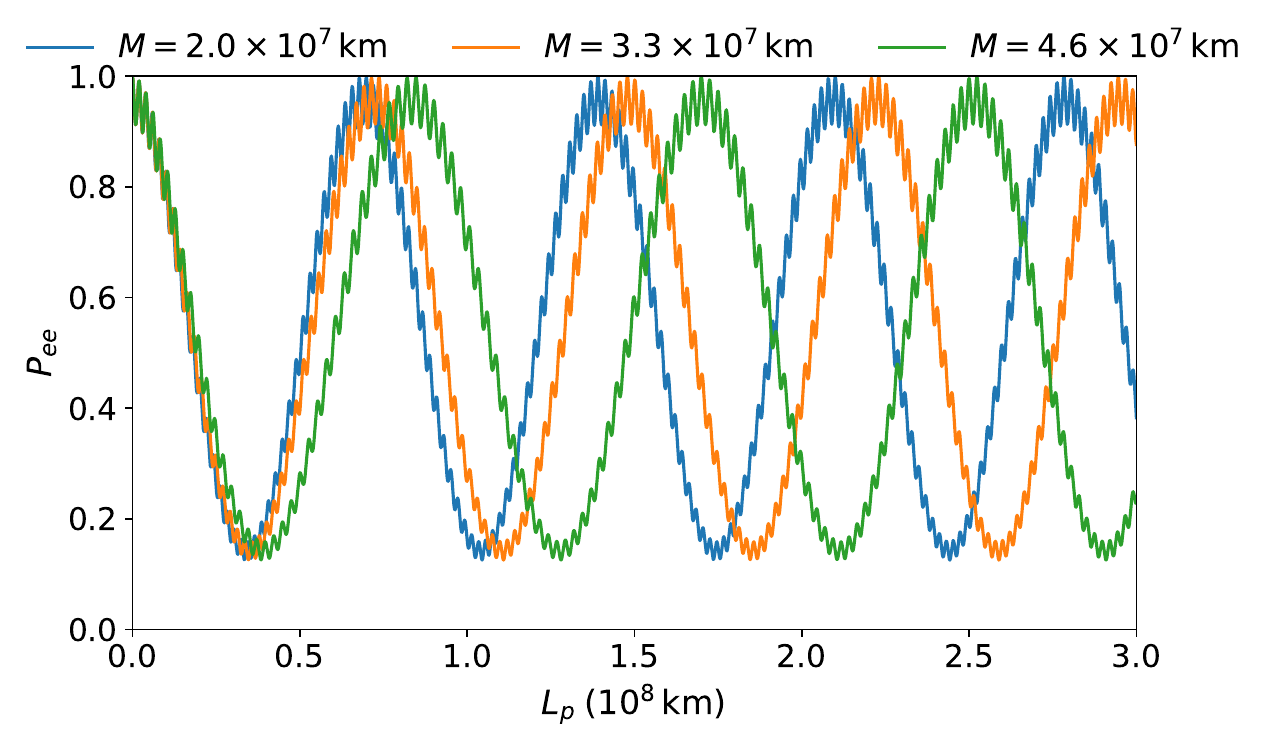}
    \end{minipage}
    \hfill
    \begin{minipage}{0.32\textwidth}
        \centering
        \includegraphics[width=\linewidth]{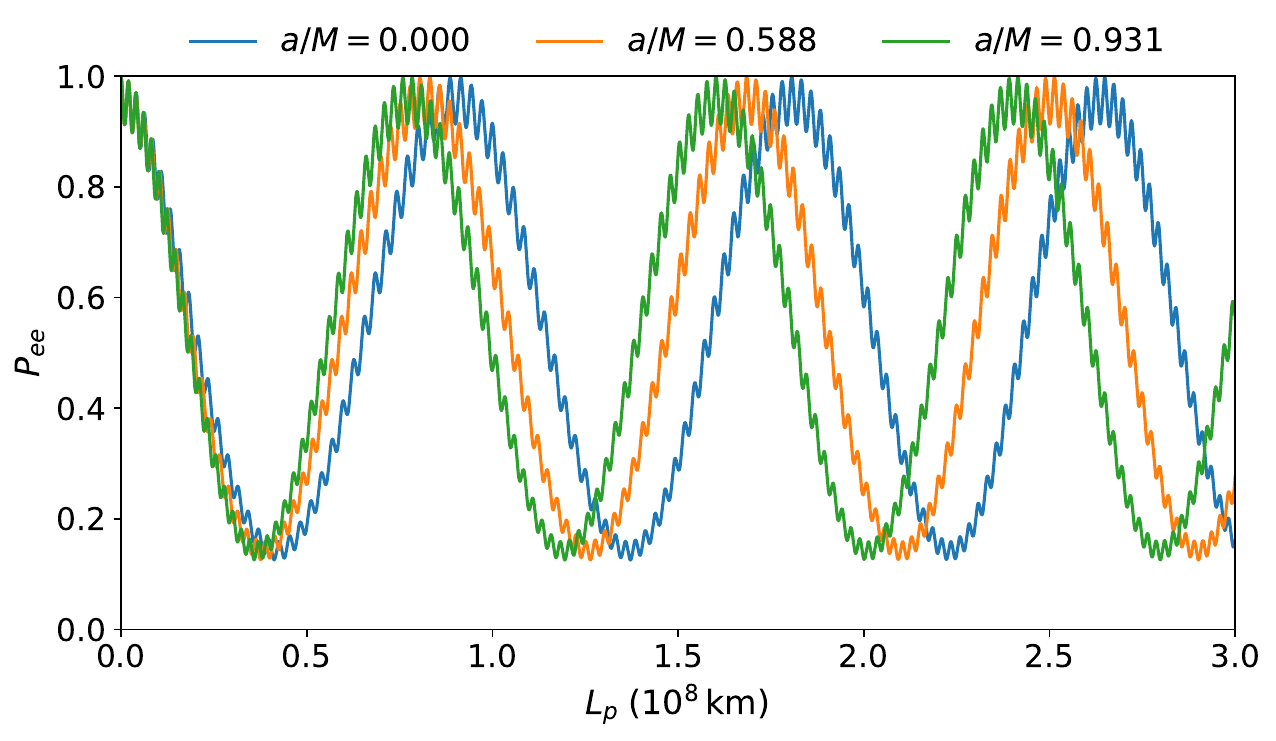}
    \end{minipage}
    \hfill
    \begin{minipage}{0.32\textwidth}
        \centering
        \includegraphics[width=\linewidth]{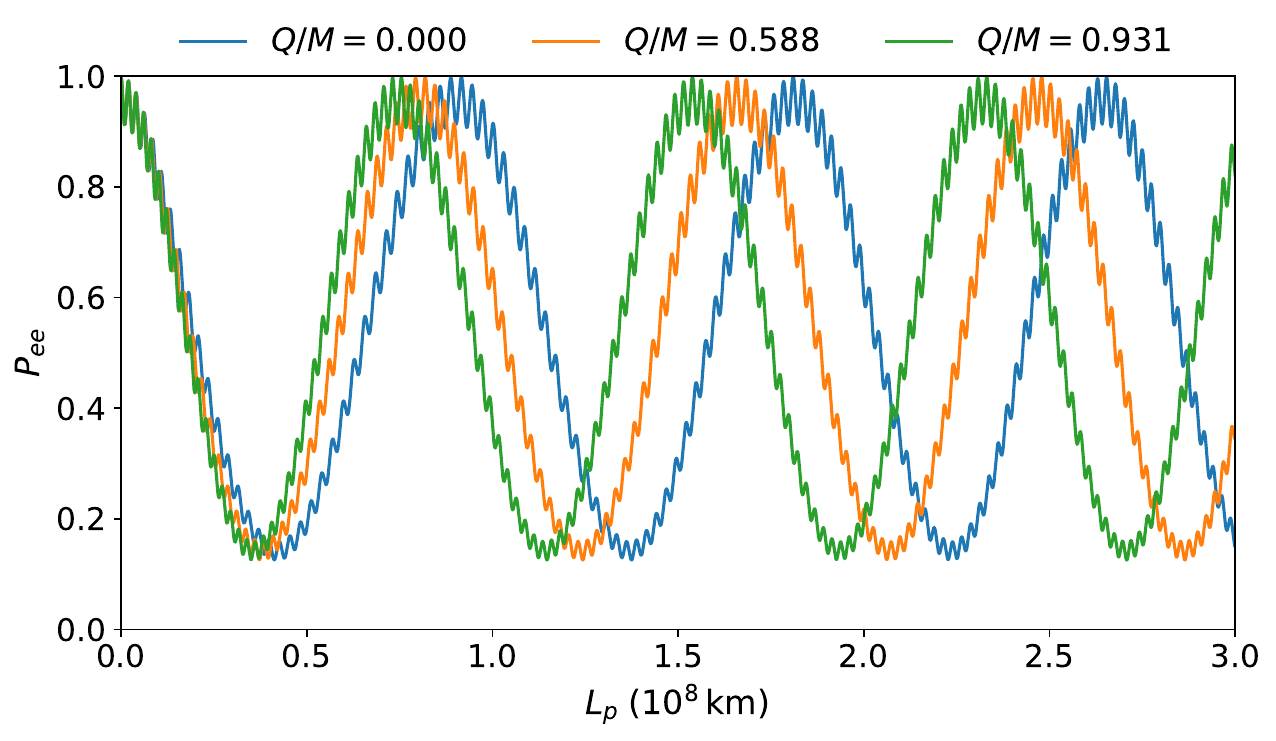}
    \end{minipage}
    \hfill
    \begin{minipage}{0.32\textwidth}
        \centering
        \includegraphics[width=\linewidth]{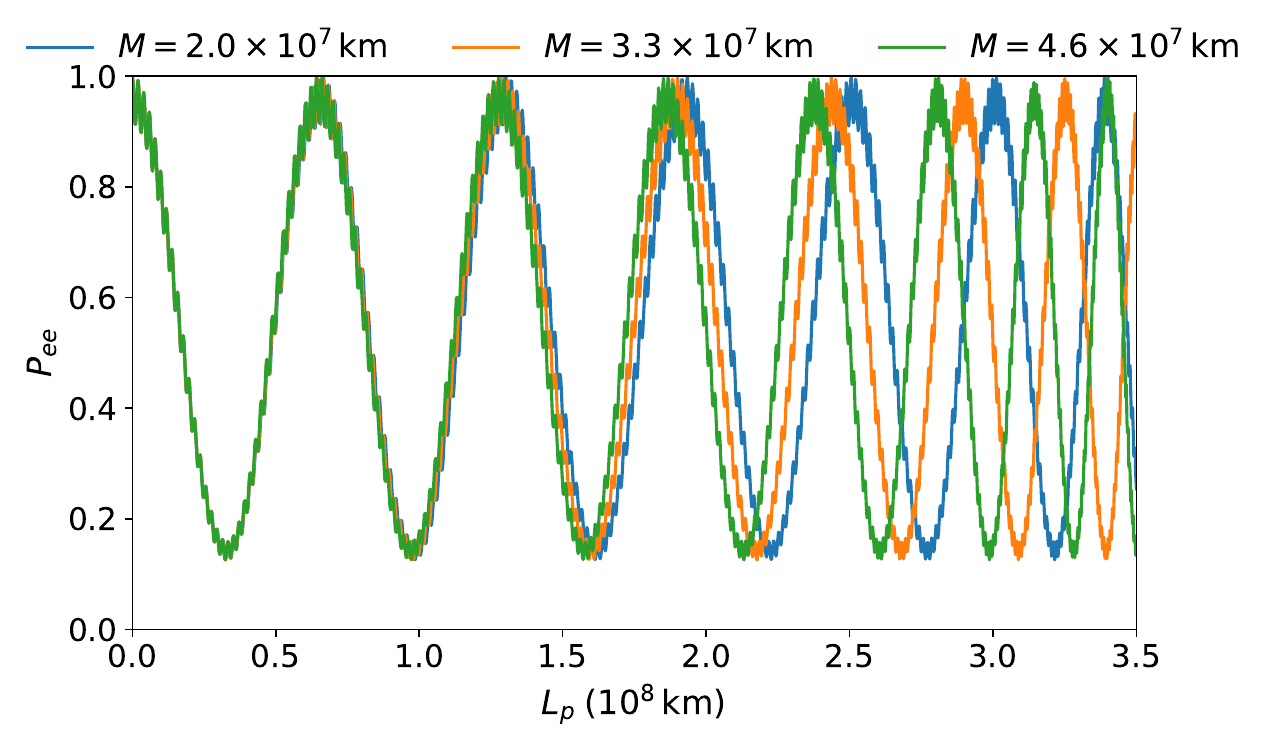}
    \end{minipage}
    \hfill
    \begin{minipage}{0.32\textwidth}
        \centering
        \includegraphics[width=\linewidth]{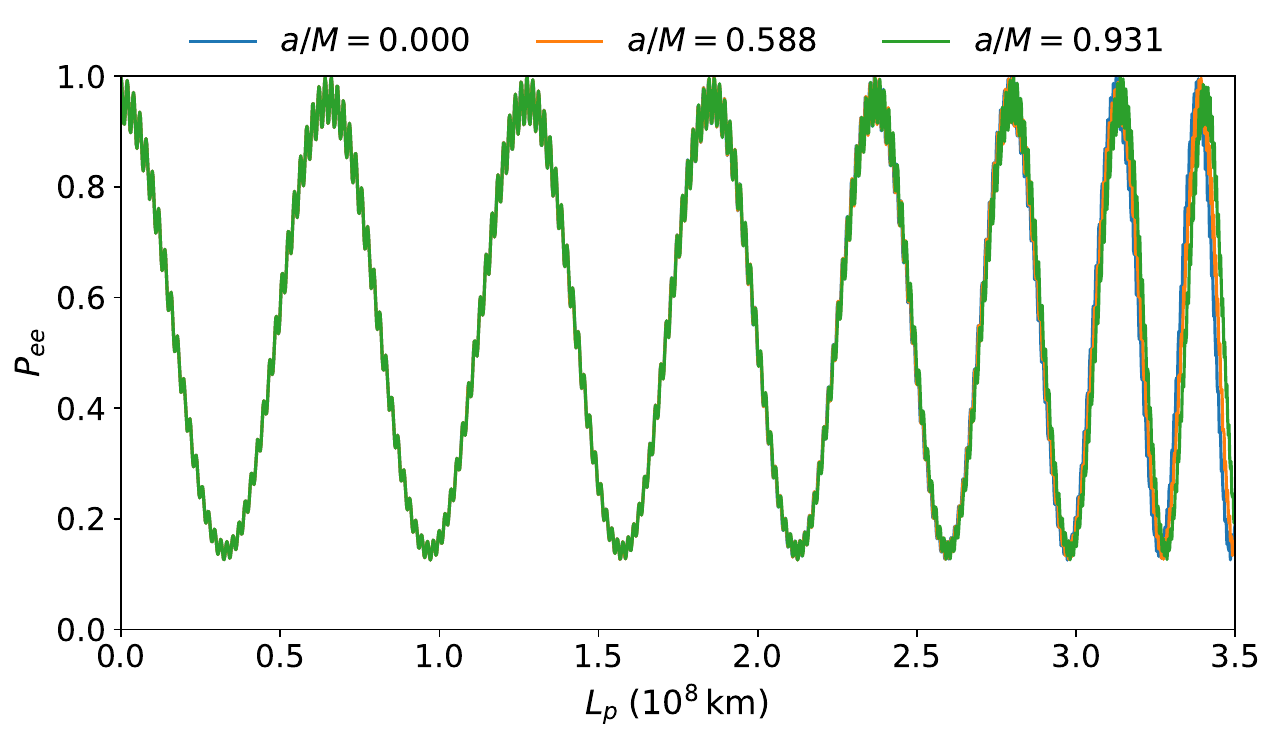}
    \end{minipage}
    \hfill
    \begin{minipage}{0.32\textwidth}
        \centering
        \includegraphics[width=\linewidth]{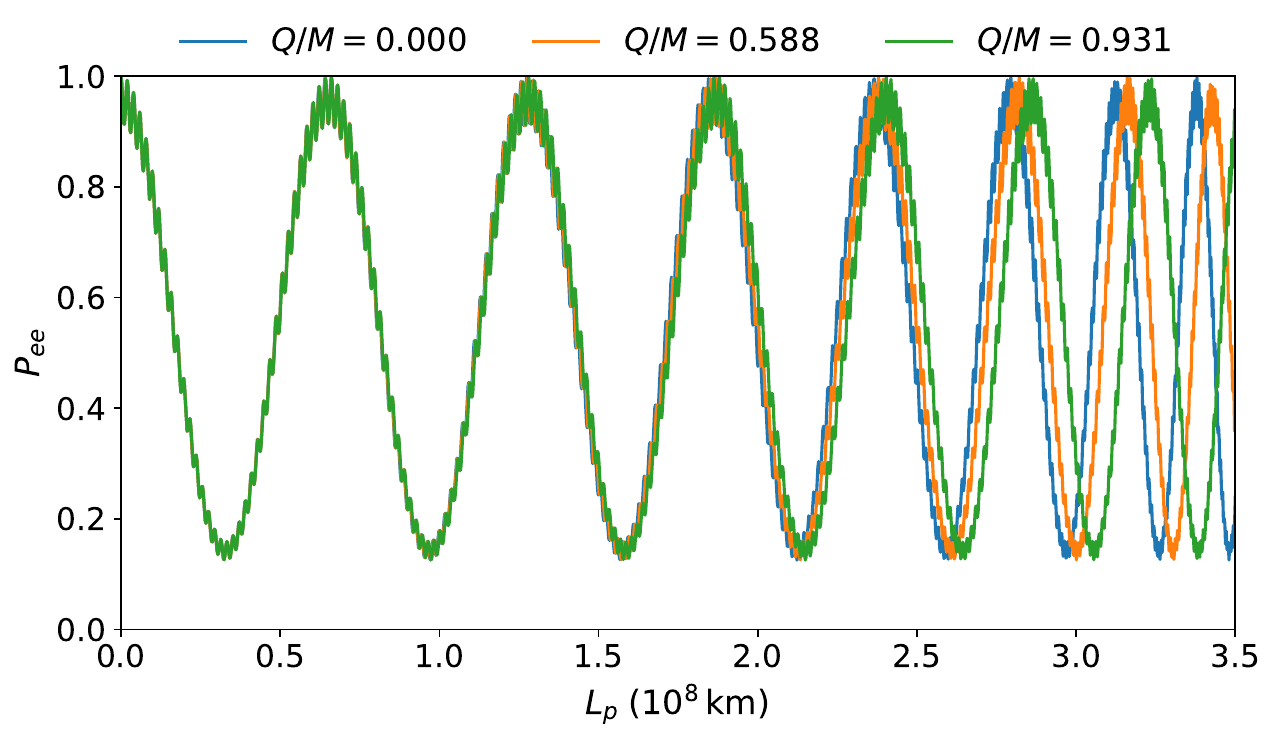}
    \end{minipage}
    \caption{Neutrino oscillation probability \(P_{\nu_e \to \nu_e}\) in the Kerr--Newman metric. The top and bottom rows correspond to outward and inward propagation, respectively.}
  \label{fig:survival_prob_2}
\end{figure}

We show the oscillation probability \(P_{\nu_e \to \nu_e}\) in flat, Schwarzschild, and Kerr--Newman spacetimes in Figs.~\ref{fig:survival_prob_1} and~\ref{fig:survival_prob_2}.\\
\hspace*{2em}As shown in Figs.~\ref{fig:survival_prob_1}
and~\ref{fig:survival_prob_2}, the metric parameters primarily affect the phase accumulation and oscillation period, whereas the overall range of the survival probability remains nearly unchanged. For outward propagation, the oscillation period remains approximately constant over
the propagation range considered. Increasing \(M\) lengthens the oscillation period, whereas increasing either \(a\) or \(Q\) shortens it. Consequently, the Kerr--Newman spacetime exhibits an intermediate oscillation period compared with the flat and Schwarzschild spacetimes, because the period-lengthening effect of \(M\) is partially counteracted by the rotation and charge parameters. For inward propagation of neutrinos, at fixed local energy
\(E^{\mathrm{loc}}(r_B)\), the path integral increases while \(E_0\) decreases as the endpoint approaches the black hole, causing the phase to accumulate progressively more rapidly. The oscillation period therefore decreases with increasing \(L_p\), leading to a gradual separation of the curves. The parameter dependences are opposite to those found for outward propagation: increasing \(M\) accelerates the reduction in the oscillation period, whereas increasing \(a\) or \(Q\) suppresses it. The dependence
on \(a\) is comparatively weak.
\vspace{-16pt}

  \vspace{-16pt}

\subsection{Entanglement of neutrinos for zero-angular-momentum propagation}

\begin{figure}[H]
  \centering
  \begin{minipage}{0.48\textwidth}
    \centering
    \includegraphics[width=\linewidth]{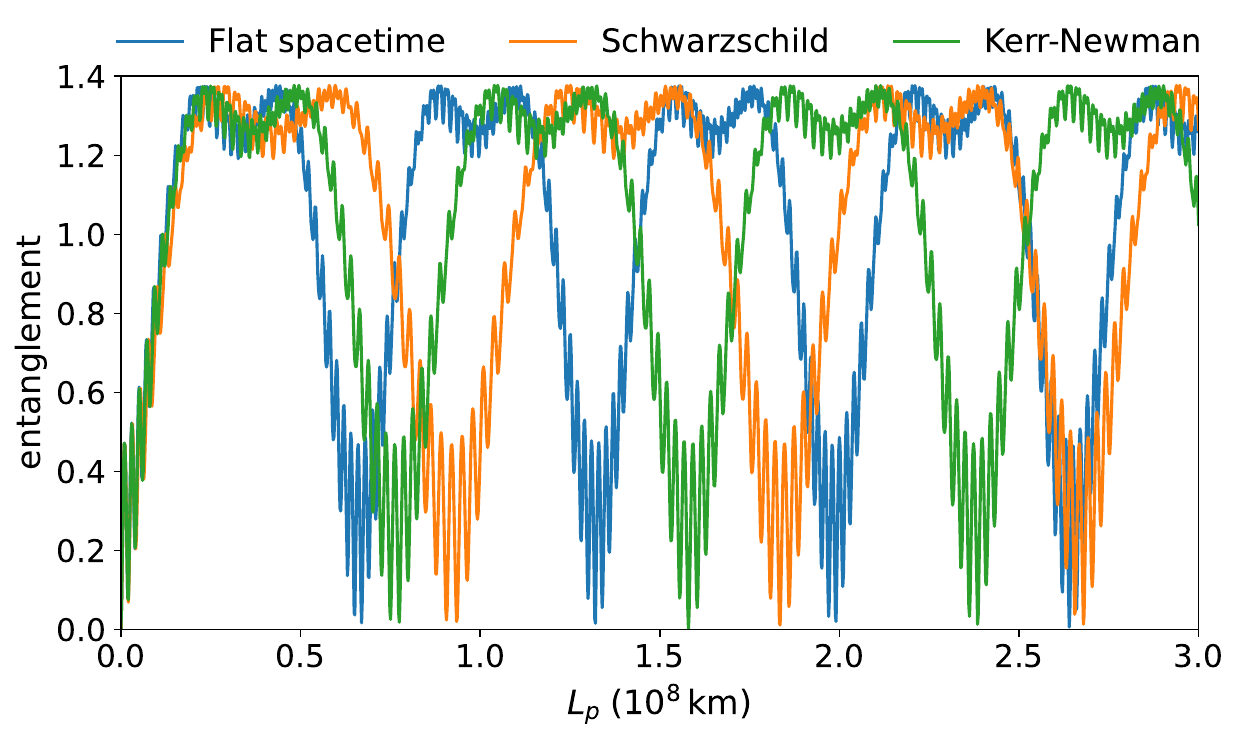}
  \end{minipage}
  \hfill
  \begin{minipage}{0.48\textwidth}
    \centering
    \includegraphics[width=\linewidth]{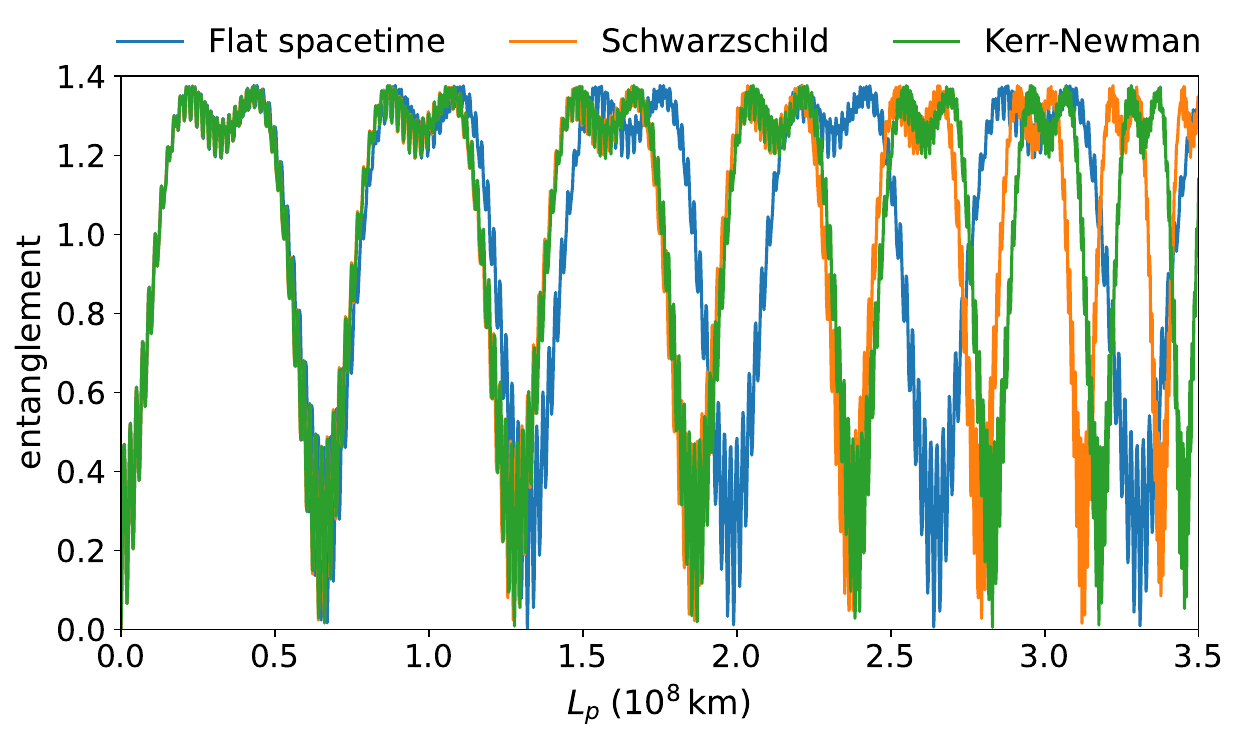}
  \end{minipage}
   \vspace{-3pt}
  \caption{Tripartite entanglement of neutrinos in different types of curved spacetime. Left panel: outward propagation; right panel: inward propagation.}
  \label{fig:EOF1}
  \end{figure}

\begin{figure}[H]
    \centering
    \begin{minipage}{0.32\textwidth}
        \centering
        \includegraphics[width=\linewidth]{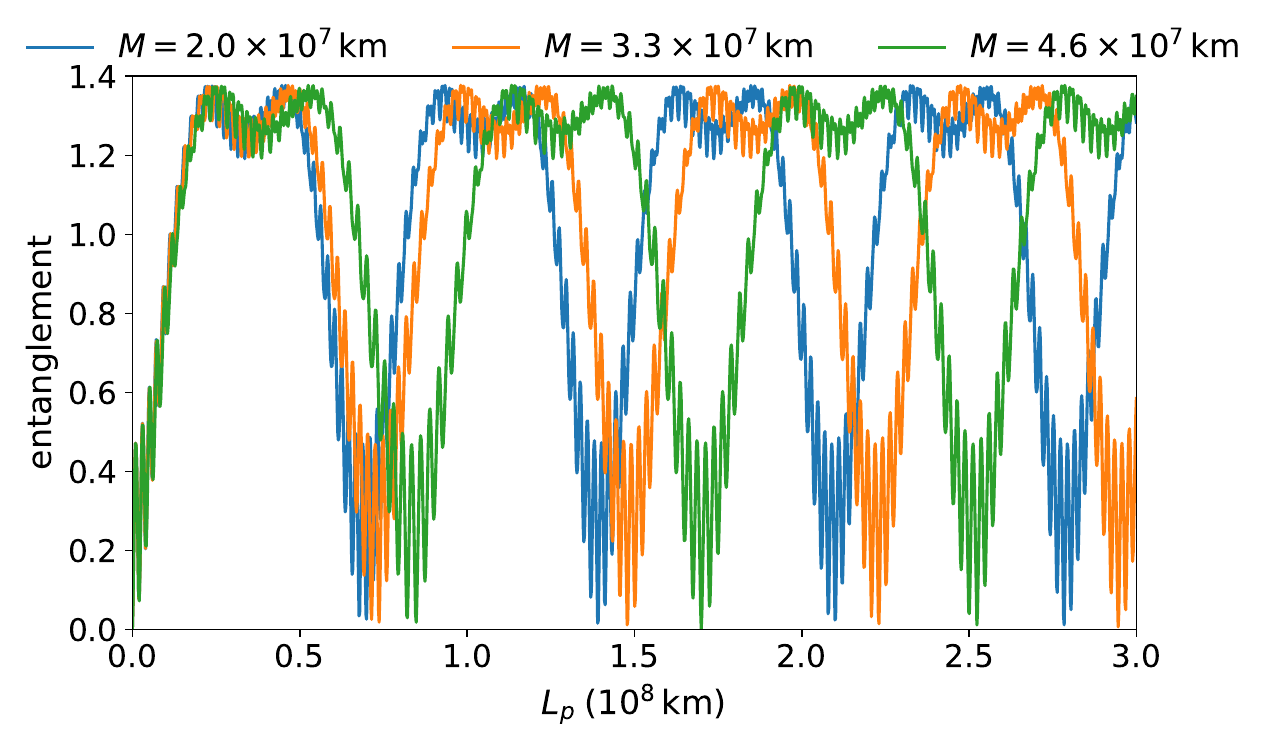}
    \end{minipage}
    \hfill
    \begin{minipage}{0.32\textwidth}
        \centering
        \includegraphics[width=\linewidth]{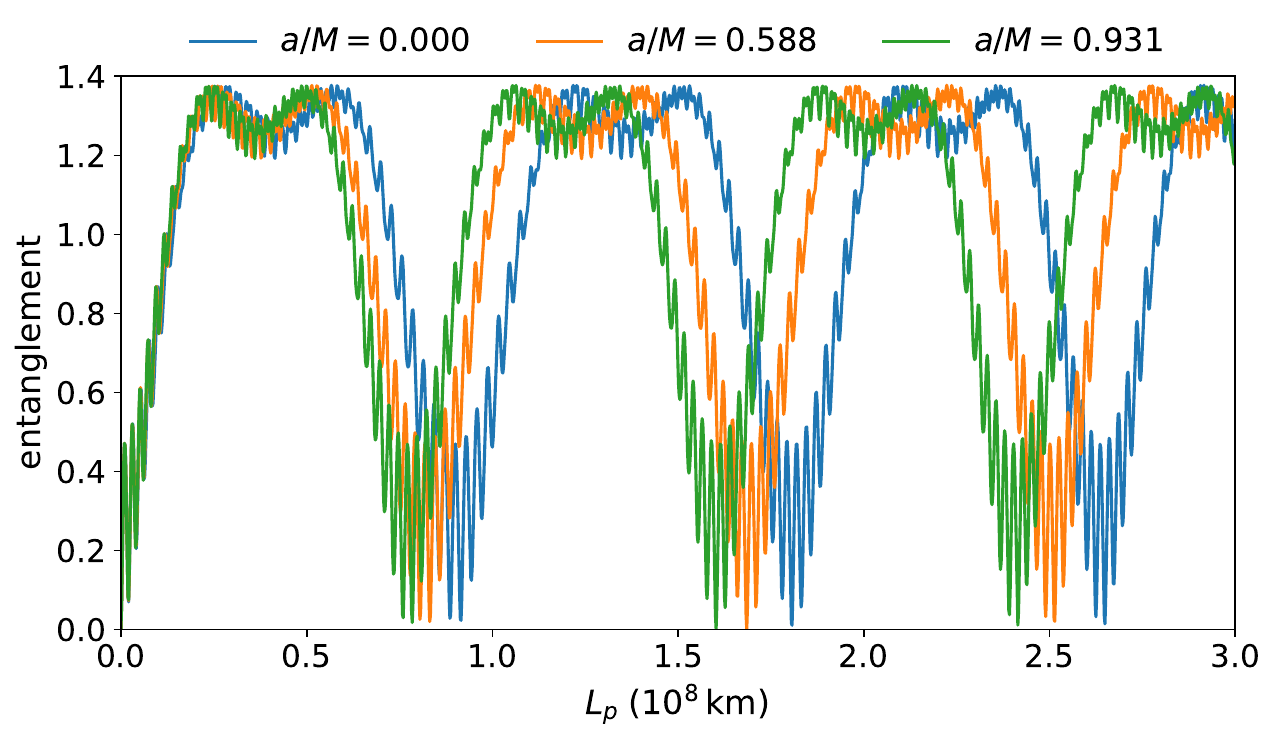}
    \end{minipage}
    \hfill
    \begin{minipage}{0.32\textwidth}
        \centering
        \includegraphics[width=\linewidth]{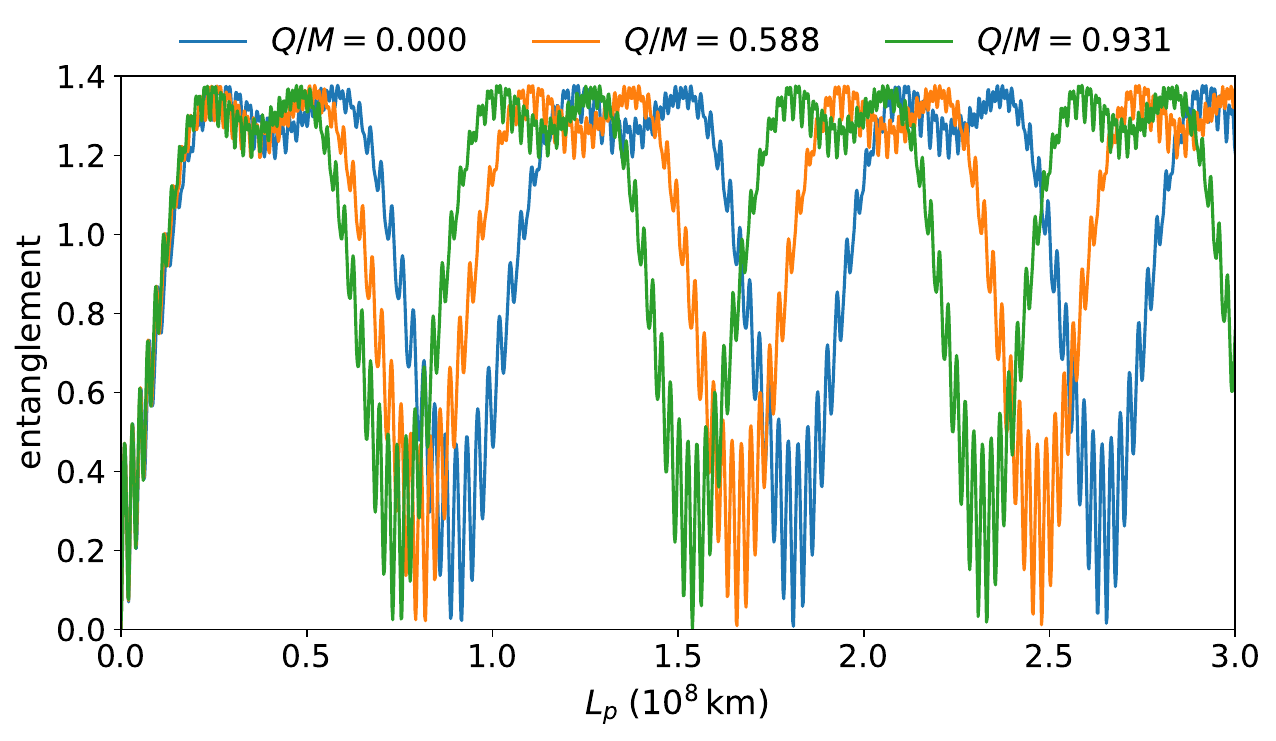}
    \end{minipage}
    \hfill
    \begin{minipage}{0.32\textwidth}
        \centering
        \includegraphics[width=\linewidth]{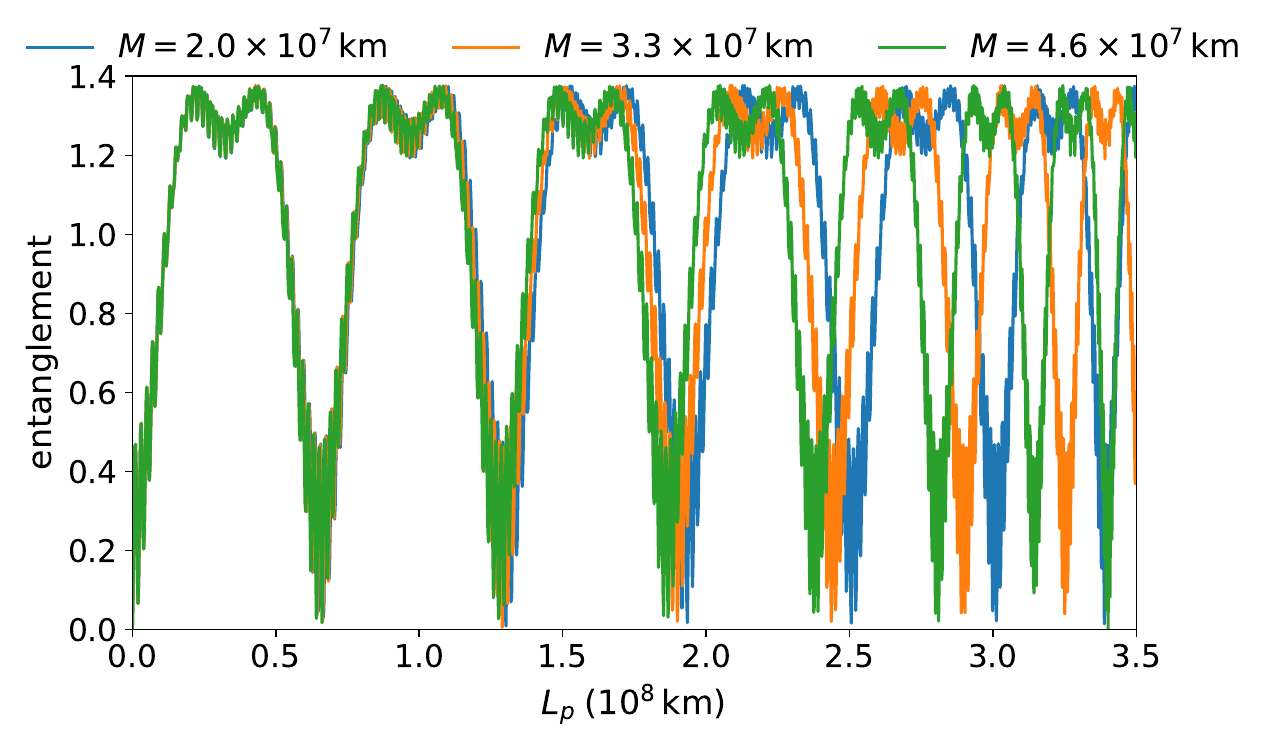}
    \end{minipage}
    \hfill
    \begin{minipage}{0.32\textwidth}
        \centering
        \includegraphics[width=\linewidth]{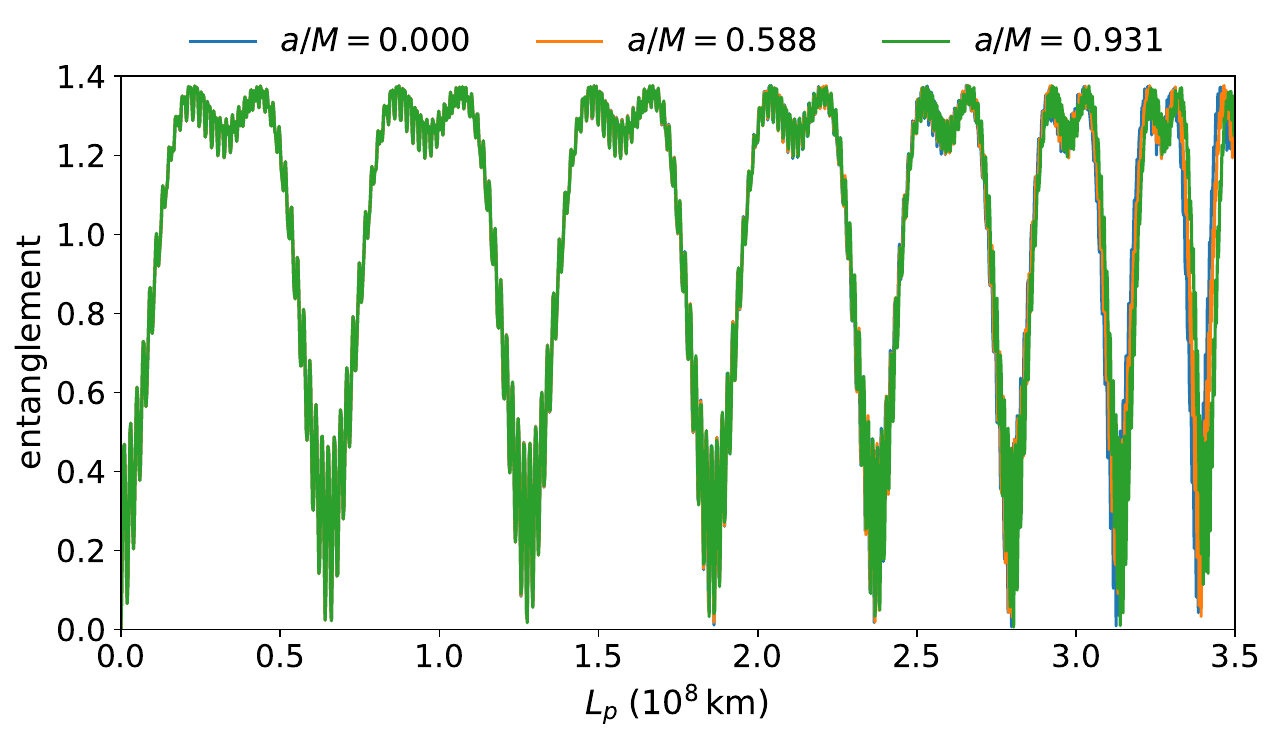}
    \end{minipage}
    \hfill
    \begin{minipage}{0.32\textwidth}
        \centering
        \includegraphics[width=\linewidth]{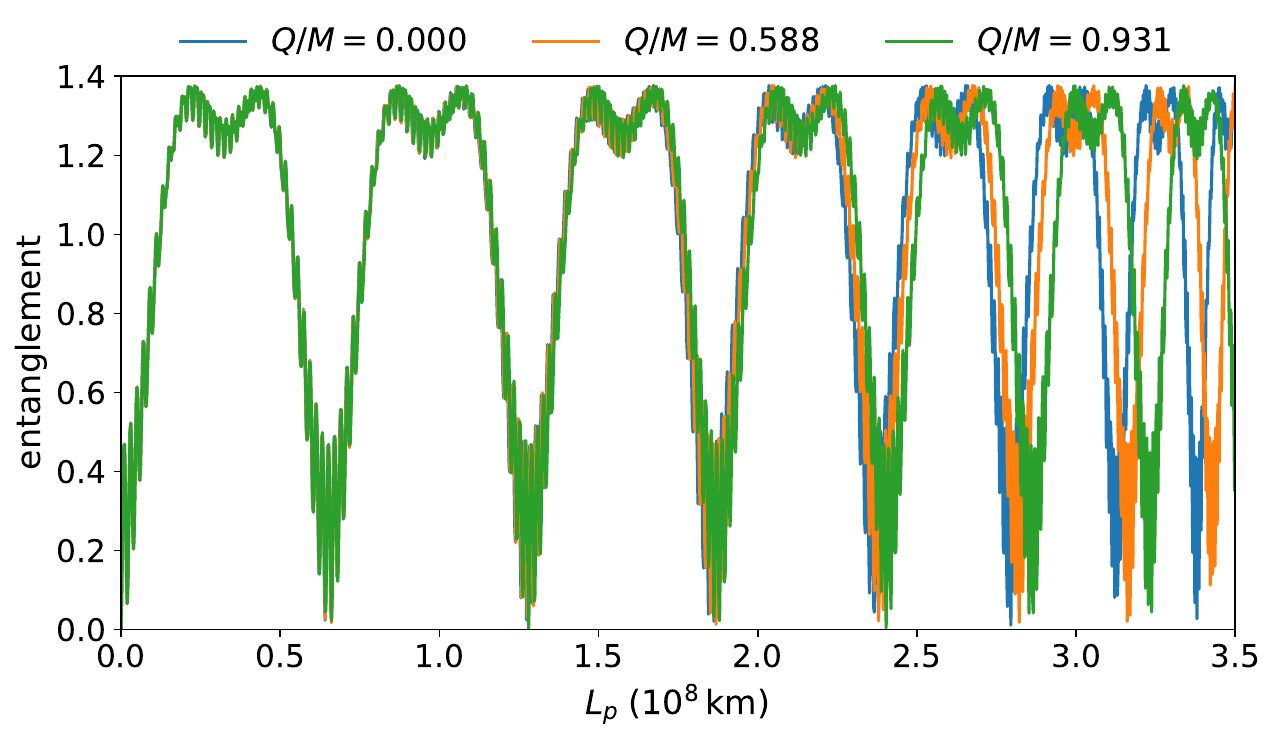}
    \end{minipage}
    \caption{Tripartite entanglement of neutrinos for different values of the Kerr-Newman metric parameters. The top and bottom rows correspond to outward and inward propagation, respectively.}
     \label{fig:EOF2}
\end{figure}

An initial flavor state \(|\nu_e\rangle\), after traveling a proper distance \(L_p\) in curved spacetime, can be expressed as:
\begin{equation}\label{eq:formula40}
|\psi_e(L_p)\rangle = a_{ee}(L_p)|100\rangle + a_{e\mu}(L_p)|010\rangle + a_{e\tau}(L_p)|001\rangle.
\end{equation}
Here, \(a_{e\beta}(L_p)\) $(\beta = e, \mu, \tau)$ denotes an oscillation amplitude, with \(P_{e\beta} = |a_{e\beta}(L_p)|^2\).
To quantify entanglement, we adopt the tripartite entanglement of formation (EOF), which is expressed as follows \cite{Guo:2019mue}:
\begin{equation}\label{eq:formula41}
E(\rho) = \frac{1}{2} \left[ S(\rho_A) + S(\rho_B) + S(\rho_C) \right],
\end{equation}
where \( S(\rho_i) = -\mathrm{Tr}\left(\rho_i \log \rho_i\right) \) denotes the von Neumann entropy of the reduced density matrix \( \rho_i = \mathrm{Tr}_{kj}(\rho) \). Therefore, according to Eqs.~(\ref{eq:formula40}) and (\ref{eq:formula41}), neutrino entanglement can be calculated as \cite{Wang:2024tfh}:\\
\begin{equation}
\begin{aligned}
E(\rho_{e\mu\tau}^e) &= -\frac{1}{2} \Big[
P_{ee}(L_p) \log_2 P_{ee}(L_p)
+ \bigl(P_{ee}(L_p) + P_{e\tau}(L_p)\bigr) \log_2 \bigl(P_{ee}(L_p) + P_{e\tau}(L_p)\bigr) \\
&\quad + P_{e\mu}(L_p) \log_2 P_{e\mu}(L_p)
+ \bigl(P_{e\mu}(L_p) + P_{ee}(L_p)\bigr) \log_2 \bigl(P_{e\mu}(L_p) + P_{ee}(L_p)\bigr) \\
&\quad + P_{e\tau}(L_p) \log_2 P_{e\tau}(L_p)
+ \bigl(P_{e\mu}(L_p) + P_{e\tau}(L_p)\bigr) \log_2 \bigl(P_{e\mu}(L_p) + P_{e\tau}(L_p)\bigr) \Big].
\end{aligned}
\label{eq:formula42}
\end{equation}
Using this equation and the aforementioned parameters to calculate the survival probability \(P_{\text{ee}}\), we obtain the numerical results for the tripartite entanglement of neutrinos shown in Figs.~\ref{fig:EOF1} and~\ref{fig:EOF2}.\\
\hspace*{2em}The tripartite entanglement of neutrinos oscillates between 0 and 1.4, with periodic features similar to those of the survival probability \(P_{\nu_e\rightarrow\nu_e}\) shown in
Figs.~\ref{fig:survival_prob_1} and~\ref{fig:survival_prob_2}. The parameters \(M\), \(a\), and \(Q\) primarily modify the oscillation phase and period without systematically enhancing or suppressing the
entanglement. Because the entanglement is determined by the flavor-transition probabilities, the absence of systematic enhancement or suppression in these probabilities leads to the same behavior in the entanglement. This consistent periodic behavior reflects the intrinsic dependence of
neutrino entanglement on the oscillation wavelength.
  \vspace{-6pt}

  \vspace{-17pt}

\subsection{Monogamy of nonlocality for neutrinos with zero-angular-momentum propagation}

We use monogamy to quantify the nonlocality of a three-qubit system. This quantification is based on the maximal violation of the CHSH inequality \cite{PhysRevA.95.032123}, i.e.,
\begin{equation}
    \Sigma_3 = \max \langle B_{AB} \rangle^2 + \max \langle B_{AC} \rangle^2 + \max \langle B_{BC} \rangle^2,
    \label{eq:formula45}
\end{equation}
where
\begin{equation}
    \max \langle B_{CHSH} \rangle_\rho = \max |\mathrm{Tr}(\rho B_{CHSH})| = 2\sqrt{M(\rho)},
   \label{eq:formula46}
\end{equation}
where $M(\rho) = \max_{j<k} \{\mu_j + \mu_k\}$, $j,k \in \{1,2,3\}$, and $\mu_j$ and $\mu_k$ are the two largest eigenvalues of the real symmetric matrix $T^T T$. The matrix \(T\) has entries $t_{ij} = \mathrm{Tr}\,\rho \left( \sigma_i \otimes \sigma_j \right)$ \cite{Horodecki:1995nsk}.
$B_{CHSH}$ denotes the CHSH operator and is expressed as follows:
\begin{equation}
    B_{CHSH} = A_1 \otimes B_1 + A_1 \otimes B_2 + A_2 \otimes B_1 - A_2 \otimes B_2.
    \label{eq:formula47}
\end{equation}

We consider the initial flavor state \(|\nu_e\rangle\). By combining Eq.~(\ref{eq:formula40}) with Eqs.~(\ref{eq:formula45})--(\ref{eq:formula47}), we obtain the following results \cite{Horodecki:1995nsk}:
\begin{equation}
\begin{split}
\max \langle B_{e\mu}^e \rangle &= 2\sqrt{4 P_{ee}(L_p) P_{e\mu}(L_p) + \max\left[4 P_{ee}(L_p) P_{e\mu}(L_p),  \left(2 P_{e\tau}(L_p) - 1\right)^2\right]}, \\
\max \langle B_{e\tau}^e \rangle &= 2\sqrt{4 P_{ee}(L_p) P_{e\tau}(L_p) + \max\left[4 P_{ee}(L_p) P_{e\tau}(L_p), \left(2 P_{e\mu}(L_p) - 1\right)^2\right]},\\
\max \langle B_{\mu\tau}^e \rangle &= 2\sqrt{4 P_{e\mu}(L_p) P_{e\tau}(L_p) + \max\left[4 P_{e\mu}(L_p) P_{e\tau}(L_p), \left(2  P_{ee}(L_p) - 1\right)^2\right]}.
\end{split}
\label{eq:formula48}
\end{equation}
Substituting Eq.~(\ref{eq:formula48}) into Eq.~(\ref{eq:formula45}), we obtain the monogamy relation for nonlocality.
Since the CHSH inequality is given by $|\mathrm{Tr}\left(\rho B_{CHSH}\right)| \leq 2$, the quantity $\Sigma_3$ is bounded by $0 \leq \Sigma_3 \leq 12$.

\begin{figure}[H]
    \centering

    \begin{minipage}{0.32\textwidth}
        \centering
        \includegraphics[width=\linewidth]{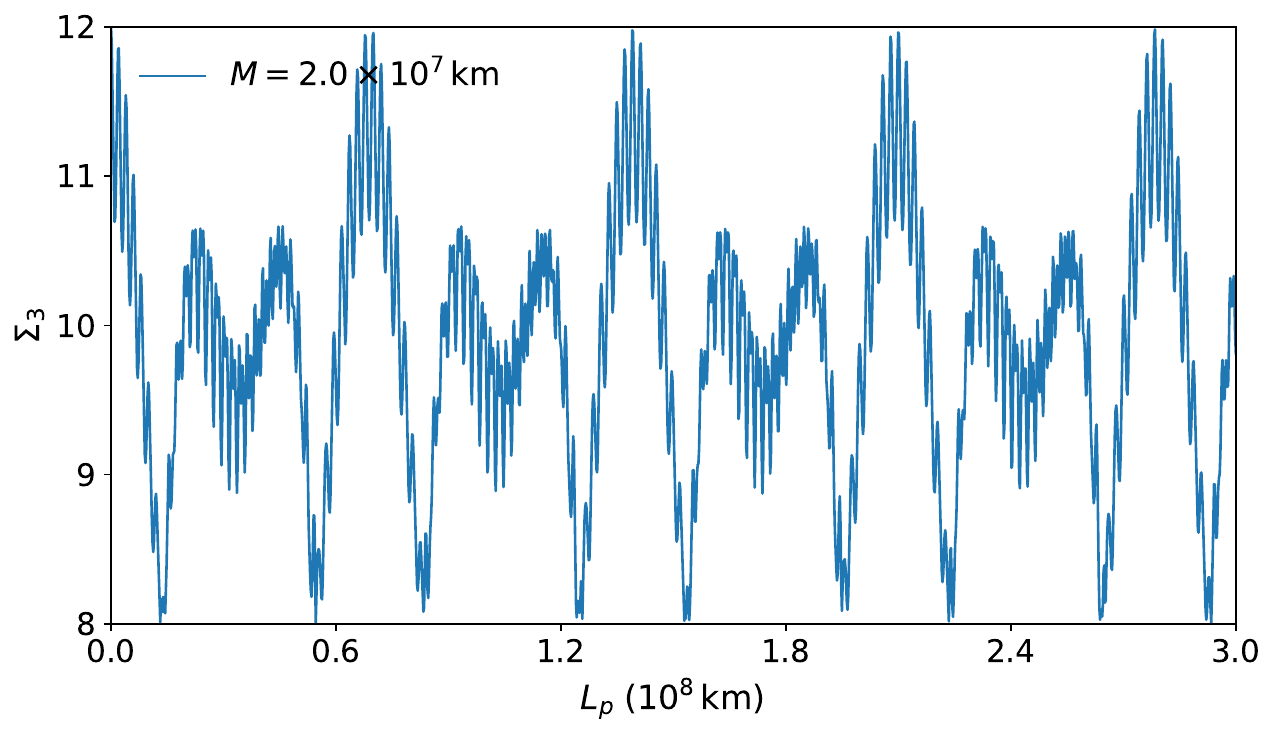}
    \end{minipage}
    \hfill
    \begin{minipage}{0.32\textwidth}
        \centering
        \includegraphics[width=\linewidth]{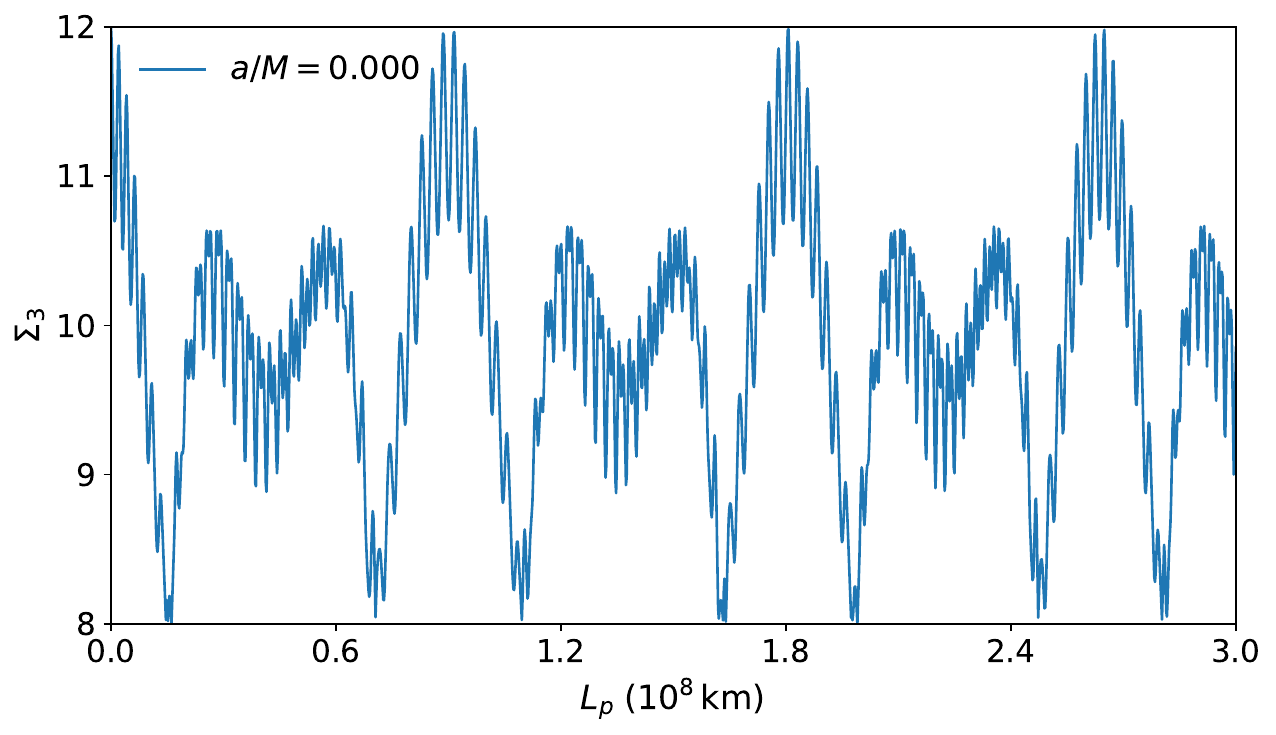}
    \end{minipage}
    \hfill
    \begin{minipage}{0.32\textwidth}
        \centering
        \includegraphics[width=\linewidth]{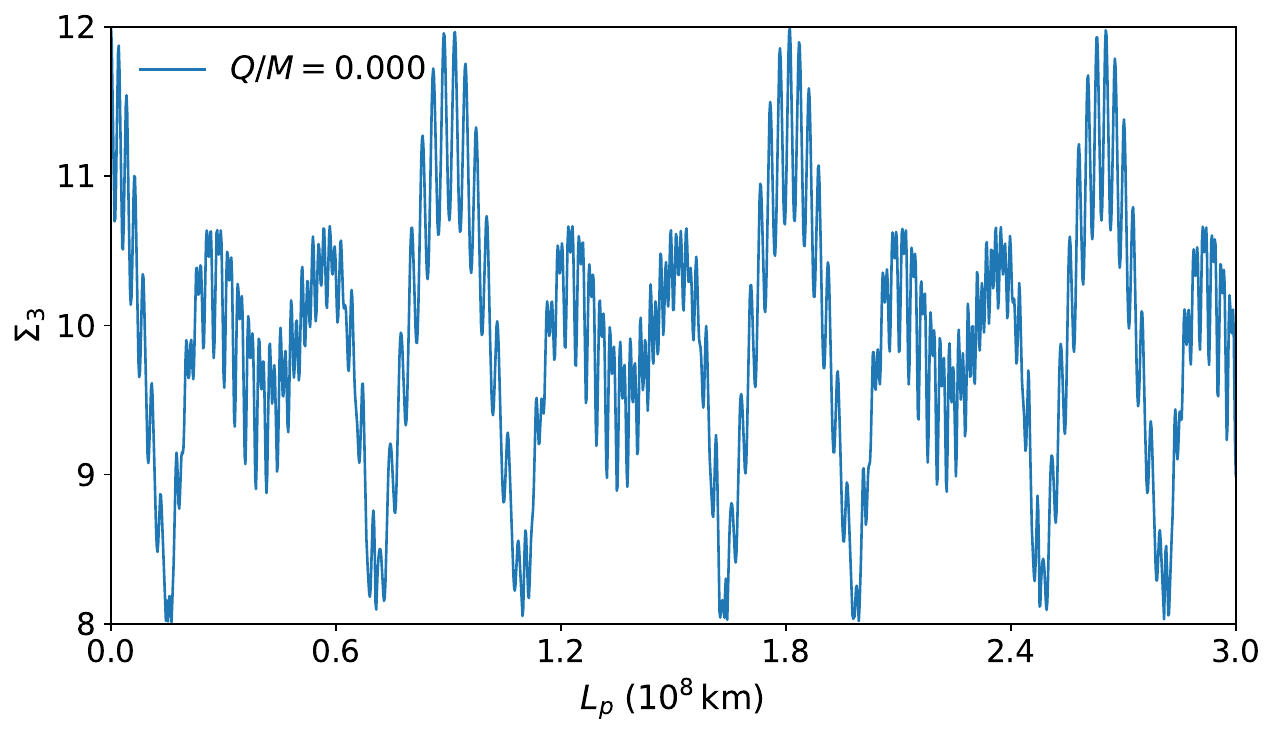}
    \end{minipage}

    \par\vspace{2pt}

    \begin{minipage}{0.32\textwidth}
        \centering
        \includegraphics[width=\linewidth]{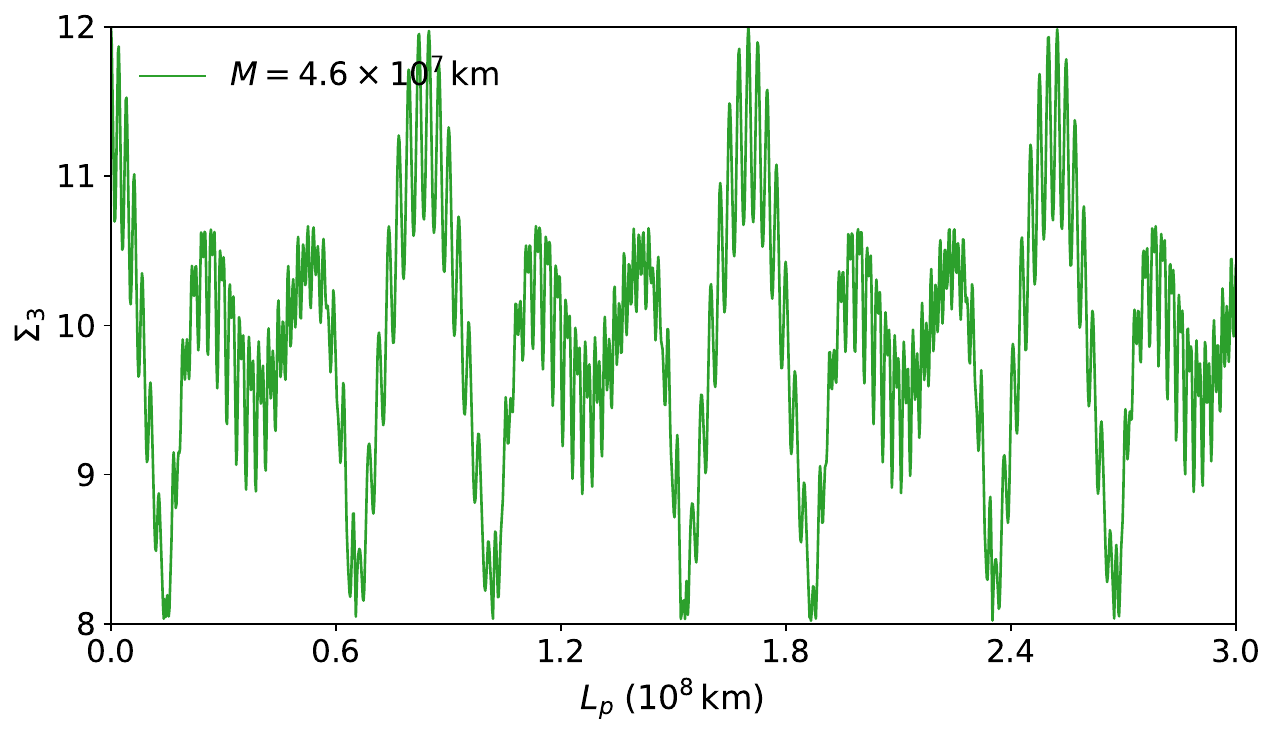}
    \end{minipage}
    \hfill
    \begin{minipage}{0.32\textwidth}
        \centering
        \includegraphics[width=\linewidth]{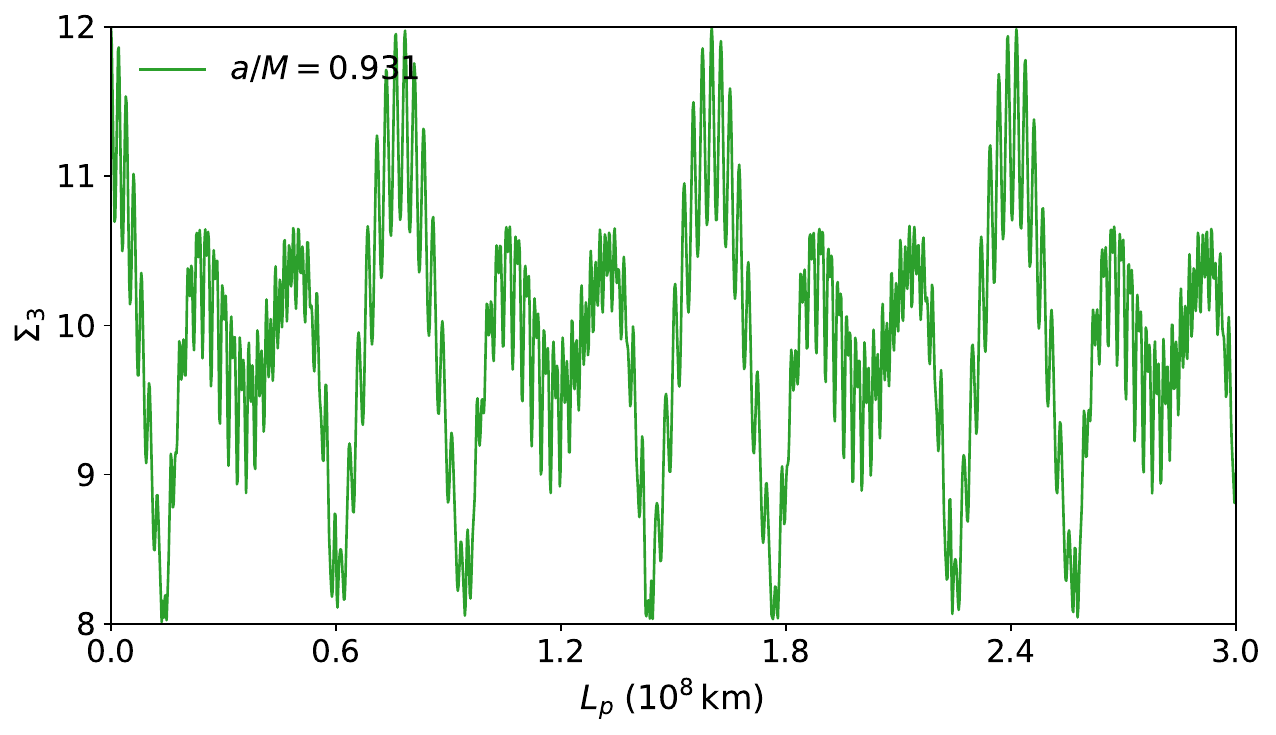}
    \end{minipage}
    \hfill
    \begin{minipage}{0.32\textwidth}
        \centering
        \includegraphics[width=\linewidth]{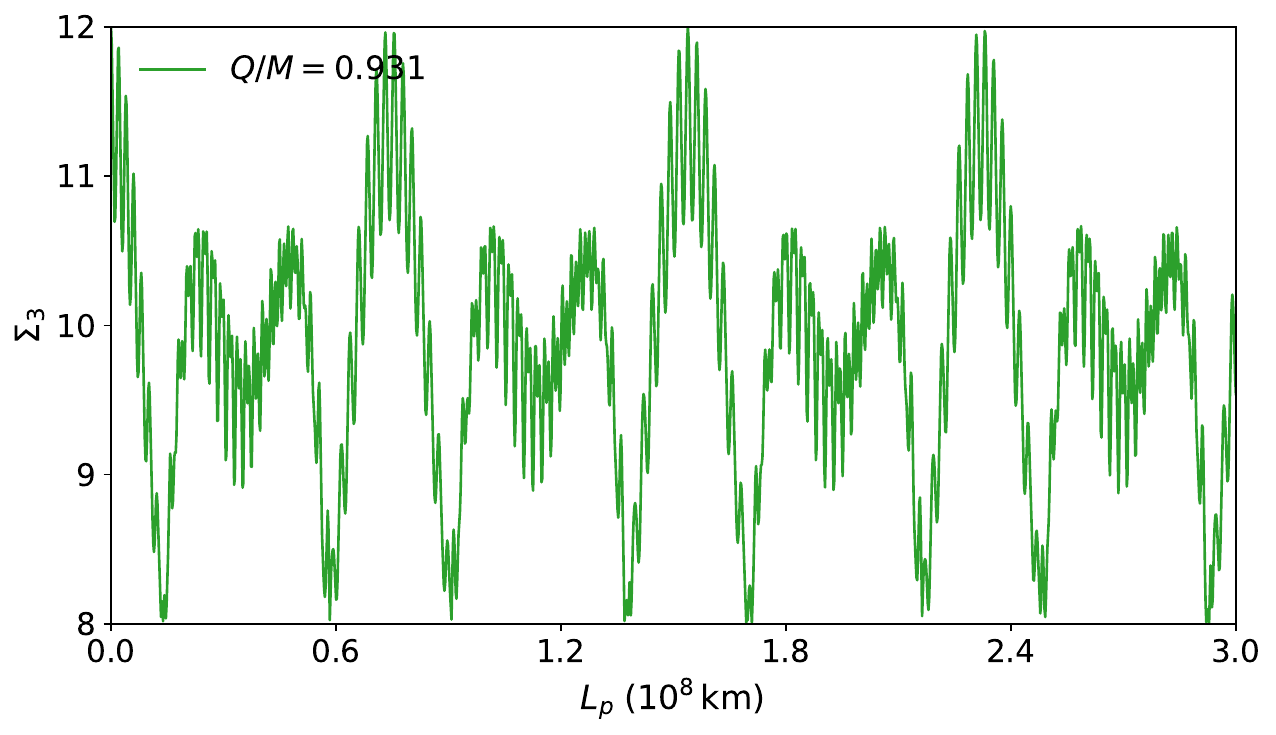}
    \end{minipage}

    \caption{Monogamy of nonlocality for neutrinos propagating outward in the Kerr--Newman spacetime. The left, middle, and right columns show variations in \(M\), \(a\), and \(Q\), respectively.}
    \label{fig:nonlocality_outward}
\end{figure}

\begin{figure}[H]
    \centering

    \begin{minipage}{0.32\textwidth}
        \centering
        \includegraphics[width=\linewidth]{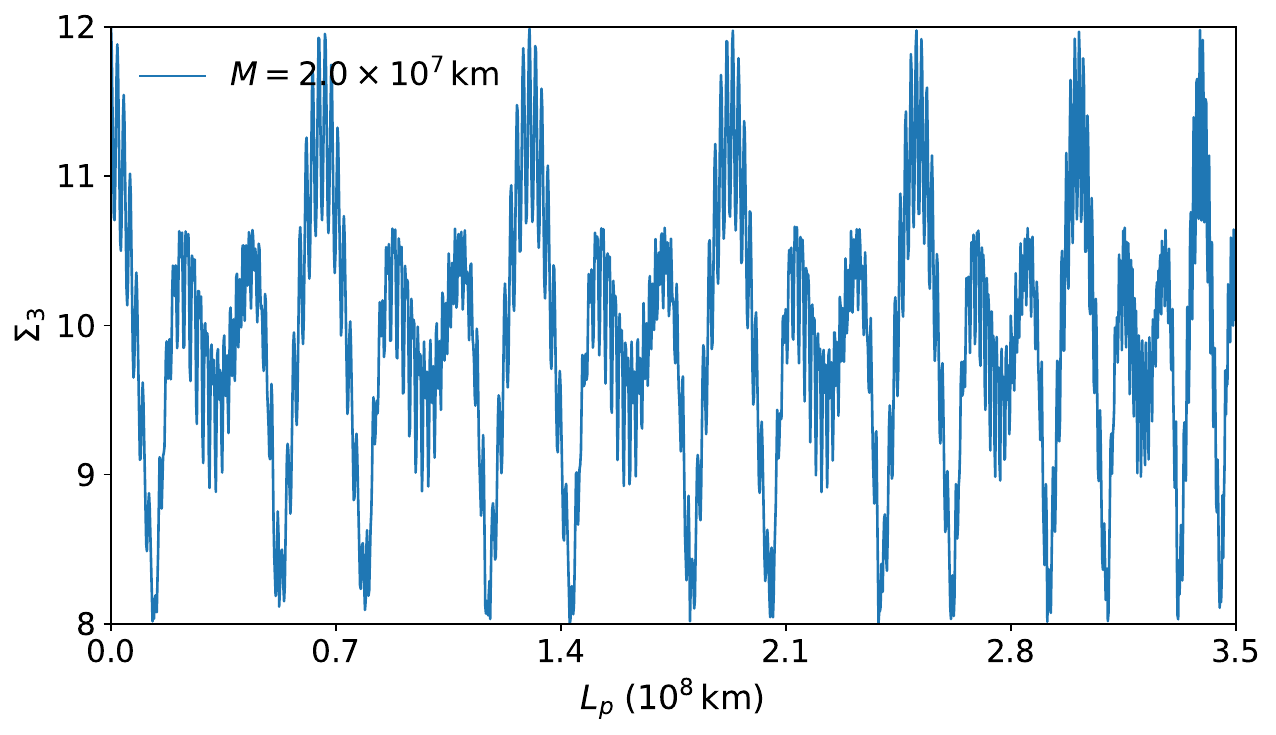}
    \end{minipage}
    \hfill
    \begin{minipage}{0.32\textwidth}
        \centering
        \includegraphics[width=\linewidth]{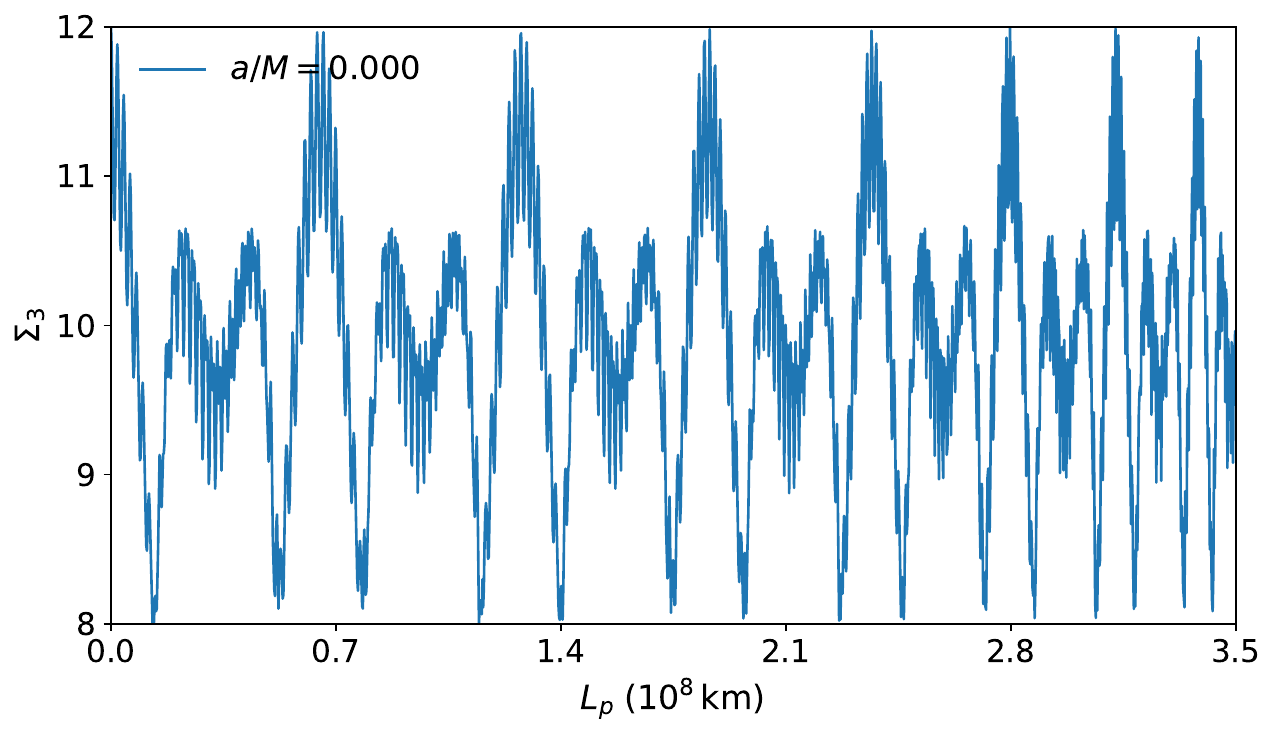}
    \end{minipage}
    \hfill
    \begin{minipage}{0.32\textwidth}
        \centering
        \includegraphics[width=\linewidth]{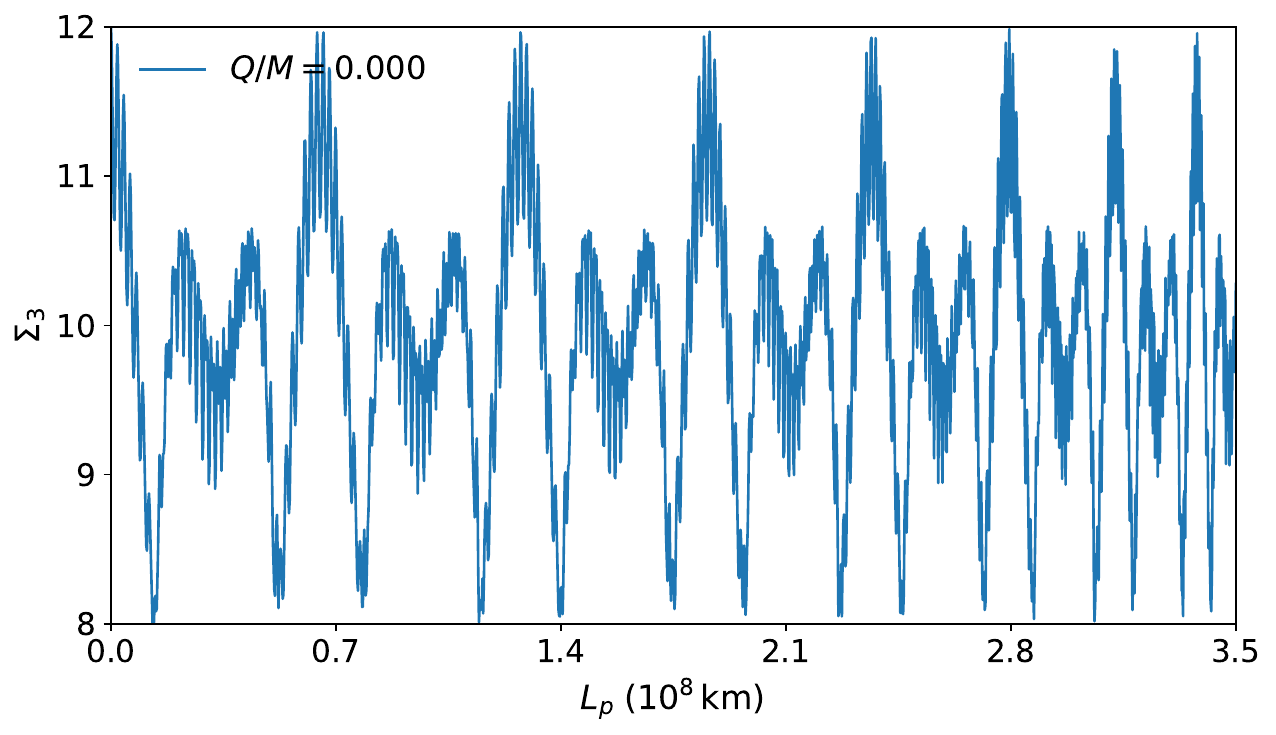}
    \end{minipage}

    \par\vspace{2pt}

    \begin{minipage}{0.32\textwidth}
        \centering
        \includegraphics[width=\linewidth]{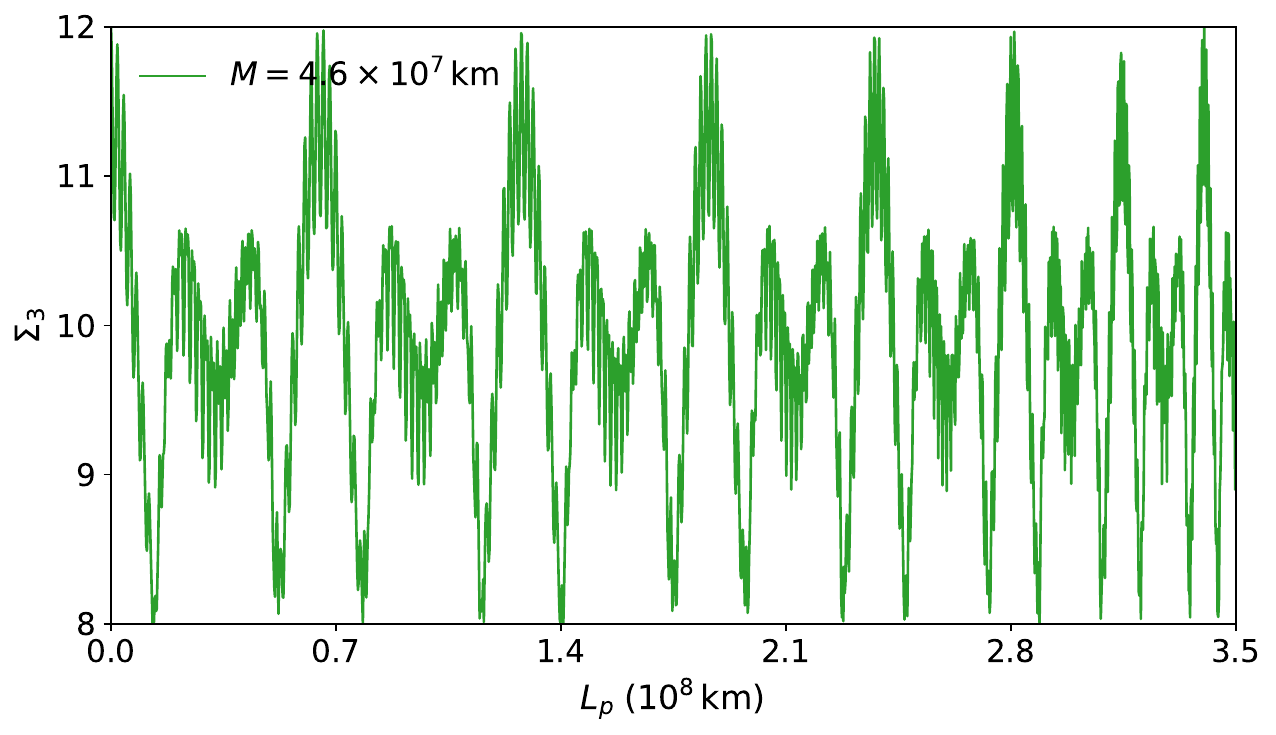}
    \end{minipage}
    \hfill
    \begin{minipage}{0.32\textwidth}
        \centering
        \includegraphics[width=\linewidth]{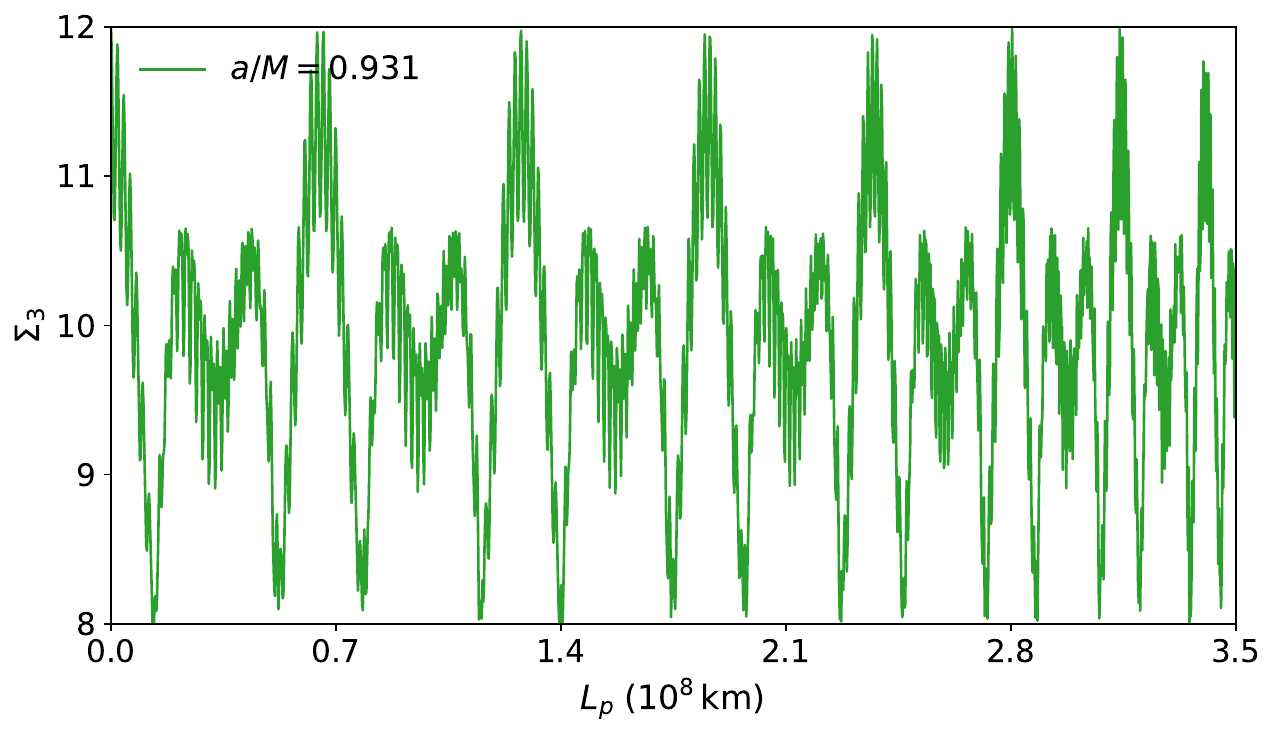}
    \end{minipage}
    \hfill
    \begin{minipage}{0.32\textwidth}
        \centering
        \includegraphics[width=\linewidth]{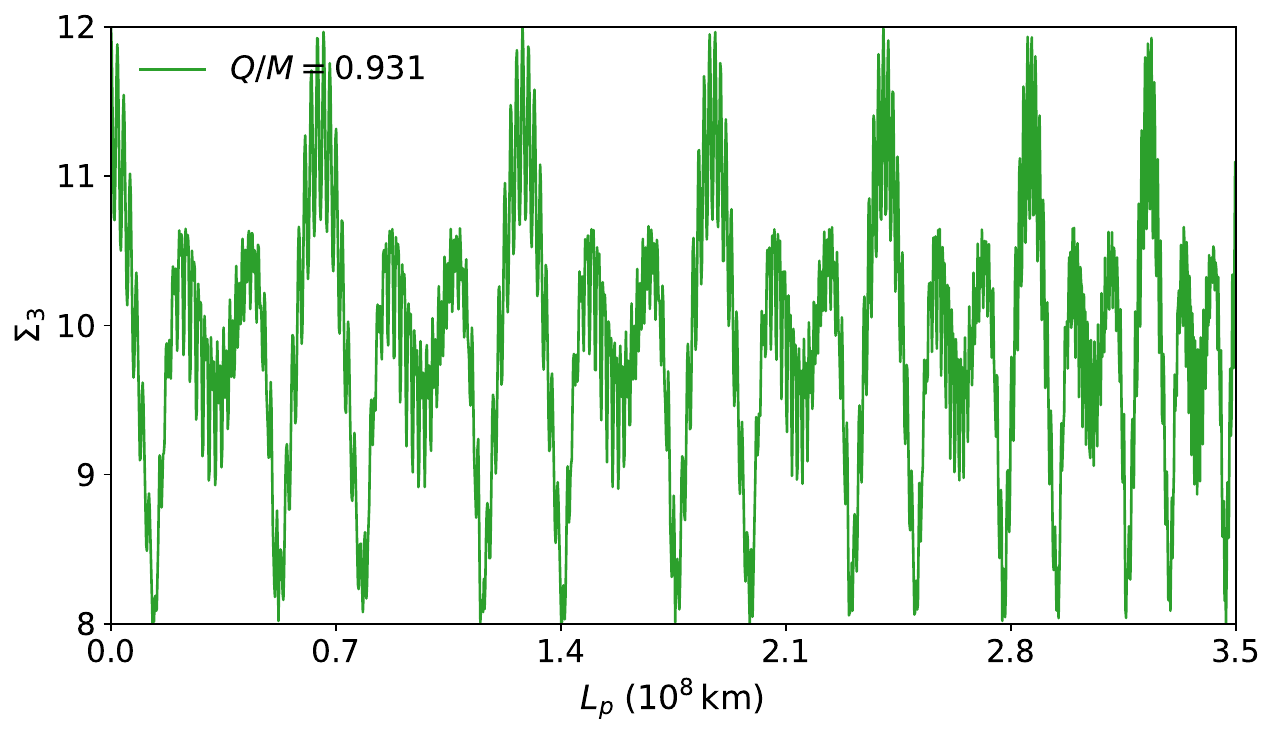}
    \end{minipage}

    \caption{Monogamy of nonlocality for neutrinos during inward propagation in Kerr--Newman spacetime. The left, middle, and right columns show variations in \(M\), \(a\), and \(Q\), respectively.}
    \label{fig:nonlocality_inward}
\end{figure}

Using the same parameter values as in the previous section, the numerical results are presented in Figs.~\ref{fig:nonlocality_outward}--\ref{fig:nonlocality_inward}. For neutrinos propagating with zero angular momentum, the dependence of the nonlocality measure on \(M\), \(a\), and \(Q\) is qualitatively consistent with that of the survival probability
\(P_{\nu_e\rightarrow\nu_e}\) and tripartite entanglement. Therefore, Figs.~\ref{fig:nonlocality_outward} and~\ref{fig:nonlocality_inward} present representative results corresponding only to the minimum and maximum values of each parameter. These results show that variations in
\(M\), \(a\), and \(Q\) mainly modify the oscillation phase and period, while leaving the overall range of nonlocality nearly unchanged.

\vspace{-6pt}

\vspace{-6pt}

\section{Quantum correlations for propagation with nonzero angular momentum}\label{4}
\subsection{Numerical results on oscillation probabilities}

For neutrino propagation with nonzero angular momentum, two scenarios are considered. In the first, a neutrino is produced in a gravitational field and subsequently propagates outward. In the second, a neutrino emitted from the source at $r_A$ approaches the massive object, reaches the point of closest approach $r_0$, and then propagates outward to the detector at $r_B$. The latter scenario corresponds to gravitational lensing. For the first scenario, the phase is given by Eq.~(\ref{eq:formula30}). For the lensing scenario, the phase accumulated along the two trajectory segments is
\begin{equation}\label{eq:formula49}
\Phi_k^{\mathrm{lens}}(b)=\frac{m_k^2}{2E_0}\left[\int_{r_0}^{r_A}\frac{r^2}{\sqrt{R(r,b)}}\,dr+\int_{r_0}^{r_B}\frac{r^2}{\sqrt{R(r,b)}}\,dr\right].
\end{equation}
The distance of closest approach, $r_0$, is determined by the turning-point condition.
\begin{equation}\label{eq:formula50}
\left.\left(\frac{dr}{d\phi}\right)_0\right|_{r=r_0}
=
\left.
\frac{\Delta\sqrt{R(r,b)}}
{-a(Q^2-r_s r)+b(r^2+Q^2-r_s r)}
\right|_{r=r_0}
=
0.
\end{equation}
It follows that
\begin{equation}\label{eq:formula51}
R(r_0,b)
=
r_0^4-(b^2-a^2)r_0^2+(r_s r_0-Q^2)(a-b)^2
=
0.
\end{equation}

Neutrino oscillations may arise not only when different mass eigenstates propagate along the same path but also when mass eigenstates propagate along distinct paths. Therefore, the neutrino flavor state at the detection point can be expressed as follows \cite{Chakrabarty:2021bpr}:
\begin{equation}
|\nu_{\alpha}(t_B, r_B)\rangle = N \sum_i U_{\alpha i}^* \sum_{\mathrm{path}=p,q} \exp\left(-i \Phi_i^{\mathrm{path}}\right) |\nu_i(t_B, r_B)\rangle.
\end{equation}
Here, \(p\) and \(q\) denote two distinct propagation paths. The corresponding probability that a neutrino produced in flavor state \(\alpha\) at source point \(A\) is detected in flavor state \(\beta\) at detection point \(B\) is given by~\cite{Chakrabarty:2021bpr}:
\begin{equation}
\begin{split}
P_{\alpha \beta}^{\text{lens}} &= \left|\langle \nu_{\beta} \mid \nu_{\alpha}(t_B, r_B) \rangle\right|^2 \\
&= |N|^2 \sum_{i,j} U_{\beta i} U_{\beta j}^* U_{\alpha j} U_{\alpha i}^* \sum_{p,q} \exp\left(-i \Delta \Phi_{ij}^{pq}\right),
\end{split}
\label{eq:formula53}
\end{equation}
where $|N|^2$ denotes the normalization constant.
\begin{equation}
|N|^2 = \left( \sum_i |U_{\alpha i}|^2 \sum_{p,q} \exp\left(-i \Delta \Phi_{ii}^{pq}\right) \right)^{-1}.
\label{eq:formula54}
\end{equation}
Neutrino propagation is schematically illustrated in Fig.~\ref{lens}:
\begin{figure}[htbp]
  \centering
  \vspace{-10pt}
  \includegraphics[width=0.8\textwidth]{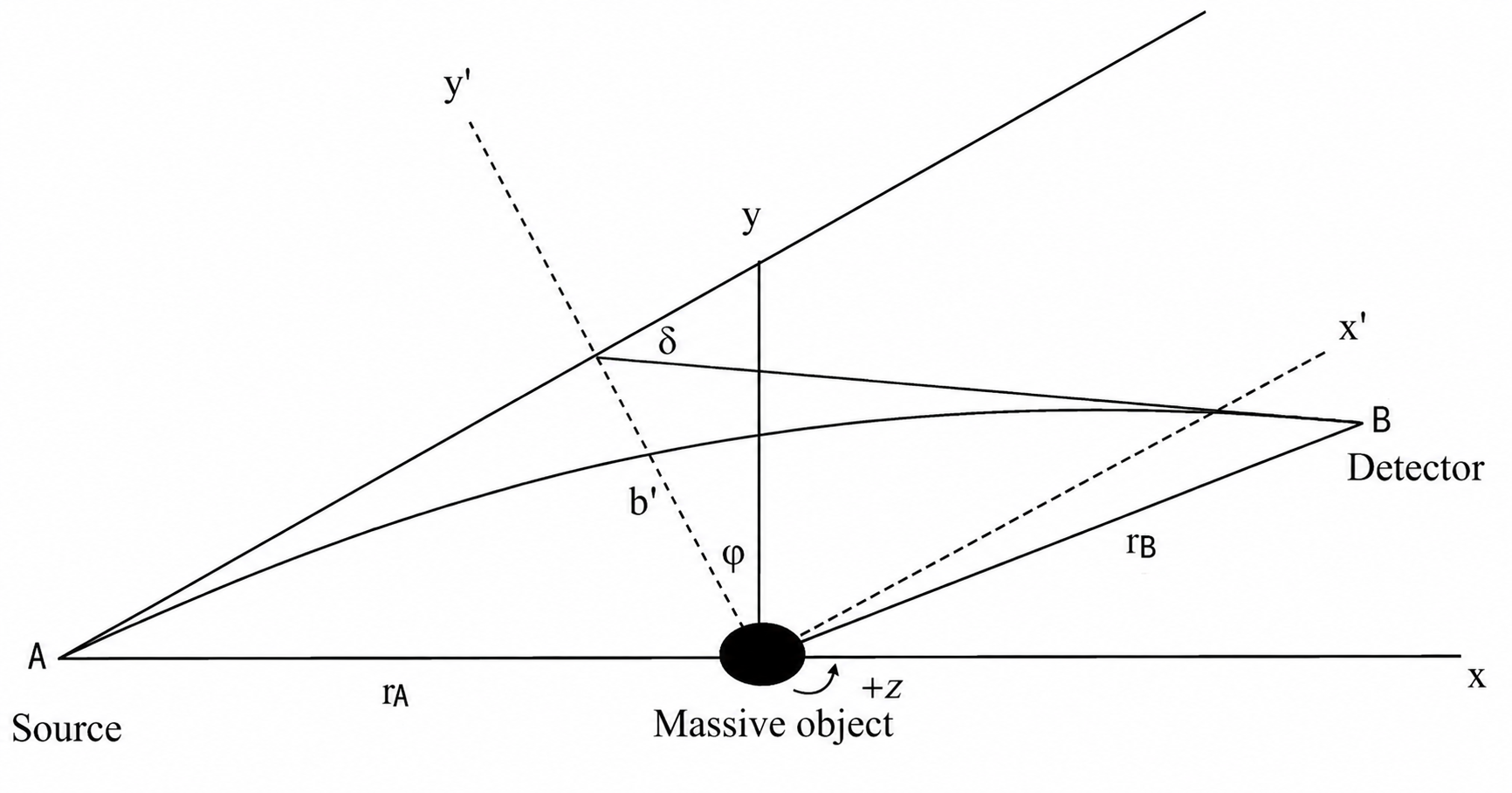}
  \vspace{-5pt}
  \caption{Schematic diagram of weak lensing of neutrinos in Kerr-Newman spacetime. Neutrinos propagate from the source $A$ to the detector $B$ in the exterior of a charged, rotating massive object described by the Kerr-Newman metric.}
  \label{lens}
\end{figure}

Before evaluating the neutrino oscillation probability under the gravitational lensing effect in the Kerr--Newman metric, the impact parameter \(b\), defined by \(J_k=E_k b\,v_k^{(\infty)}\) in Eq.~(\ref{eq:formula22}), must first be determined. As shown in Fig.~\ref{lens}, in the \((x,y)\) coordinate system, the physical distances from the lens to the neutrino source and detector are denoted by \(r_A\) and \(r_B\), respectively. We further introduce a rotated coordinate system, \(x'=x\cos\varphi+y\sin\varphi\) and \(y'=-x\sin\varphi+y\cos\varphi\), in which the \(x'\)-axis is parallel to the incident asymptote. Thus, the incident trajectory has the constant coordinate \(y'=b'\) in this system. Accordingly, \(b'\) is introduced as the geometric impact parameter.
Since the asymptotic neutrino momentum is directed along \(+x'\), we have
\[
J_k=(\boldsymbol{r}\times\boldsymbol{p})_z
=x'p_{y'}-y'p_{x'}
=-E_kv_k^{(\infty)}b'.
\]
Comparing with \(J_k=E_kb\,v_k^{(\infty)}\) gives
\[
b'=-b.
\]
Thus, for \(a>0\), \(b<0\) denotes the upper retrograde trajectory, whereas \(b>0\) denotes the lower prograde trajectory. Because the parameter regime considered for gravitational lensing satisfies the weak-field conditions, the corresponding signed deflection angle, expressed in terms of the geometric impact parameter \(b'\), can be approximated as~\cite{Li:2019mqw,Li:2021xhy}:
\begin{equation}\label{eq:deflection_angle}
\delta(b')
\simeq
-\frac{y'_B-b'}{x'_B}
=
\frac{4M}{b'}
+\frac{4aM}{b'^2}
-\frac{3\pi Q^2}{4b'|b'|}.
\end{equation}
Here, \(\delta\) and \(\varphi\) are assumed to be extremely small angles. From Fig.~\ref{lens}, the geometric relation \(\sin\varphi = b'/r_A\) follows. Thus, using Eq.~(\ref{eq:deflection_angle}), we obtain the following result:
\begin{equation}\label{eq:formula56}
-b'-\frac{b'x_B}{r_A}
+\frac{b'\left(\frac{4M}{b'}+\frac{4aM}{b^{\prime 2}}-\frac{3\pi Q^2}{4b'|b'|}\right)y_B}{r_A}
+\sqrt{1-\frac{b^{\prime 2}}{r_A^2}}
\left[
\left(\frac{4M}{b'}+\frac{4aM}{b^{\prime 2}}-\frac{3\pi Q^2}{4b'|b'|}\right)x_B+y_B
\right]
=0.
\end{equation}
We assume that the detector follows a circular trajectory around the central massive object, with \(x_B = r_B \cos \phi\) and \(y_B = r_B \sin \phi\). For each value of \(\phi\), Eq.~\eqref{eq:formula56} is solved separately for the \(b'>0\) and \(b'<0\) branches using Brent's method\footnote{Brent's method is a bracketed one-dimensional root-finding algorithm that combines bisection and inverse quadratic interpolation. See
\url{https://docs.scipy.org/doc/scipy/reference/generated/scipy.optimize.brentq.html}.}. This yields two roots that satisfy the weak-lensing conditions, \(b'_+>0\) and \(b'_-<0\). These roots are then converted into the corresponding impact parameters using \(b=-b'\). For each \(b_\pm\), \(r_0\) is determined from Eq.~\eqref{eq:formula51} as the largest positive real root satisfying \(r_+<r_0<\min(r_A,r_B)\). The leptonic mixing parameters are set to the same values as those used for radial propagation. The remaining parameters are set to \(r_B = 10^8\,\text{km}\), \(r_A = 10^5 r_B\), and \(E_0 = 10\,\text{MeV}\). Using Eqs.~(\ref{eq:formula49}), (\ref{eq:formula53}), and~(\ref{eq:formula54}), we calculate the neutrino oscillation probabilities with nonzero angular momentum in the Kerr--Newman space-time as functions of \(\phi \in [0,0.0035]\). Representative numerical values of the two impact parameters \(b_\pm\), their corresponding closest-approach radii \(r_{0\pm}\), and the outer event-horizon radius \(r_+\) are provided in Appendix~\ref{app:exterior-horizon-lensing} for all three parameter scans.

\begin{figure}[H]
  \centering
  \begin{minipage}{0.32\textwidth}
    \centering
    \includegraphics[width=\linewidth]{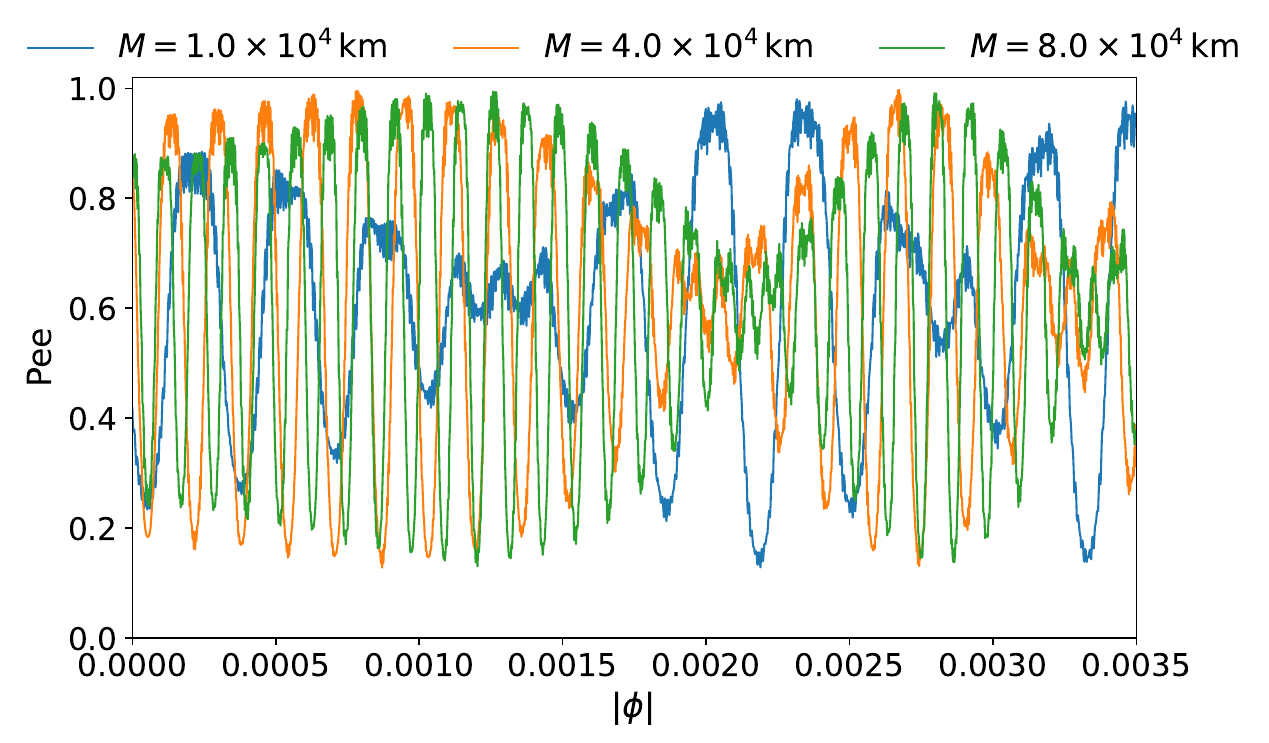}
  \end{minipage}
  \hfill
  \begin{minipage}{0.32\textwidth}
    \centering
    \includegraphics[width=\linewidth]{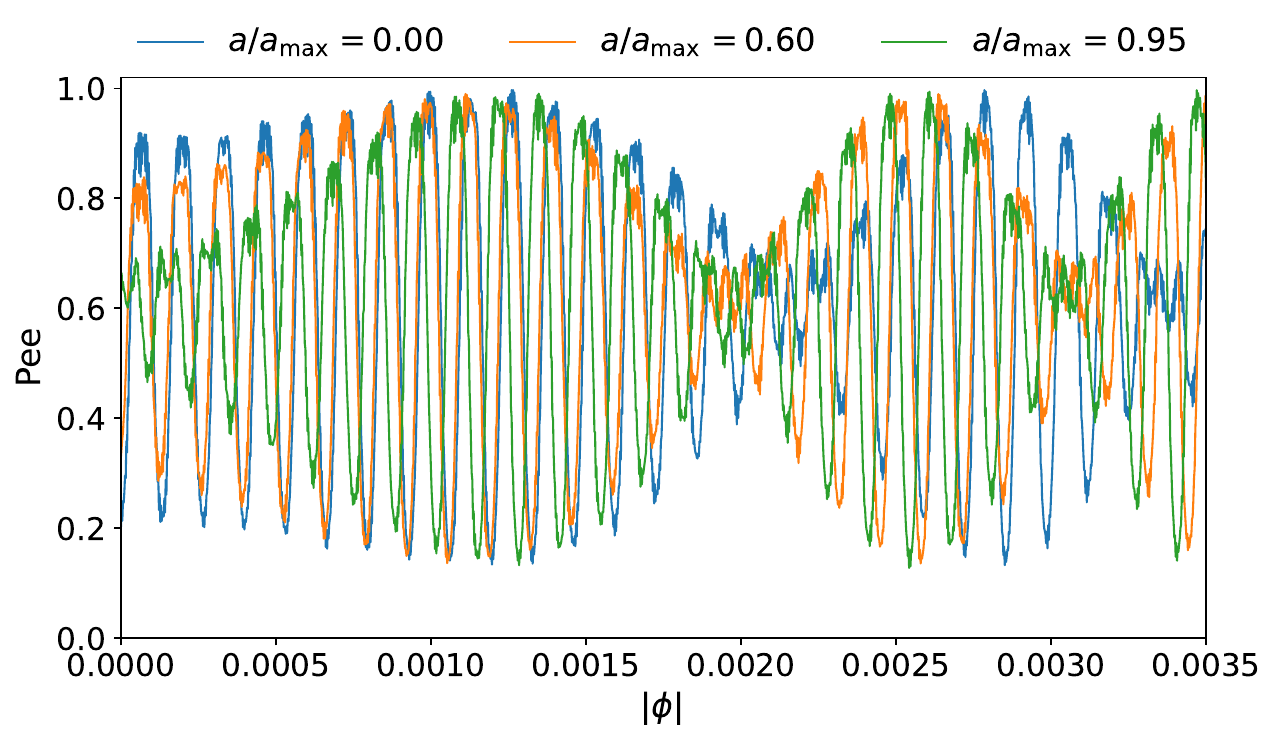}
  \end{minipage}
  \hfill
  \begin{minipage}{0.32\textwidth}
    \centering
    \includegraphics[width=\linewidth]{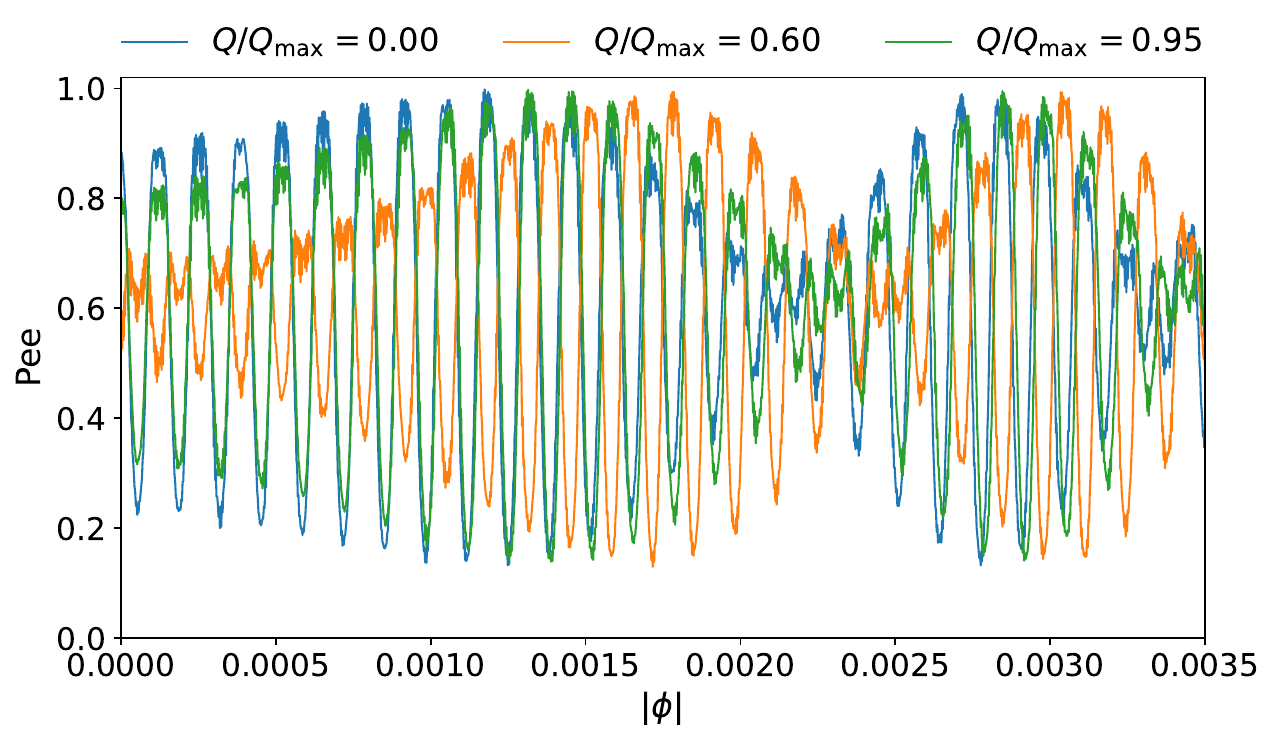}
  \end{minipage}
  \hfill
 \caption{Neutrino oscillation probability \(P_{\nu_e \to \nu_e}\) for propagation with nonzero angular momentum under different metric parameters.}
  \label{fig:fig12}
\end{figure}

For the gravitational-lensing case, the Kerr--Newman parameters must also satisfy the horizon condition. The specific parameter settings are summarized in Table~\ref{tab:lensing_parameters}.
\begin{table}[H]
    \centering
    \caption{Metric parameters used in the gravitational-lensing calculations. All values are expressed in units of \(10^{4}\,\mathrm{km}\).}
    \label{tab:lensing_parameters}
    \renewcommand{\arraystretch}{1.15}
    \begin{tabular}{lccc}
        \hline
        Case & \(M\) & \(a\) & \(Q\) \\
        \hline
        Variation of \(M\)
        & 1.0,\;4.0,\;8.0
        & 0.5
        & 0.5 \\

        Variation of \(a\)
        & 6.0
        & 0,\;3.5875,\;5.6802
        & 0.5 \\

        Variation of \(Q\)
        & 6.0
        & 0.5
        & 0,\;3.5875,\;5.6802 \\
        \hline
    \end{tabular}
\end{table}
The numerical results for the nonzero-angular-momentum oscillation probability \(P_{ee}\) are presented in Fig.~\ref{fig:fig12}.

As shown in Fig.~\ref{fig:fig12}, within the parameter ranges considered, increasing \(M\) leads to more closely spaced oscillations, whereas \(a\) and \(Q\) primarily induce phase shifts. The overall range of \(P_{\nu_e\rightarrow\nu_e}\) remains nearly unchanged. Furthermore, compared with the zero-angular-momentum case, the effects of the metric parameters in the gravitational-lensing case are not limited to rescaling the oscillation period; they also include local profile modulations.
\vspace{-6pt}

\subsection{Entanglement of neutrinos for nonzero-angular-momentum propagation}
\begin{figure}[H]
  \centering
  \begin{minipage}{0.32\textwidth}
    \centering
    \includegraphics[width=\linewidth]{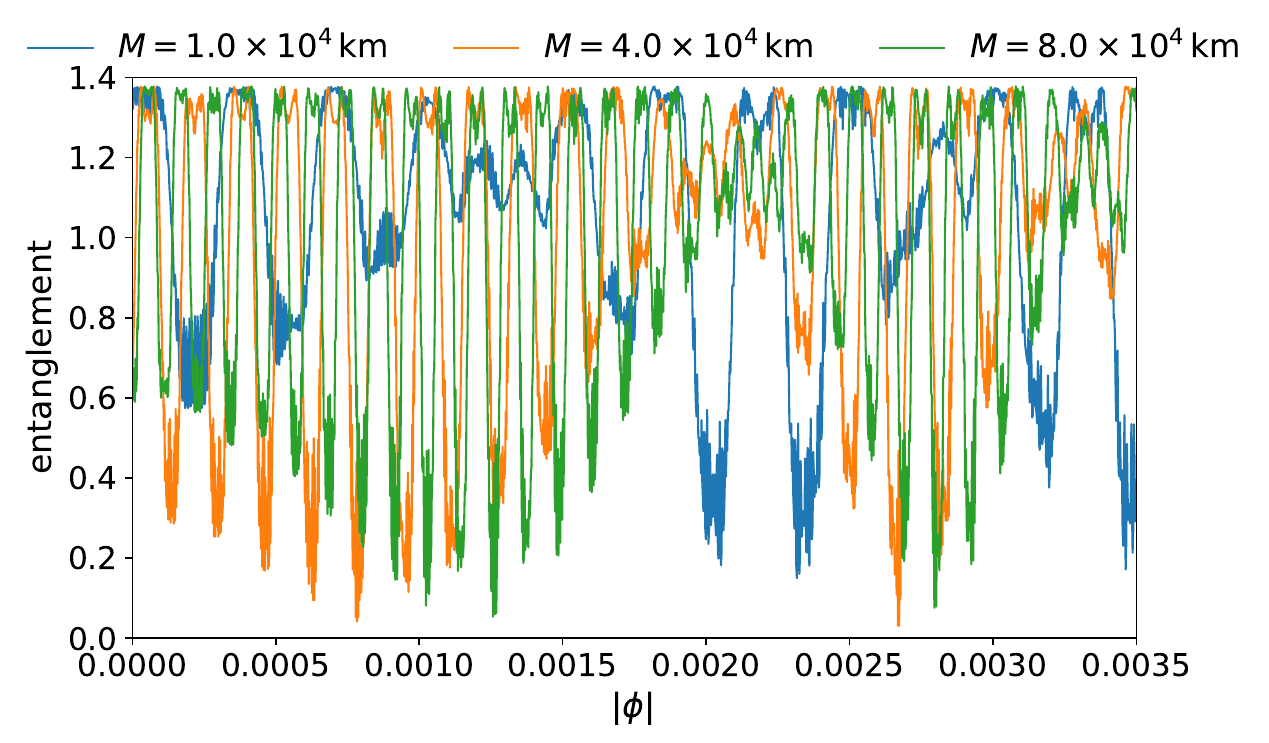}
  \end{minipage}
  \hfill
  \begin{minipage}{0.32\textwidth}
    \centering
    \includegraphics[width=\linewidth]{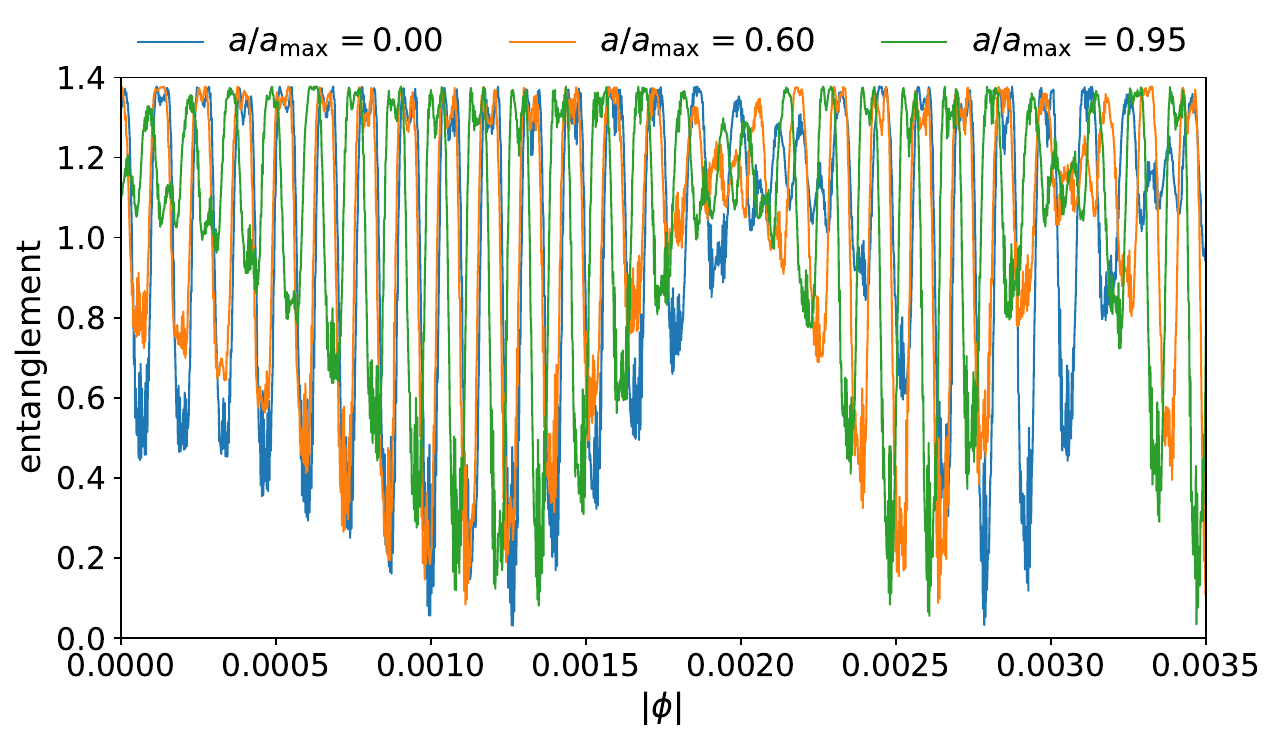}
  \end{minipage}
  \hfill
  \begin{minipage}{0.32\textwidth}
    \centering
    \includegraphics[width=\linewidth]{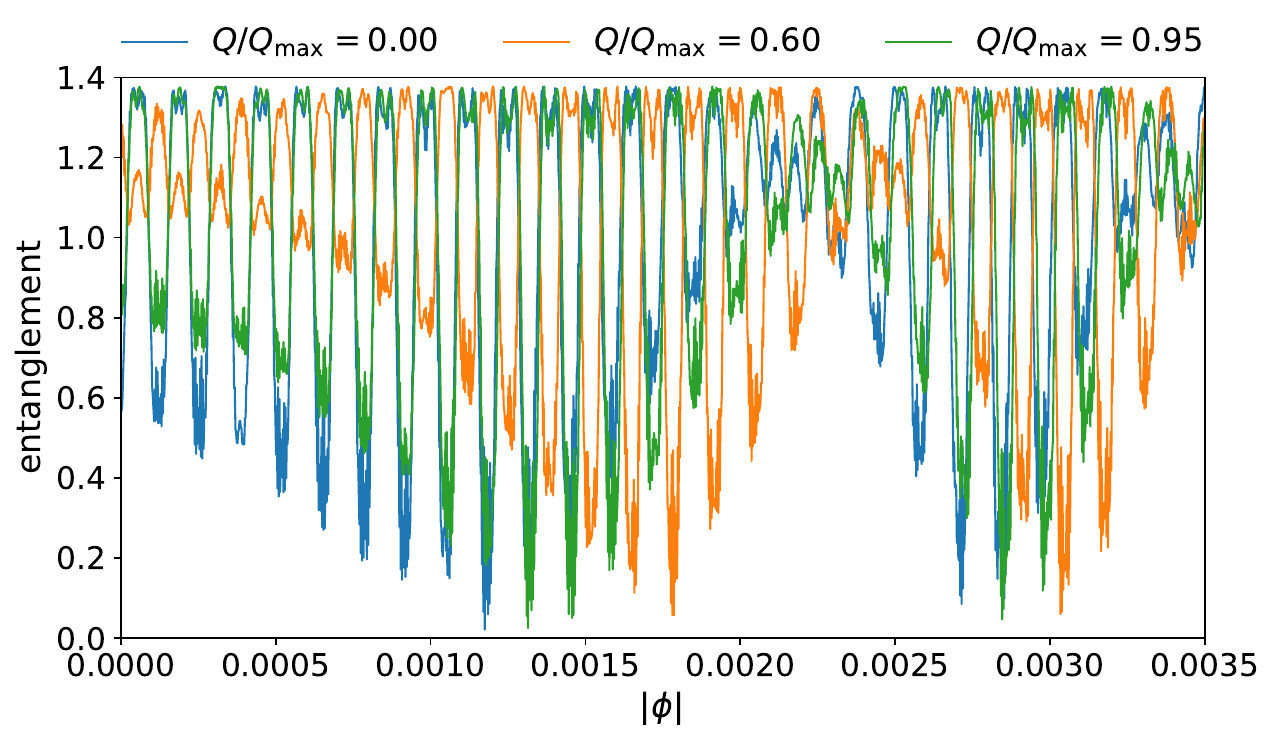}
  \end{minipage}
  \hfill
 \caption{Tripartite entanglement of neutrinos propagating with nonzero angular momentum under different metric parameters.}
  \label{fig:fig13}
\end{figure}

We employ the same quantification method and initial neutrino state as those used in the zero-angular-momentum case, as given in Eq.~(\ref{eq:formula42}). Using the parameter values adopted in Fig.~\ref{fig:fig12}, the corresponding results are presented in Fig.~\ref{fig:fig13}.

As shown in Fig.~\ref{fig:fig13}, for propagation with nonzero angular momentum, the effects of \(M\), \(a\), and \(Q\) on the oscillatory behavior of tripartite entanglement mirror those observed for \(P_{\nu_e\rightarrow\nu_e}\). Moreover, none of these parameters enhances or suppresses the entanglement. The additional lensed path mainly introduces a relative phase, thereby redistributing the normalized flavor-transition probabilities with \(\lvert\phi\rvert\) while leaving the overall entanglement range nearly unchanged.
\vspace{-6pt}

\subsection{Monogamy of nonlocality for neutrinos with nonzero-angular-momentum propagation}

\begin{figure}[H]
    \centering

    \begin{minipage}{0.32\textwidth}
        \centering
        \includegraphics[width=\linewidth]{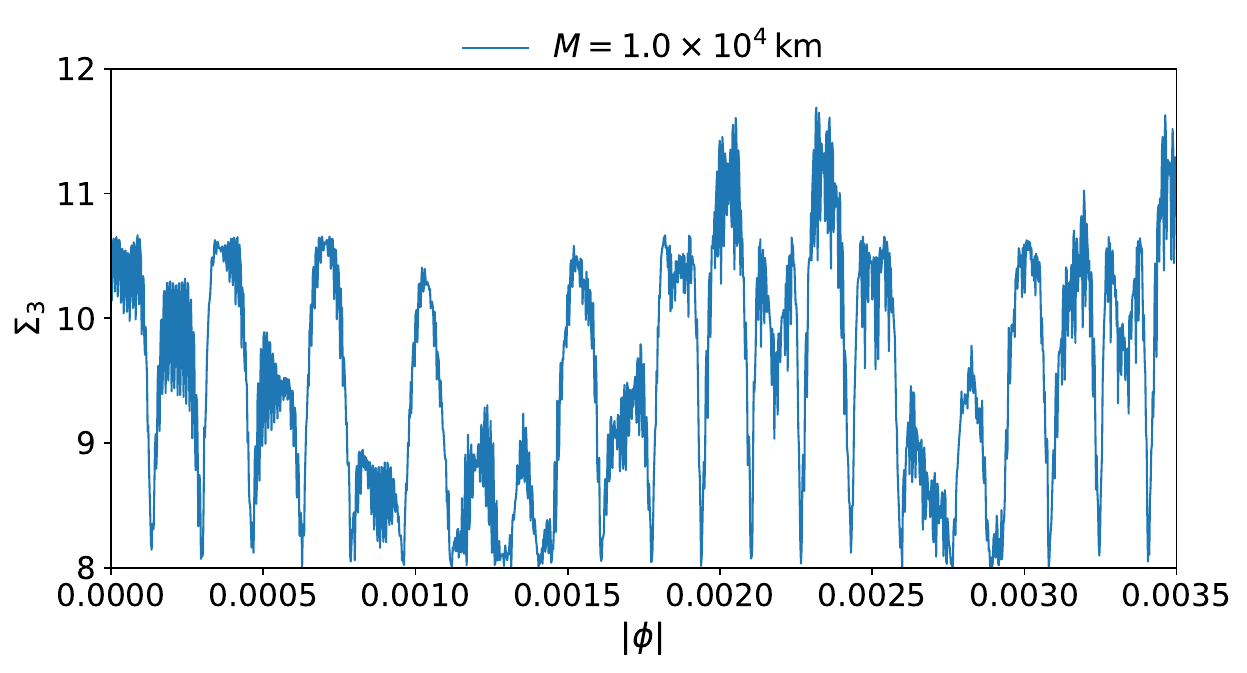}
    \end{minipage}
    \hfill
    \begin{minipage}{0.32\textwidth}
        \centering
        \includegraphics[width=\linewidth]{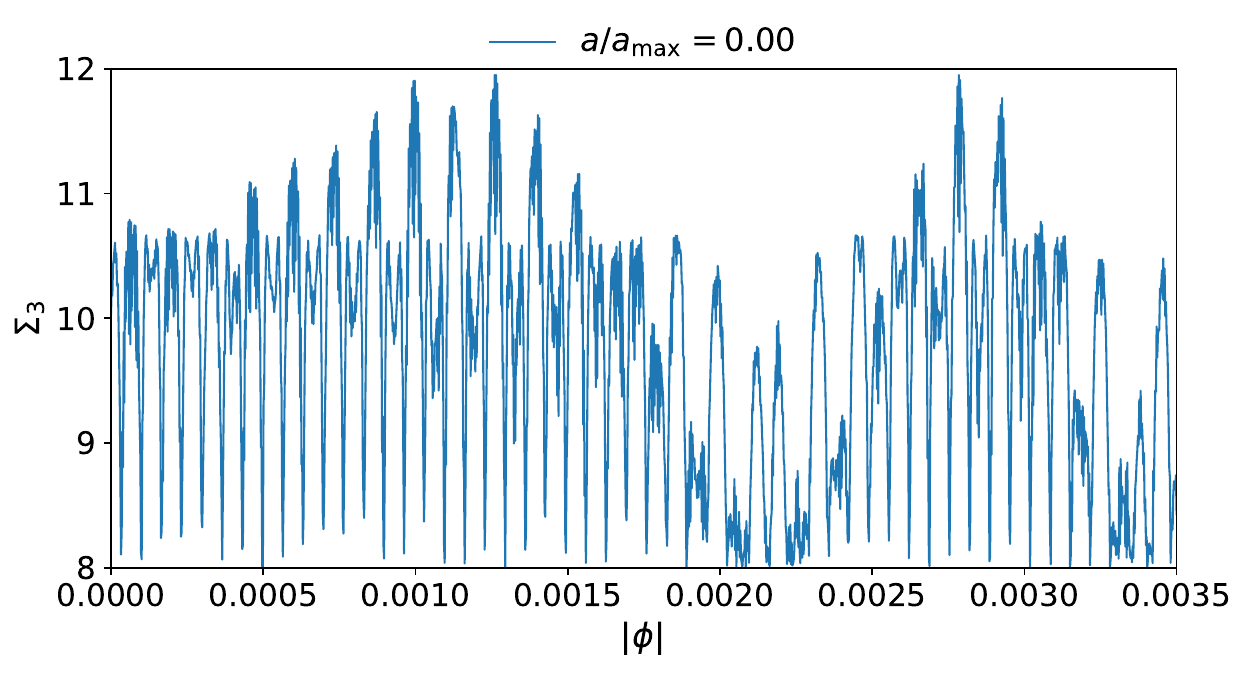}
    \end{minipage}
    \hfill
    \begin{minipage}{0.32\textwidth}
        \centering
        \includegraphics[width=\linewidth]{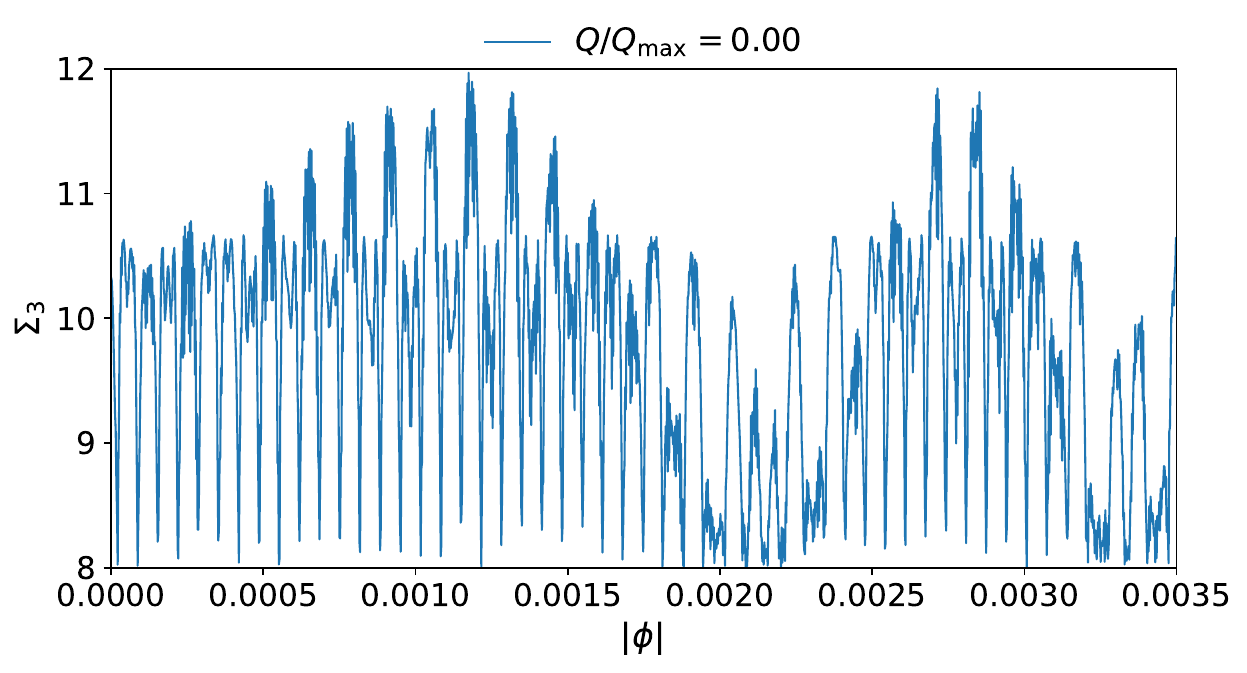}
    \end{minipage}

    \par\vspace{2pt}

    \begin{minipage}{0.32\textwidth}
        \centering
        \includegraphics[width=\linewidth]{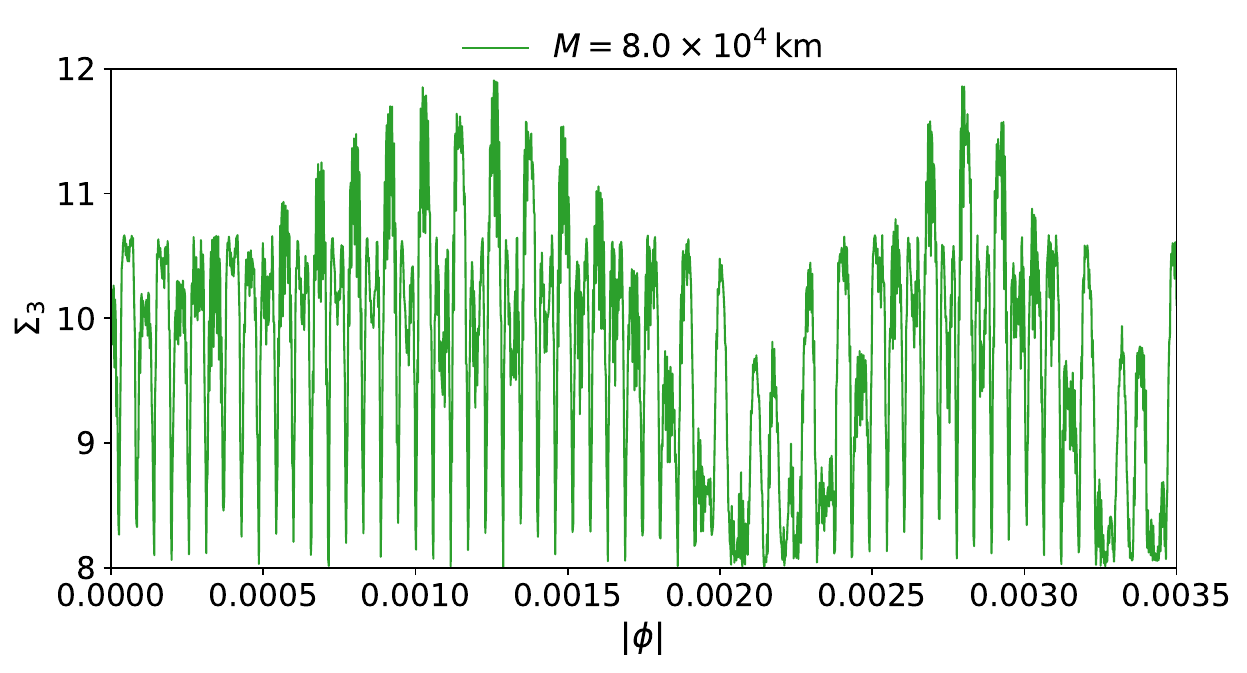}
    \end{minipage}
    \hfill
    \begin{minipage}{0.32\textwidth}
        \centering
        \includegraphics[width=\linewidth]{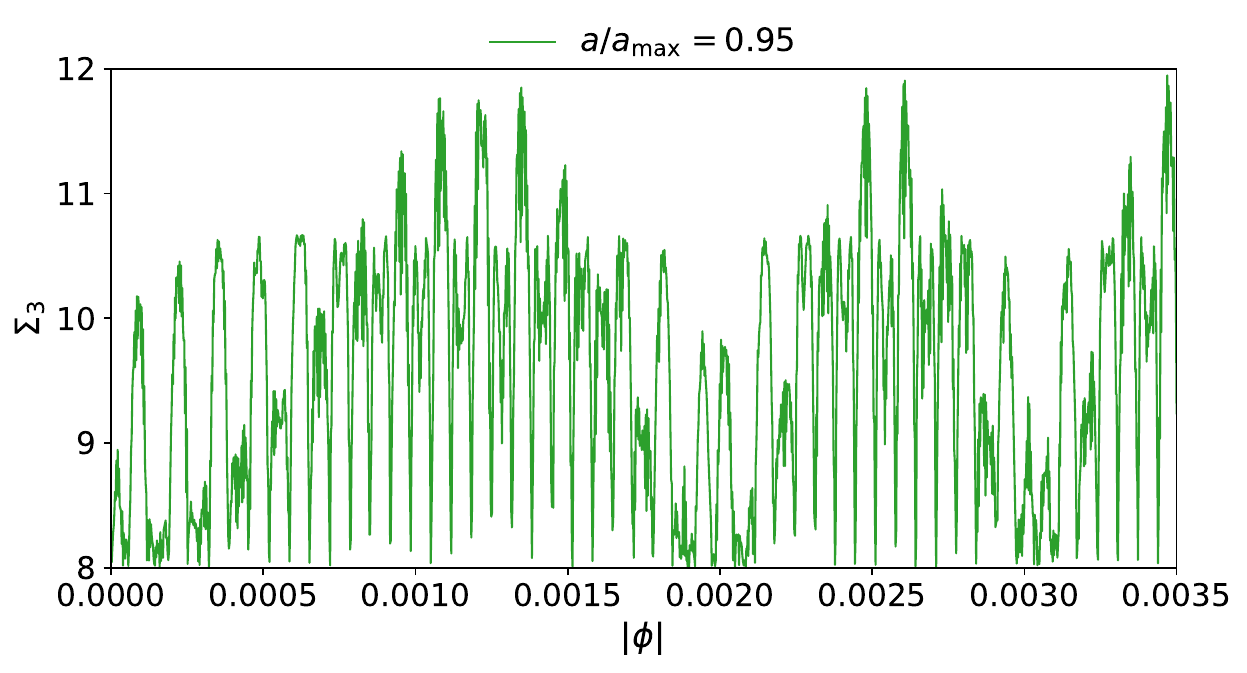}
    \end{minipage}
    \hfill
    \begin{minipage}{0.32\textwidth}
        \centering
        \includegraphics[width=\linewidth]{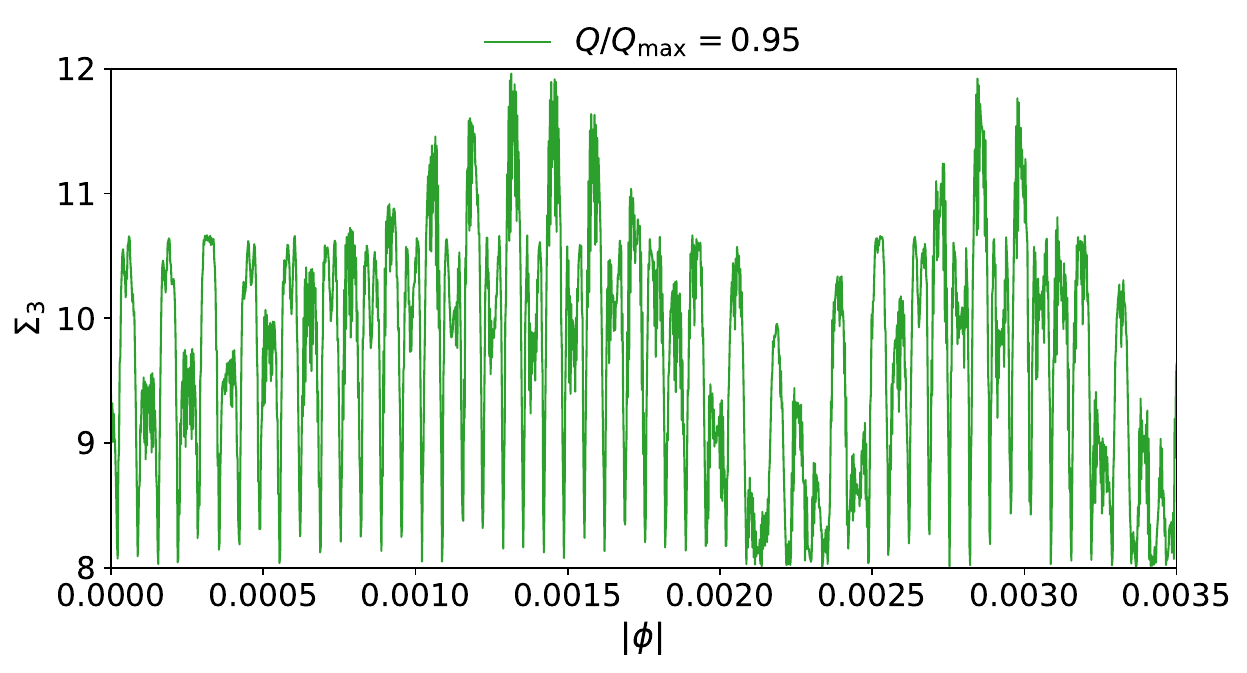}
    \end{minipage}

    \caption{Monogamy of nonlocality for neutrinos propagating with nonzero angular momentum under different metric parameters.}
    \label{fig:fig15}
\end{figure}

We now evaluate the monogamy of nonlocality for neutrinos with nonzero angular momentum in Kerr--Newman space-time. The initial neutrino state and the method used to quantify nonlocality are the same as those in the zero-angular-momentum case, as given in Eq.~(\ref{eq:formula48}). Using the parameter settings adopted in Fig.~\ref{fig:fig12}, the corresponding results are presented in Fig.~\ref{fig:fig15}.

As shown in Fig.~\ref{fig:fig15}, the variations in nonlocality induced by \(M\), \(a\), and \(Q\) for neutrinos with nonzero angular momentum are qualitatively analogous to those observed for the survival probability and tripartite entanglement. To avoid redundancy, Fig.~\ref{fig:fig15} presents only representative results for the minimum and maximum values of each parameter.

\vspace{-6pt}

For comparison, we also calculate the quantum coherence of neutrinos for propagation with zero and nonzero angular momentum. The detailed procedure and corresponding results are presented in Appendix~\ref{coherence}.

\section{Conclusions}\label{5}
In this paper, we investigated the properties of QCs, including entanglement, monogamy of nonlocality, and coherence, in the Kerr--Newman spacetime. We examined the interplay between the metric parameters and the QCs. Two propagation scenarios were considered: zero-angular-momentum propagation in both inward and outward directions, and nonzero-angular-momentum propagation. For the nonzero-angular-momentum case, we focused primarily on gravitational lensing effects.

For zero-angular-momentum propagation, the three QCs exhibit similar oscillatory behavior, whereas their period evolution depends on the propagation direction. The oscillation periods remain nearly constant during outward propagation but progressively decrease during inward propagation, with the oscillation periods in Kerr--Newman spacetime lying between those in flat and Schwarzschild spacetimes. The parameters \(M\), \(a\), and \(Q\) primarily modify the oscillation phase and period without producing a systematic enhancement or suppression of the QCs. For outward propagation, increasing \(M\) lengthens the period, whereas increasing \(a\) or \(Q\) shortens it. These dependencies are reversed for inward propagation: increasing \(M\) accelerates the period reduction, whereas increasing \(a\) or \(Q\) weakens this reduction, with the influence of \(a\) being comparatively weak. As a result, the oscillation curves corresponding to different values of the selected parameter gradually separate as \(L_p\) increases during inward propagation. The same trends are observed in \(P_{\nu_e\rightarrow\nu_e}\), reflecting the intrinsic dependence of the QCs on the neutrino oscillation wavelength. These results provide a theoretical basis for understanding how the electric charge and angular momentum of massive celestial bodies affect neutrino oscillations and their QCs.

For nonzero-angular-momentum propagation under gravitational lensing, the metric parameters significantly modify the oscillation patterns of \(P_{\nu_e\rightarrow\nu_e}\) and the three QCs. Increasing \(M\) produces denser oscillations, whereas \(a\) and \(Q\) primarily induce phase shifts. As in the zero-angular-momentum case, none of these parameters produces a systematic enhancement or suppression of the QCs. The key distinction is that their effects are not limited to rescaling the oscillation period but also include local profile modulations. This additional structure arises from the interference terms associated with the two lensed trajectories, which generate additional path-dependent phase differences that vary with \(\lvert\phi\rvert\) and the metric parameters, thereby modifying the local oscillation pattern.

Overall, our work provides a comprehensive understanding of the influence of compact celestial objects on neutrino oscillation probabilities and QCs. In addition, by extending the investigation to nonzero-angular-momentum propagation of neutrinos, this study offers a new perspective for exploring other QCs of neutrinos.

\section*{Acknowledgements}
We thank Xue-Ke Song for helpful discussions. This work was supported by the National Natural Science Foundation of China under Grant No. 12065007 and the Natural Science Foundation of Guangxi under Grant No. 2026GXNSFAA00640923.

\appendix
\section{Verification That the Propagation Paths Remain Outside the Outer Horizon}
\label{app:exterior-horizon-verification}
\subsection{Zero-Angular-Momentum Propagation}
\label{app:exterior-horizon-zero-angular}
For each parameter scan, the labels ``scan 1,'' ``scan 2,'' and ``scan 3'' denote the three parameter values listed in Table~\ref{tab:zero_angular_parameters}, ordered from smallest to largest. For outward propagation, the minimum radius is simply the initial radius, \(r_A=1\times10^{8}\,\mathrm{km}\). For inward propagation, the minimum radius is the final radius, \(r_B\), which is obtained by numerically inverting the proper-distance relation in Eq.~\eqref{eq:formula37}. All radii are expressed in units of \(10^{7}\,\mathrm{km}\).

\begin{table}[H]
    \centering
    \caption{Minimum radii of zero-angular-momentum propagation paths and the corresponding outer-horizon radii.}
    \label{tab:radial-horizon-check}
    \begin{tabular*}{\textwidth}{@{\extracolsep{\fill}}lccc@{}}
        \hline
        Case
        & \(r_{\min}^{\mathrm{out}}\)
        & \(r_{\min}^{\mathrm{in}}\)
        & \(r_{+}\) \\
        \hline
        Schwarzschild
            & 10.00 & 13.35 & 9.60 \\
        Kerr--Newman
            & 10.00 & 12.53 & 6.69 \\
        \(M\) scan 1
            & 10.00 & 8.64 & 3.41 \\
        \(M\) scan 2
            & 10.00 & 10.85 & 6.28 \\
        \(M\) scan 3
            & 10.00 & 12.95 & 8.98 \\
        \(a\) scan 1
            & 10.00 & 13.31 & 9.50 \\
        \(a\) scan 2
            & 10.00 & 12.98 & 8.56 \\
        \(a\) scan 3
            & 10.00 & 12.47 & 6.27 \\
        \(Q\) scan 1
            & 10.00 & 13.31 & 9.50 \\
        \(Q\) scan 2
            & 10.00 & 12.98 & 8.56 \\
        \(Q\) scan 3
            & 10.00 & 12.47 & 6.27 \\
        \hline
    \end{tabular*}
\end{table}

\subsection{Nonzero-Angular-Momentum Lensing Propagation}
\label{app:exterior-horizon-lensing}
The representative detector angles \(\phi=0\), \(0.00175\), and \(0.0035\) correspond to the beginning, midpoint, and end of the plotted interval, respectively. In all three tables, \(M\), \(a\), \(Q\), and \(r_+\) are given
in units of \(10^4\,\mathrm{km}\), whereas \(b_\pm\) and \(r_{0\pm}\) are given in units of \(10^6\,\mathrm{km}\).

\begingroup
\setlength{\intextsep}{6pt plus 1pt minus 1pt}
\setlength{\textfloatsep}{6pt plus 1pt minus 1pt}
\setlength{\floatsep}{6pt plus 1pt minus 1pt}
\setlength{\abovecaptionskip}{2pt}
\setlength{\belowcaptionskip}{3pt}
\renewcommand{\arraystretch}{0.95}

\begin{table}[H]
\centering
\caption{Representative physical lensing solutions for the
\(M\)-scan with
\(a=Q=0.5\times10^4\,\mathrm{km}\).}
\label{tab:appendix_b_M_scan}
\small
\begin{tabular*}{\textwidth}
{@{\extracolsep{\fill}}cccccccc@{}}
\hline
\(M\) & \(r_+\) & \(\phi\) & \(b_-\) & \(b_+\) &
\(r_{0-}\) & \(r_{0+}\) \\
\hline
1.0000 & 1.7071 & 0.0000  & -2.0018 & 1.9967 & 1.9916 & 1.9867 \\
1.0000 & 1.7071 & 0.00175 & -2.0911 & 1.9110 & 2.0810 & 1.9010 \\
1.0000 & 1.7071 & 0.0035  & -2.1842 & 1.8291 & 2.1741 & 1.8191 \\
4.0000 & 7.9370 & 0.0000  & -4.0023 & 3.9973 & 3.9616 & 3.9568 \\
4.0000 & 7.9370 & 0.00175 & -4.0907 & 3.9107 & 4.0500 & 3.8702 \\
4.0000 & 7.9370 & 0.0035  & -4.1810 & 3.8260 & 4.1403 & 3.7855 \\
8.0000 & 15.9687 & 0.0000  & -5.6592 & 5.6542 & 5.5773 & 5.5726 \\
8.0000 & 15.9687 & 0.00175 & -5.7474 & 5.5674 & 5.6655 & 5.4857 \\
8.0000 & 15.9687 & 0.0035  & -5.8368 & 5.4818 & 5.7550 & 5.4002 \\
\hline
\end{tabular*}
\end{table}

\begin{table}[H]
\centering
\caption{Representative physical lensing solutions for the
\(a\)-scan with
\(M=6.0\times10^4\,\mathrm{km}\) and
\(Q=0.5\times10^4\,\mathrm{km}\).
Here, \(a_{\max}=\sqrt{M^2-Q^2}
=5.9791\times10^4\,\mathrm{km}\).}
\label{tab:appendix_b_a_scan}
\small
\begin{tabular*}{\textwidth}
{@{\extracolsep{\fill}}cccccccc@{}}
\hline
\(a/a_{\max}\) & \(a\) & \(r_+\) & \(\phi\) &
\(b_-\) & \(b_+\) & \(r_{0-}\) & \(r_{0+}\) \\
\hline
0.0000 & 0.0000 & 11.9791 & 0.0000  & -4.8988 & 4.8988 & 4.8377 & 4.8377 \\
0.0000 & 0.0000 & 11.9791 & 0.00175 & -4.9871 & 4.8121 & 4.9260 & 4.7510 \\
0.0000 & 0.0000 & 11.9791 & 0.0035  & -5.0769 & 4.7269 & 5.0158 & 4.6658 \\
0.6000 & 3.5875 & 10.7833 & 0.0000  & -4.9167 & 4.8808 & 4.8545 & 4.8204 \\
0.6000 & 3.5875 & 10.7833 & 0.00175 & -5.0046 & 4.7937 & 4.9425 & 4.7334 \\
0.6000 & 3.5875 & 10.7833 & 0.0035  & -5.0942 & 4.7083 & 5.0321 & 4.6479 \\
0.9500 & 5.6802 & 7.8670 & 0.0000  & -4.9270 & 4.8702 & 4.8641 & 4.8101 \\
0.9500 & 5.6802 & 7.8670 & 0.00175 & -5.0148 & 4.7829 & 4.9519 & 4.7229 \\
0.9500 & 5.6802 & 7.8670 & 0.0035  & -5.1041 & 4.6972 & 5.0413 & 4.6372 \\
\hline
\end{tabular*}
\end{table}

\begin{table}[H]
\centering
\caption{Representative physical lensing solutions from the
\(Q\) scan, with
\(M=6.0\times10^4\,\mathrm{km}\) and
\(a=0.5\times10^4\,\mathrm{km}\).
Here, \(Q_{\max}=\sqrt{M^2-a^2}
=5.9791\times10^4\,\mathrm{km}\).}
\label{tab:appendix_b_Q_scan}
\small
\begin{tabular*}{\textwidth}
{@{\extracolsep{\fill}}cccccccc@{}}
\hline
\(Q/Q_{\max}\) & \(Q\) & \(r_+\) & \(\phi\) &
\(b_-\) & \(b_+\) & \(r_{0-}\) & \(r_{0+}\) \\
\hline
0.0000 & 0.0000 & 11.9791 & 0.0000  & -4.9015 & 4.8965 & 4.8402 & 4.8354 \\
0.0000 & 0.0000 & 11.9791 & 0.00175 & -4.9897 & 4.8097 & 4.9284 & 4.7487 \\
0.0000 & 0.0000 & 11.9791 & 0.0035  & -5.0795 & 4.7245 & 5.0182 & 4.6634 \\
0.6000 & 3.5875 & 10.7833 & 0.0000  & -4.8951 & 4.8901 & 4.8340 & 4.8292 \\
0.6000 & 3.5875 & 10.7833 & 0.00175 & -4.9835 & 4.8032 & 4.9224 & 4.7423 \\
0.6000 & 3.5875 & 10.7833 & 0.0035  & -5.0734 & 4.7179 & 5.0123 & 4.6570 \\
0.9500 & 5.6802 & 7.8670 & 0.0000  & -4.8856 & 4.8805 & 4.8246 & 4.8198 \\
0.9500 & 5.6802 & 7.8670 & 0.00175 & -4.9741 & 4.7935 & 4.9132 & 4.7328 \\
0.9500 & 5.6802 & 7.8670 & 0.0035  & -5.0642 & 4.7080 & 5.0033 & 4.6473 \\
\hline
\end{tabular*}
\end{table}
\endgroup

\section{Quantum coherence of neutrinos in the Kerr- Newman space-time}
\label{coherence}
\begin{figure}[H]
  \centering
  \begin{minipage}{0.48\textwidth}
    \centering
    \includegraphics[width=\linewidth]{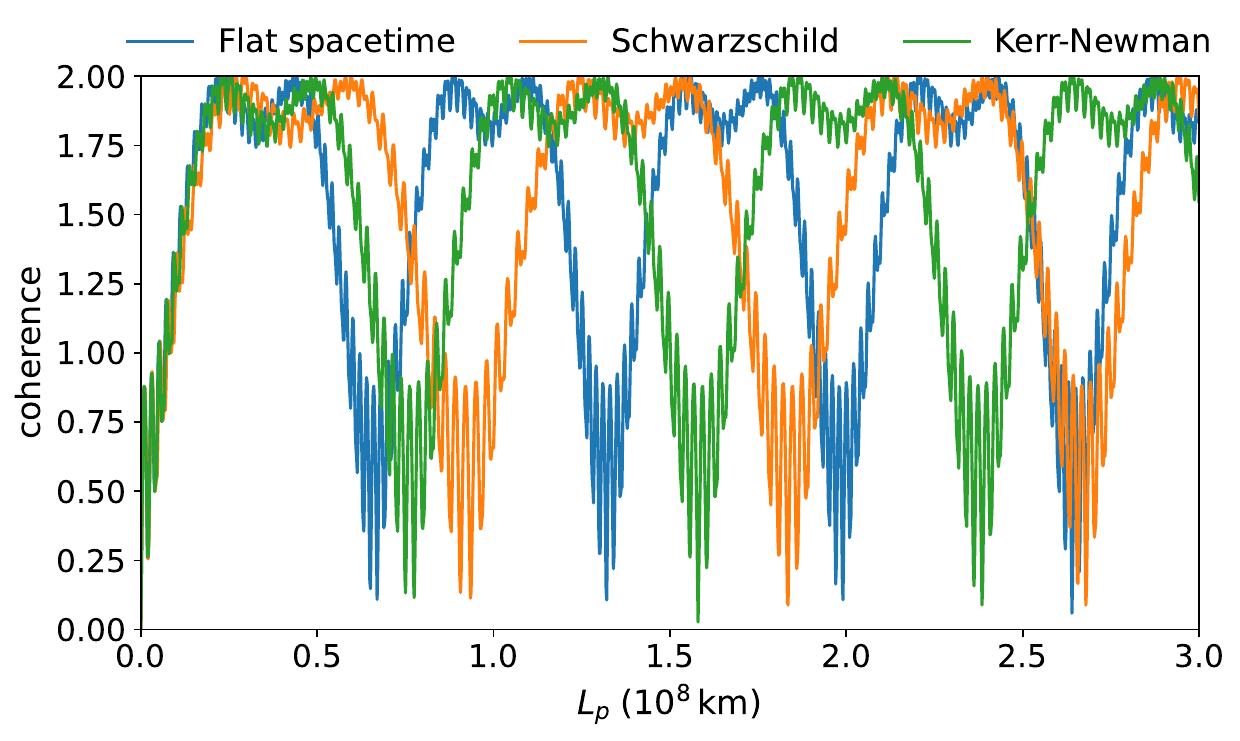}
  \end{minipage}
  \hfill
  \begin{minipage}{0.48\textwidth}
    \centering
    \includegraphics[width=\linewidth]{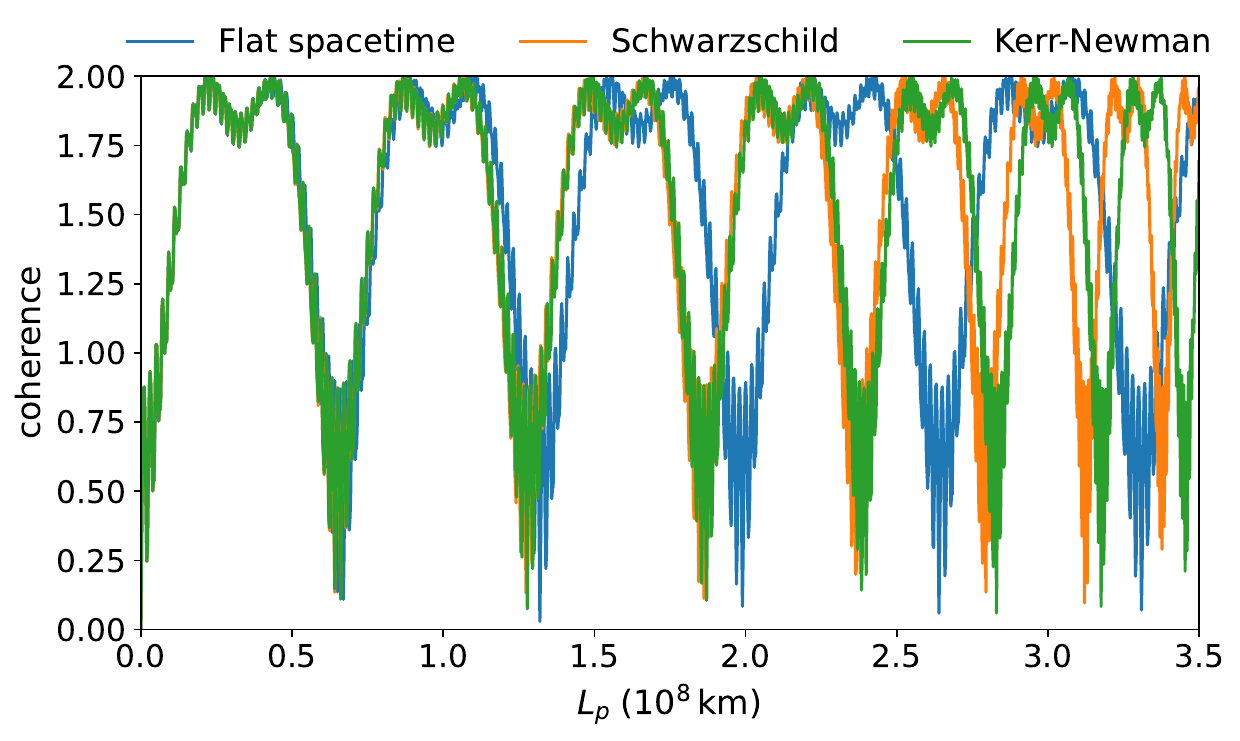}
  \end{minipage}
   \vspace{-3pt}
  \caption{ Quantum coherence of neutrinos in different types of curved spacetime. Left panel: outward propagation; right panel: inward propagation.}
  \label{fig:coherence1}
\end{figure}

\begin{figure}[H]
    \centering
    \begin{minipage}{0.32\textwidth}
        \centering
        \includegraphics[width=\linewidth]{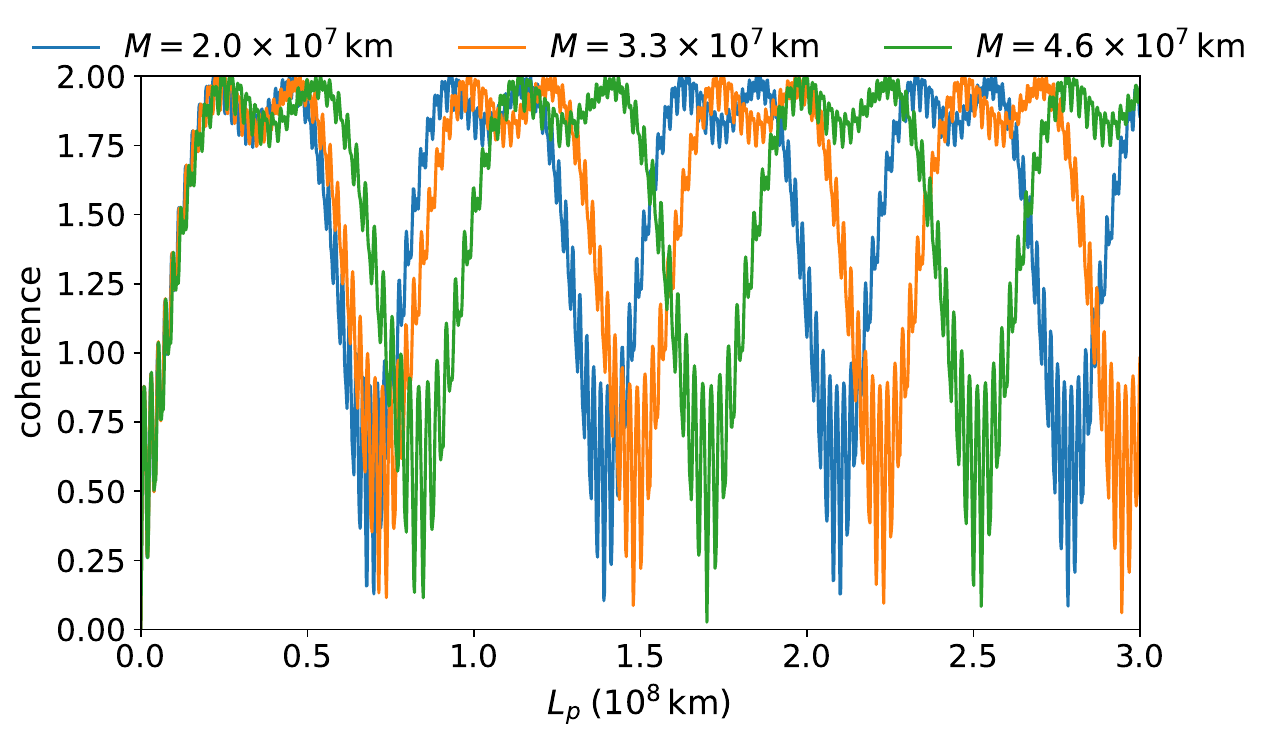}
    \end{minipage}
    \hfill
    \begin{minipage}{0.32\textwidth}
        \centering
        \includegraphics[width=\linewidth]{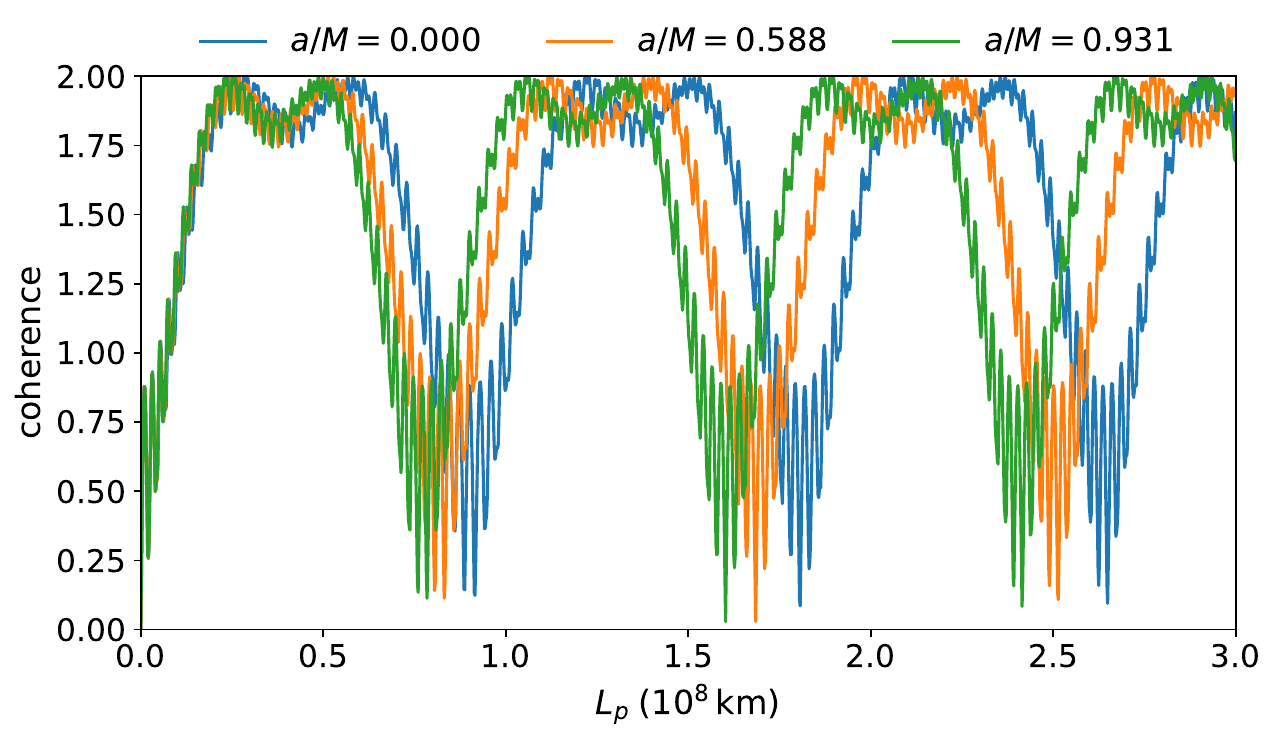}
    \end{minipage}
    \hfill
    \begin{minipage}{0.32\textwidth}
        \centering
        \includegraphics[width=\linewidth]{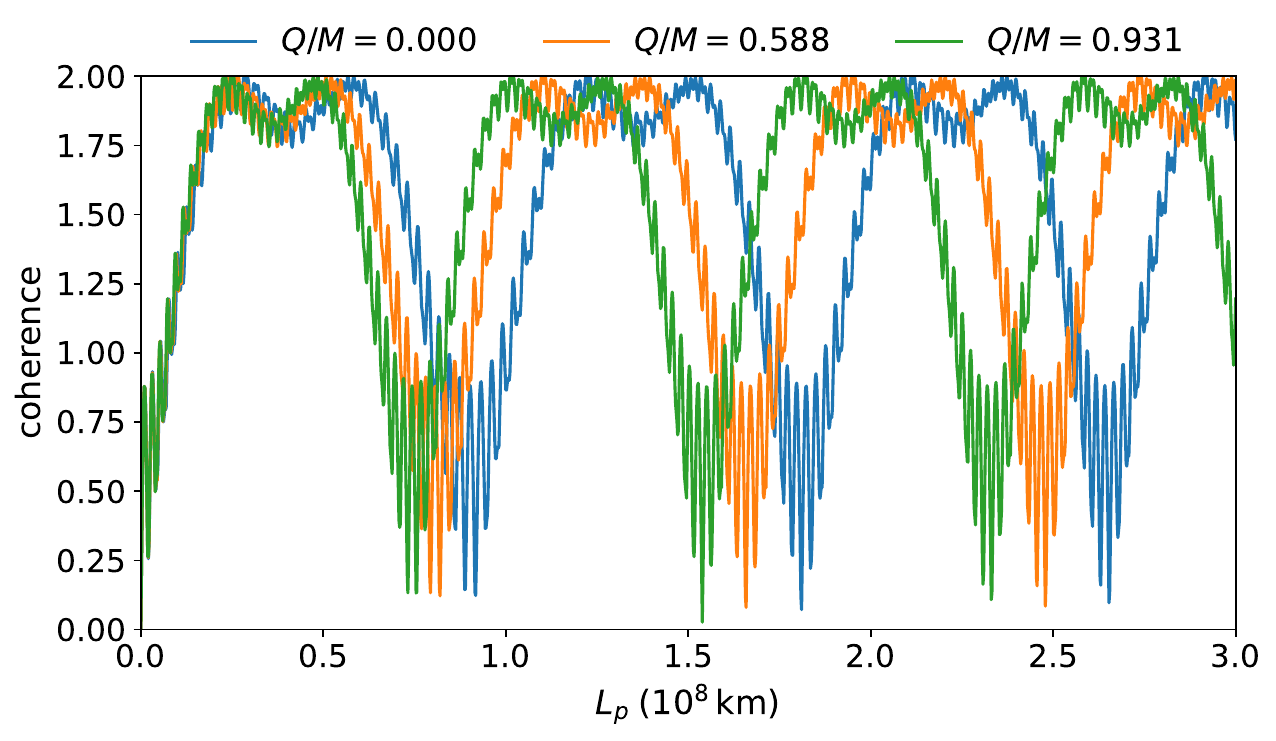}
    \end{minipage}
    \hfill
    \begin{minipage}{0.32\textwidth}
        \centering
        \includegraphics[width=\linewidth]{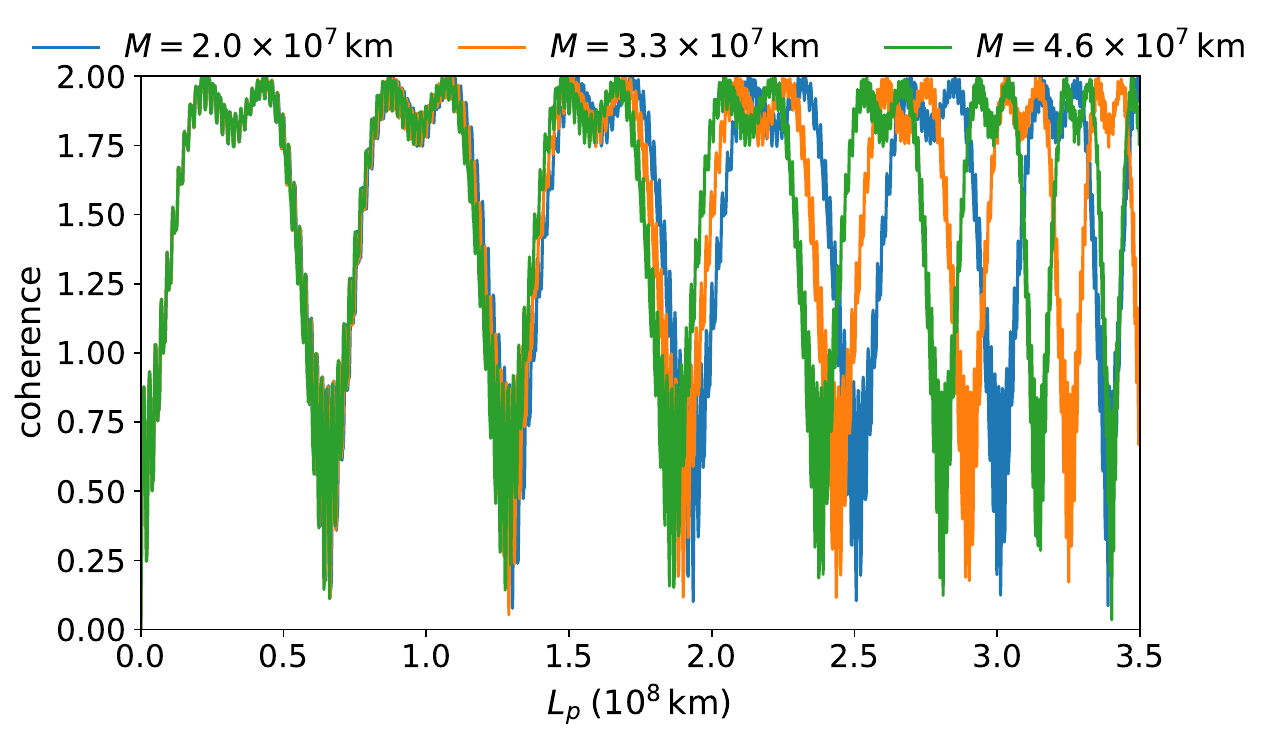}
    \end{minipage}
    \hfill
    \begin{minipage}{0.32\textwidth}
        \centering
        \includegraphics[width=\linewidth]{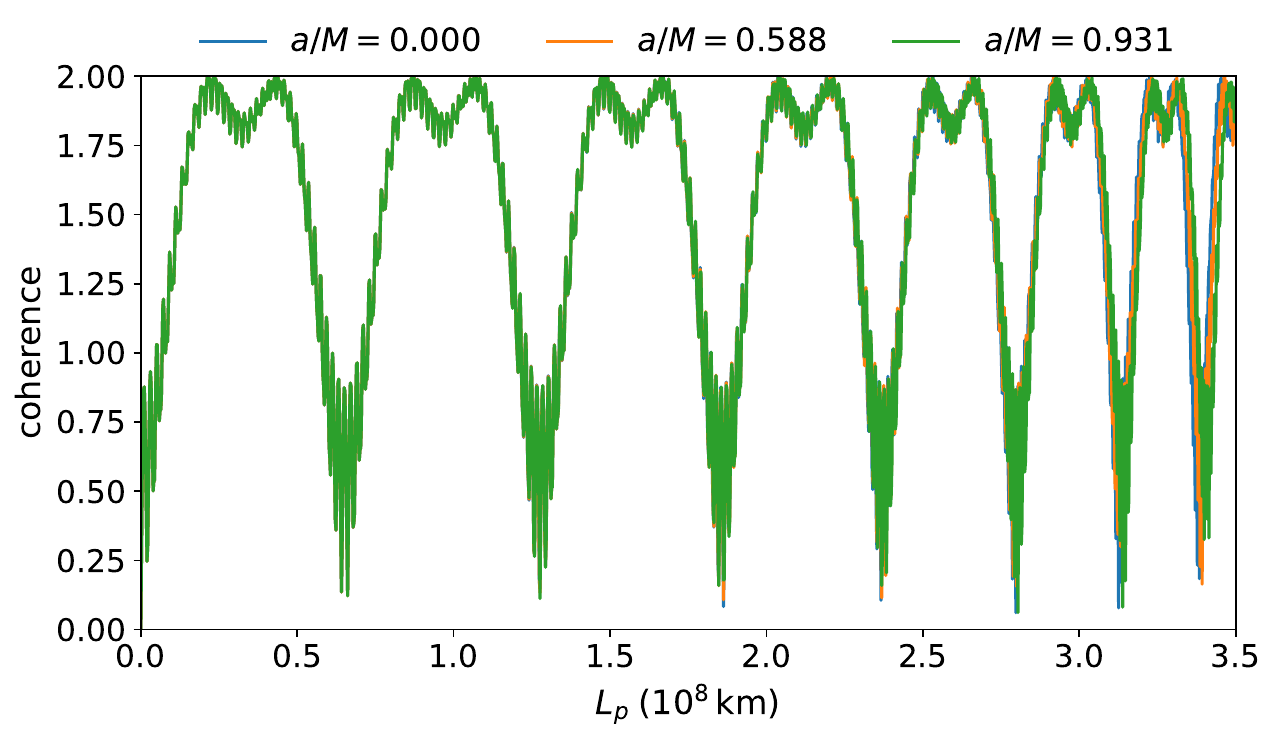}
    \end{minipage}
    \hfill
    \begin{minipage}{0.32\textwidth}
        \centering
        \includegraphics[width=\linewidth]{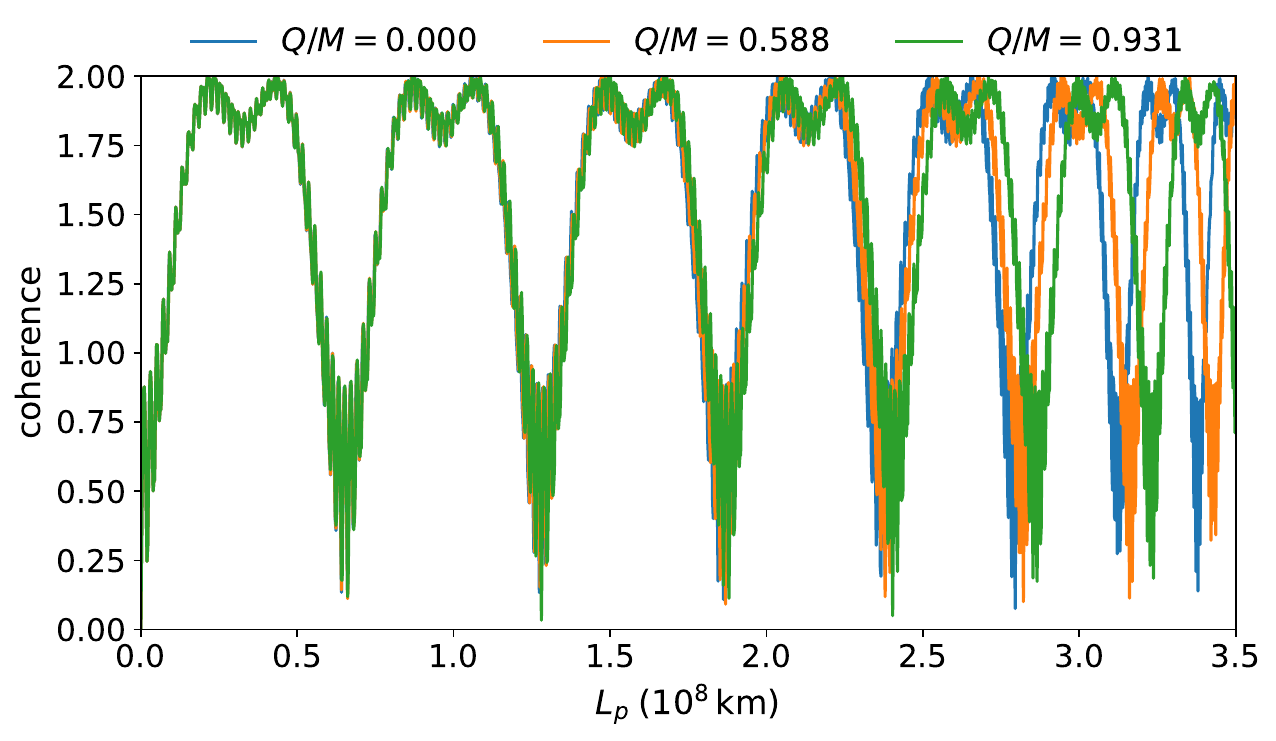}
    \end{minipage}
    \caption{Quantum coherence of neutrinos for different Kerr--Newman metric parameters. The top and bottom rows correspond to outward and inward propagation, respectively.}
     \label{fig:coherence2}
\end{figure}

\begin{figure}[H]
  \centering
  \begin{minipage}{0.3\textwidth}
    \centering
    \includegraphics[width=\linewidth]{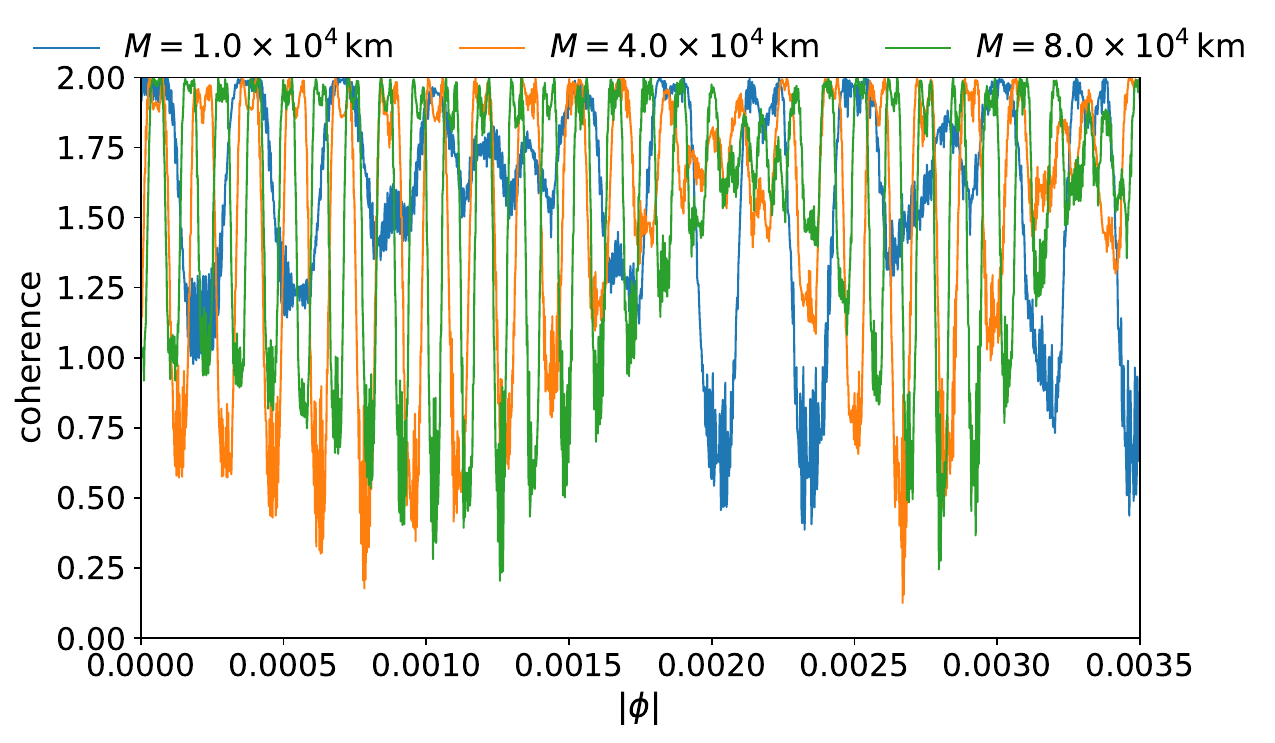}
  \end{minipage}
  \hfill
  \begin{minipage}{0.3\textwidth}
    \centering
    \includegraphics[width=\linewidth]{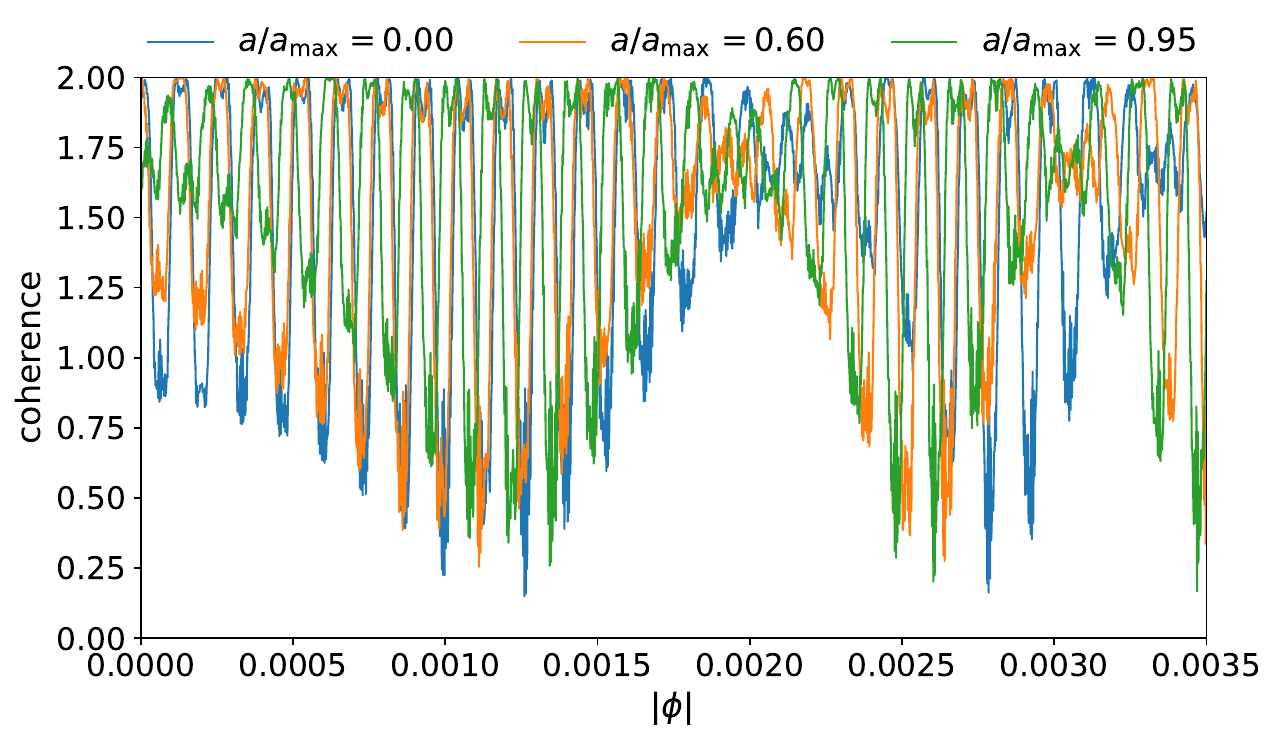}
  \end{minipage}
  \hfill
  \begin{minipage}{0.3\textwidth}
    \centering
    \includegraphics[width=\linewidth]{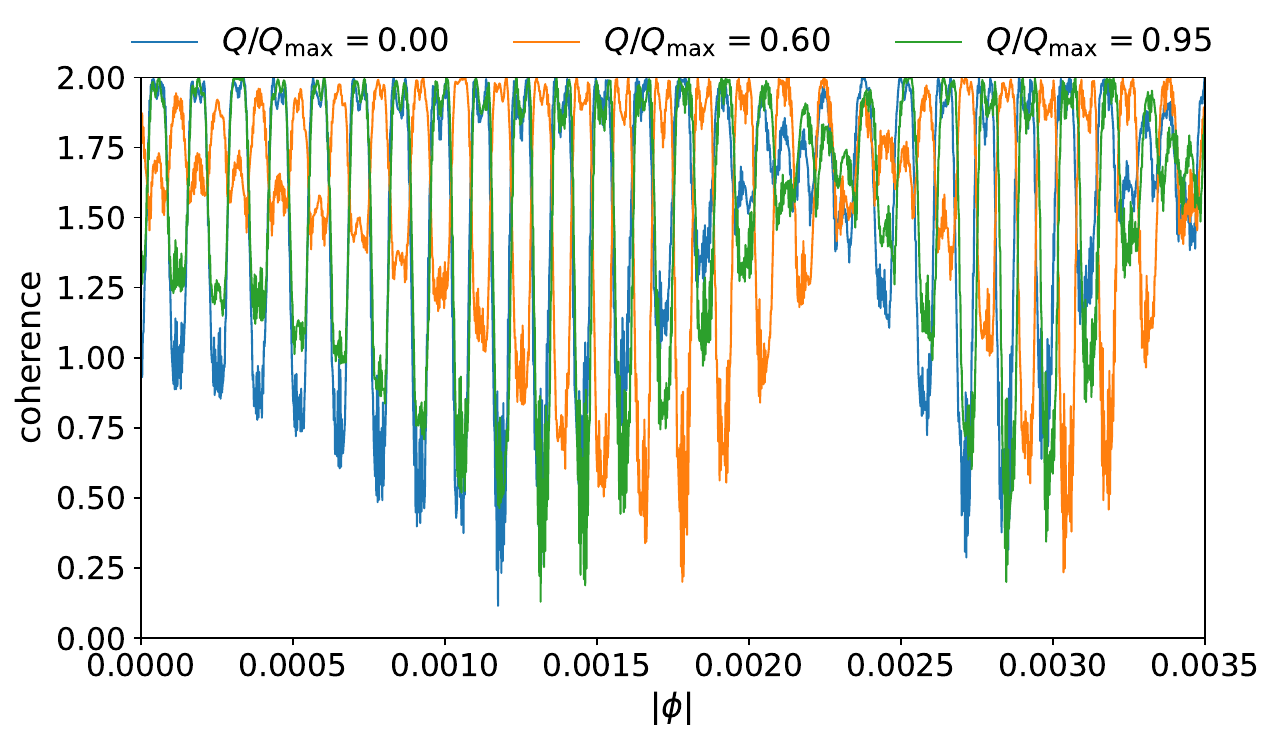}
  \end{minipage}
  \hfill
 \caption{Quantum coherence of neutrinos propagating with nonzero angular momentum under different metric parameters.}
  \label{fig:coherence3}
\end{figure}

 To quantify coherence, we use the \(l_1\)-norm, defined as the sum of the absolute values of the off-diagonal elements \(\rho_{ij}\) of the corresponding density matrix \(\rho\)~\cite{Dixit:2019swl}, namely,
\begin{equation}
    \mathcal{C}(\rho) = \sum_{i \neq j} |\rho_{ij}| \geq 0.
    \label{eq:formula43}
\end{equation}
Considering the initial flavor state \(|\nu_e\rangle\), the coherence is calculated as follows \cite{Song:2018bma}:
\begin{equation}\label{eq:formula44}
C_e = 2 \left(
\sqrt{P_{ee}(L_p) P_{e\mu}(L_p)} +
\sqrt{P_{ee}(L_p) P_{e\tau}(L_p)} +
\sqrt{P_{e\mu}(L_p) P_{e\tau}(L_p)}
\right).
\end{equation}
The parameter settings used above are adopted for both propagation cases. As shown in Figs.~\ref{fig:EOF1}, \ref{fig:EOF2}, \ref{fig:fig13}, and~\ref{fig:coherence1}--\ref{fig:coherence3}, apart from differences in oscillation range, the oscillation patterns of entanglement and coherence are nearly identical. This similarity indicates a close underlying connection between entanglement and coherence. Similar conclusions have also been reported in Ref.~\cite{Ettefaghi:2020otb}, where a natural relationship between the \(l_1\)-norm and concurrence was identified. Therefore, our calculations provide a more general theoretical basis for exploring the relationship between entanglement and coherence.
  \vspace{-6pt}
  \vspace{-20pt}
\vspace{-6pt}

\end{document}